%% file: BPS_MM.tex
\documentclass[11pt]{article}
\usepackage[dvipsnames]{xcolor}
\usepackage{graphicx} 
\usepackage{amsmath}
\usepackage{appendix}
\usepackage{authblk}
\usepackage{amssymb}
\usepackage{float}
\usepackage{slashed}
\usepackage{mathrsfs} 
\usepackage{mathtools}
\usepackage{physics}
\usepackage{subcaption}
\usepackage{cancel}
\usepackage{simpler-wick}
\usepackage{tikz}
\usetikzlibrary{matrix, positioning, arrows.meta, calc}
\usepackage{esint}
\usepackage{cite}

\definecolor{darkblue}{cmyk}{0.9,0.9,0,0}
\usepackage[colorlinks=true,linkcolor=darkblue,citecolor=darkblue,urlcolor=darkblue]{hyperref}

\def\XXint#1#2#3{{\setbox0=\hbox{$#1{#2#3}{\int}$}
     \vcenter{\hbox{$#2#3$}}\kern-.5\wd0}}

\def\nn{\nonumber}
\newcommand{\defeq}{\vcentcolon=}

\title{\textbf{(Un)solvable Matrix Models for BPS Correlators}}
\author[1]{Prokopii Anempodistov}
\author[2]{Adolfo Holguin}
\author[1]{Vladimir Kazakov}
\author[3,4]{Harish Murali}

\affil[1]{Laboratoire de Physique de l’École Normale Supérieure, CNRS, Université PSL, Sorbonne Universités, 
24 rue Lhomond, 75005 Paris, France}
\affil[2]{School of Mathematics, Trinity College Dublin, 17 Westland Row, Dublin 2, Ireland}
\affil[3]{Perimeter Institute for Theoretical Physics, Waterloo, Ontario N2L 2Y5, Canada}
\affil[4]{Department of Physics and Astronomy, University of Waterloo, Waterloo, Ontario, N2L 3G1, Canada}

\newcommand{\vol}[1]{ }

\def\tr{{\rm tr~}}
\newcommand{\la}[1]{\label{#1}}
\newcommand{\beq}{\begin{equation}}
\newcommand{\eeq}{\end{equation}}
\newcommand\beqa{\begin{eqnarray}}
\newcommand\eeqa{\end{eqnarray}}

\newcommand{\p}{\partial}

\begin{document}
\date{}
\maketitle

\begin{abstract}
   We propose and study a family of complex matrix models computing the protected two- and three-point correlation functions  in $\mathcal{N}=4$ SYM.  Our description allows us to directly relate the eigenvalue density of the matrix model for ``Huge" operators with $ \Delta \sim N^2$  to the shape of droplets in the dual Lin-Lunin-Maldacena (LLM) geometry.  We demonstrate how to determine the eigenvalue distribution for various choices of operators such as those of exponential, 
   character, or coherent state type, which then allows us to efficiently compute one-point functions of light chiral primaries in generic LLM backgrounds. In particular, we successfully match the results for light probes with the supergravity calculations of Skenderis and Taylor. We provide a large $N$ formalism for one-point functions of ``Giant" probes, such as operators dual to giant graviton branes in LLM backgrounds, and explicitly apply it for particular backgrounds. We also explicitly compute the correlator of three huge half-BPS operators of exponential type and stacks of determinant operators by reducing them to the known matrix model problems such as the Potts or $O(n)$ model on random planar graphs. Finally, we point out a curious relation between the correlators of $\frac{1}{4}$-BPS and $\frac{1}{8}$-BPS coherent state operators and the Eguchi-Kawai reduction of the Principal Chiral Model in $2D$ and $3D$ correspondingly.
\end{abstract}

\newpage
\setcounter{tocdepth}{2}
\tableofcontents


    
    



\newpage
    
\section{Introduction}   
\textit{What is the geometry dual to a highly excited state in a holographic theory?} One of the main successes of gauge-gravity duality is to be able to address questions such as this one quantitatively, in principle. The general expectation is that operators whose conformal dimensions scale with the cental charge of the model describe bulk configurations where gravitational backreaction cannot be neglected, and that correlation functions in the background of typical huge operators are able to probe gravitational physics in black hole backgrounds. However, there are no models where the field theory description is sufficiently under control to be able to test these ideas precisely. While at strong coupling a putative holographic dual provides an effective field theory description that allows us to organize correlation functions of simple operators, correlation functions involving huge probes or transition amplitudes between different backgrounds are largely inaccessible to conventional techniques such as holographic renormalization. That is just to say that the extrapolate dictionary \cite{Harlow:2011ke, Banks:1998dd} is likely insufficient for describing non-perturbative effects in AdS/CFT . This is due to the fact that operators with sufficiently large conformal dimensions no longer describe local bulk field operator on an $AdS$ background, but are often better described by extended semiclassical objects.
For clarity, let us classify the operators with different conformal dimensions as follows:
\begin{itemize}
    \item $\Delta \sim 1$ : \textit{Light} operators. These correspond to Kaluza-Klein modes in the dual supergravity description~\cite{Witten:1998qj,Gubser:1998bc};
    \item $\Delta \sim N$ : \textit{Giant} operators. These operators correspond to extended objects in the gravity dual (for example, giant gravitons, dual giant gravitons, etc.) \cite{McGreevy:2000cw, Hashimoto:2000zp};
    \item $\Delta \sim N^2$ : \textit{Huge} operators. These operators have large enough conformal dimension to substantially backreact on the background geometry. 
\end{itemize}

Not all huge operators are created equal, and their identification with a non-trivial geometry is not always straightforward. For example one can imagine situations where a state is prepared that describes a highly energetic bound state as opposed to a non-trivial bulk geometry. Examples of such situations include stable orbiting states dual to higher twist primaries, or quantum many body scars \cite{Shiraishi:2017tsj, Pakrouski:2020hym, Milekhin:2023was}.  These situations are qualitatively different from the generic huge state in that the wavefunctions of bulk fields are localized far from the center of the geometry, but without control of the dynamics it can be difficult to make these distinctions precise. Another such situation relevant to gauge theories is that of large charge configurations where the dimensions of operators scale much faster than $N$ \cite{Caetano:2023zwe, Brown:2025cbz, Budzik:2022hcd}. In this limit, the physics simplifies vastly and one can compute the semiclassical background profiles of the fields and systematically work in an EFT.

One way to avoid the difficulties of dynamics of huge operators is to consider topological (BPS) subsectors of a model where the quantum effects in the field theory are under control. The price to pay for this simplification is that the correlation functions do not capture the full gravitational dynamics and the resulting sector is described by something closer to a topological string theory~\cite{Grant:2005qc,Maoz:2005nk,Bershadsky:1993cx, Costello:2018zrm}.   These models are nevertheless still rich enough to describe quantum geometry, and for that reason are a perfect playground to develop the holographic dictionary beyond the perturbative gravitational regime. 
The simplest of such models is the half-BPS sector of $\mathcal{N}=4$ SYM. In this case, the mapping between operators and geometry can be made quite precise with the computation of holographic vevs of general half-BPS geometries serving as an important check of the duality. Comparison between both sides of the duality is made possible by a powerful non-renormalization theorem for OPE coefficients of three half-BPS operators \cite{Lee:1998bxa, Baggio:2012rr}.

The identification between a geometry and the dual half-BPS operator relies on reducing the gauge theory to the eigenvalue dynamics of a single matrix, which is equivalent to $N$ free fermions in a harmonic trap \cite{Berenstein:2004kk}. The half-BPS state wavefunctions can be associated with Fermi droplet configurations on a 2d plane. This description in terms of droplets precisely matches the boundary conditions for half-BPS solutions of type IIB supergravity \cite{Lin:2004nb}. The metric and field profiles of such solutions are uniquely determined from a potential function determined from the density of fermions on the plane. The Fermi sea of the fermions is identified with the regions of the spacetime on the boundary of which the topology change occurs \cite{Lin:2004nb} in the dual description. Since this relies on an understanding of the collective eigenvalue variables as opposed to the gauge invariant operators of the gauge theory it may well appear that constructing the state dual to a particular geometry is challenging and that it requires the knowledge of an infinite number of correlation functions with simple operators. This problem was addressed for some domains in \cite{Vazquez:2006id} using complex matrix model methods. In \cite{Berenstein:2022srd, Holguin:2023naq} a different class of coherent state operators were introduced to create controllable  semiclassical states but their geometric dual wasn't clear at the moment. Part of our goal is to determine the shape of the geometry dual to these operators.

In the field theory the most general half-BPS operator is given by an arbitrary function of multi-traces of a complex matrix field $n \cdot \phi\,(x)$. For concreteness we will restrict our attention to three different families of such operators which may be used to construct different kinds of backgrounds. These are:

\begin{itemize}
    \item  \textit{Orthogonal Schur basis}
    
The most natural  operators in \(\mathcal{N}=4 \) SYM from the point of view of conformal field theory kinematics are those that have definite dimensions and R-charges. These have to be homogeneous functions of scalar fields.
 Since many different operators can have the same charge and dimension, one usually prefers to work in a basis where the two point functions of operators are orthogonal. For the case of $U(N)$ symmetry the natural basis of $\frac12$-BPS operators is the  Schur polynomial basis represented  by homogeneous multi-trace  polynomials \cite{Corley:2001zk}
\begin{equation}
    O_{\Delta,R}(x,n)= \chi_R \left(n \cdot \Phi\,(x)\right),\qquad n\cdot n=0.
\end{equation}
where $\chi_R(X)$ is a $U(N)$ character of element (matrix) $X$ in an irreducible representation $R$. Such operators are labeled by a Young diagram $R$, and the conformal conformal dimension $\Delta$ is equal to the number of boxes in the diagram. Explicitly the Schur polynomials are given by
\beq
    \chi_R(X) = \frac{\det_{ij} x_i^{h_j}}{\Delta(x)},
\eeq
where $h_i = N-i+(\texttt{\# of boxes in the i}^{th}\texttt{ row})$, and $x_i$ are the eigenvalues of $X$. These operators have a rich literature, staring from the seminal~\cite{Corley:2001zk} and their gravity dual was found in \cite{Berenstein:2004kk, Lin:2004nb}. 

\item \textit{Exponential operators}

Another class of $\frac12$-BPS operators  which we will call {\it exponential} in what follows, is given by
\begin{equation}
    O_{\boldsymbol{t}}(Z)= \exp\left[\sum_{k>0} \frac{t_k}{k} \tr[Z^k]\right]= e^{\tr[V_{\boldsymbol{t}}(Z)]},\label{exponentialOpsDef}
\end{equation}
where $V_{\boldsymbol{t}}(Z)$ is often chosen as a polynomial.\footnote{We use all over the paper also the standard complex matrices  notations: $X=\Phi_{1}+i\Phi_{2},\,\,Y=\Phi_{3}+i\Phi_{4},\,\,Z=\Phi_{5}+i\Phi_{6}$. 
} They do not form an orthogonal basis but they can be always expanded in the Schur-type orthogonal operators and generically they lead to non-normalizable states. The non-normalizability of these states is fixed once the contour of integration is chosen appropriately. The computations of correlators of exponential 1/2-BPS operators are often easier than for other types of operators, and they have nice analogues for  other physical problems: two point correlators will be given in terms of two-matrix model  directly related to the Laplacian growth problem~\cite{Zabrodin:2002up} or to the Ising model on dynamical planar graphs~\cite{Kazakov:1986hu,Boulatov:1986sb}. The three-point functions of exponential operators are related to the 3-state Potts model on dynamical planar graphs~\cite{Kazakov:1987qg}. We will explore these analogies in in sections \ref{sec:exp-op} and \ref{sec:3point-Potts} to extract explicit results on correlators of these operators.

Orthogonal states can be made out of exponential operators by introducing bi-orthogonal polynomials associated with the potential $V(z)$ (see Appendix~\ref{coord_exp} for the details):
\begin{align}\label{OrtExp}
O^{V}_{\{m\}}=e^{\tr V(Z)}\,\,\Upsilon_{\{m\}}(Z),\qquad\text{where}\,\,\, \Upsilon_{\{m\}}(Z)= \frac{\underset{1<k,j<N}{\det} P_{m_{k}}(z_{j})}{\Delta(z)} .\end{align}

\item \textit{Coherent state operators}

Another interesting class of $\frac{1}{2}$-BPS states are those generated via coherent states.  For the matrix model formulation of correlators this amounts to the introduction of a background matrix field $\Lambda$ (which can be always chosen diagonal)\cite{Berenstein:2022srd, Lin:2022wdr, Holguin:2023orq, Holguin:2022drf,Holguin:2022zii}: 
\begin{equation}\label{half-BPS CS}
    O_{CS}(Z, \Lambda)= \int_{U(N)} \, dU\,e^{\Tr[UZU^\dagger \Lambda]}.
\end{equation}
 The unitary integral over $U$ can be done by the Itzykson-Zuber-Harish-Chandra formula and the answer represents a complicated function of multi-traces of $Z$ and $\Lambda$ matrices  (separately!). In section \ref{coherentStatesSection}, we explicitly determine the shape of the droplets for some choices of $\Lambda$.

\end{itemize}

A more complicated story are the cases of $\frac{1}{4}$-and $\frac{1}{8}$-BPS operators in $\mathcal{N}=4$ SYM. There is a proposed geometric picture for $\frac{1}{4}$-and $\frac{1}{8}$-BPS states \cite{Chen:2007du, Donos:2006iy, Lin:2010nd}, but a precise dictionary between gauge theory operators and their corresponding geometries is still lacking. This is because the construction of such states on both sides of the duality is technically difficult. At strong coupling the corresponding geometries arise from solving fully non-linear partial differential equations of Monge-Ampere type \cite{Donos:2006iy,Chen:2007du}, while at weak coupling constructing an orthogonal basis of such operators involves intricate combinatorics \cite{Pasukonis:2010rv, Lewis-Brown:2020nmg, Kimura:2010tx, deMelloKoch:2024sdf}. It has also been argued that the effective dynamics of the matrix degrees of freedom should reduce to a strongly coupled Coulomb gas. On the gauge theory side the difficulty of the problem can be circumvented by giving up exact orthogonality of the states.  
The idea is to introduce a set of coherent states that are annihilated by the one-loop dilation operator of the model \cite{Berenstein:2022srd}:
\begin{align}\label{CSoperators}
     O_{CS}(X,Y,Z; \Vec{P})= \int [dU]_{U(N)} e^{\Tr[UXU^\dagger P_{X}+UYU^\dagger P_{Y}+UZU^\dagger P_{Z}]},
\end{align}
where $P_{X,Y,Z}$ are a set of three commuting matrices of coherent state parameters. 
It is trivially annihilated by the one-loop dilatation operator of \(\mathcal{N}=4 \) SYM \cite{Beisert:2004ry} and it has been checked that the $\frac{1}{2}$ BPS case is annihilated by the two-loop dilatation operator \cite{Lin:2022wdr}.

General three-point correlation functions involving $\frac{1}{4}$- and $\frac{1}{8}$-BPS operators are not protected by supersymmetry. But there exist special cases of protected three-point functions,  such as the correlators of one such operator with two arbitrary $\frac12$-BPS operators, or extremal correlators, or those descending from the underlying chiral algebra of the theory~\cite{Bissi:2021hjk}. Nevertheless we expect that their two-point functions carry non-trivial information about the shape of their geometric dual.

The simplest way of probing a geometry is to measure the linear response of light fields moving on the curved background \cite{Balasubramanian:1998de}. These are associated to asymptotic values of bulk fields. This is in contrast to correlation functions involving giant gravitons and light fields which were studied in \cite{Lee:1998bxa, Bissi:2011dc, Caputa:2012yj}, and more recently in \cite{Chen:2019gsb, Yang:2021kot, Holguin:2022zii}, which can be addressed using Witten diagram methods. 
In the black hole context we can associate the information of these correlators with horizon physics. This may be thought of as measuring the overall macroscopic description of the state. In the case of bubbling geometries we do not have a horizon and we should instead think of one-point functions of light probes as encoding information about the classical shape of the bubbles \cite{Skenderis:2006di, Skenderis:2007yb}. Sharper probes can be taken to be huge modes that move as semi-classical objects that couple to the background fields \cite{Fidkowski:2003nf, Louko:2000tp}. Recently these have been argued to probe the time to reach the black hole singularity in the context of thermal physics \cite{Grinberg:2020fdj}.
If we are to make any of these ideas precise we should develop techniques to address the relevant questions in the large $N$ limit. In this paper we demonstrate these ideas in the context of protected correlation functions in $\mathcal{N}=4$ SYM. We do this by studying protected two- and three-point correlation functions of BPS operators in the large $N$ limit and sometimes at finite $N$ by mapping them to a family of matrix models. In section \ref{CorrelatorsDef} we explain how to map BPS gauge theory correlators matrix integrals. 
In section \ref{Two Point Functions Half BPS} we review the relevant complex matrix model techniques needed to evaluate two-point functions of $\frac{1}{2}$-BPS operators in the large $N$ limit. Particularly we explain how to the different types of operators can be tuned to create arbitrary eigenvalue distributions. 
In section \ref{Light probes half BPS} we show how to compute the expectation value of arbitrary light probes on a general half-BPS background. We give a explicit formulas for single-trace chiral primaries of arbitrary dimension, perfectly reproducing the results of \cite{Skenderis:2007yb}. 
In section  \ref{Giant Probes} we consider correlators involving giant graviton probes. We show how to compute a special class of these correlators at finite $N$ and connect the large $N$ asymptotics to a saddle point problem. In section \ref{HHH Correlators} we study saddle point configurations for some simple correlators of three huge operators using the tools introduced in \cite{Eynard:1992cn, Eynard:1995nv, Eynard:1995zv, Eynard:1999gp}. In  particular we solve the case of three stacks of determinants i.e. $\langle\det (n_1\cdot\Phi(x_1))^K\det (n_2\cdot\Phi(x_2))^K\det (n_3\cdot\Phi(x_3))^K\rangle$ when $K=O(N)$. The solution is very explicit and is governed by a cubic spectral curve with singularities. Another example of explicitly calculable HHH correlator is the case of three cubic exponential operators \eqref{exponentialOpsDef}. We show that the problem reduces to the matrix model of three-state Potts spins on planar graphs~\cite{Kazakov:1987qg}   which  is also governed by a cubic spectral curve~\cite{KOSTOV1989295,Daul:1994qy,Eynard:1995nv}.
In section \ref{Quarter BPS} we turn our attention to two-point functions of the $\frac{1}{4}$- and $\frac{1}{8}$-BPS coherent state operators. Generally, they lead to complicated unitary matrix integrals which seem to be technically inaccessible with standard techniques. However, we make a curious observation which gives certain hope: the two-point function of 1/4- and 1/8-BPS operators appear to be equivalent to Eguchi-Kawai quenched reduction of the principal chiral model (PCM) in 2 and 3 dimensions, respectively. The former one is known to be integrable~\cite{Polyakov:1983tt,Polyakov:1984et,Wiegmann:1984ec,Fateev:1994dp,Fateev:1994ai,Kazakov:2019laa,Kazakov:2023imu}. This may open a path (still long to go) to solving the PCM via its gravity dual~\cite{Donos:2006iy,Chen:2007du}!

\section{Matrix Integrals for Protected BPS Correlators}
\label{CorrelatorsDef}

In this section, we will construct complex matrix models for protected BPS correlators made out of only bosonic degrees of freedom. Since these are protected by SUSY, we can work at zero coupling. We will work in the Euclidean signature. Starting with the $\mathcal N=4$ Super Yang-Mills action, we set the gauge coupling to zero, and integrate out the gauge fields and fermions to get
\begin{align}
    \langle O_1(x_1) O_2(x_2)\ldots\rangle = \frac1{\mathcal Z}\int \prod_{I=1}^6 \left[D\Phi_I\right]\ e^{-N\int \frac{d^4 x}{(2\pi)^4} \tr \partial_\mu \Phi^I(x)\partial^\mu \Phi_I(x)} \Bigl(O_1(x_1)O_2(x_2)\ldots\Bigr)\,,\label{freeQFTcorrelator}
\end{align}
where $\mathcal Z$ is the normalization and $O_i(x_i)$ are the BPS operators including in this case only six scalar fields $\Phi_{I}(x)$ of $\mathcal{N}=4$ SYM. We have reduced the full $\mathcal N=4$ SYM correlators to a free QFT computation with matrix degrees of freedom. We can compute these by doing Wick contractions and evaluating the double line Feynman graphs. 

For the case of two- and three-point functions of primary operators the spacetime dependence can be factored out completely and the non-trivial part is just the matrix Feynman graphs which we can equivalently compute in a zero-dimensional matrix model. For more general, non-primary operators like the coherent states and exponential operators considered in the introduction, we should treat the space-time part more carefully. Below, we consider some examples to illustrate this. A word about conventions -- in this text, in addition to the usual definition of trace ($\tr M = \sum_i M_{ii}$), we sometimes find it convenient for large $N$ counting to use the normalized trace defined as
\[\Tr M = \frac1N \tr M.\]\,


\subsection{Two-point functions}
All two-point functions of BPS operators are protected.  First let us consider the simplest case of two $\frac12$-BPS operators : $\langle O_1(n_1 \cdot  \Phi(x_1))O_2(n_2\cdot \Phi(x_2))\rangle$, where $n_{1,2} \in \mathbb{C}^6$ are null vectors. The propagator between these two matrices is given by
\begin{equation}
    \wick{\c{n_1\cdot\Phi(x_1)_{ab}}\ \c{n_2\cdot\Phi(x_2)_{cd}}} = \frac{n_1\cdot n_2}{(x_1-x_2)^2} \ \frac{\delta_{bc} \delta_{ad}} N\,,\label{wicks1/2bps}
\end{equation}
where $ab$ and $cd$ are gauge indices and we took the gauge group to be $U(N)$. The computation of the two-point function is therefore equivalent to the following complex  matrix  integral\footnote{We could equivalently think of this as a Hermitian two-matrix model by rotating contours and adding a regulator \cite{Eynard:2005wg}. Note also the $O_{1,2}(Z)$ appearing on the RHS are to be understood as matrix valued functions}
\beq
    \langle O_1(n_1 \cdot \Phi(x_1))O_1(n_2 \cdot \Phi(x_2))\rangle = \int dZ\,d\bar Z\ e^{-N\,\left(\frac{n_1\cdot n_2}{(x_1-x_2)^2}\right)^{-1} \tr Z\bar Z} O_1(Z) O_2(\bar Z)\,,
\eeq
where $Z$ and $\bar Z$ are auxiliary complex conjugate matrices with a kinetic term that follows from \eqref{wicks1/2bps}. Note that when $O_1$ and $O_2$ are primaries, i.e. when they are homogeneous functions of $Z$ and $\bar Z$, we can factor out the spacetime dependence by rescaling the matrices $Z\rightarrow \left(\frac{n_1\cdot n_2}{(x_1-x_2)^2}\right)^{\frac12} Z$ and similarly for $\bar Z$. We will encounter many examples of these $\frac12$BPS two point functions in Section \ref{Two Point Functions Half BPS}.

We can also cast two point functions of $\frac14$ and $\frac18$-BPS operators as matrix integrals. For instance, consider a $\frac18$-BPS operator $O_1(n_1,n_2,n_3;x)$ where by this notation we mean that $O_1$ is made out of the letters $n_i\cdot \Phi(x)$~\footnote{Such operators are not arbitrary functions of these fields and their construction is not a trivial matter. One construction, based on generalizations of characters, is proposed in~\cite{Lewis-Brown:2020nmg}. We will discuss in this paper the $\frac{1}{4}$-BPS and $\frac{1}{8}$-BPS coherent state operators~\eqref{CSoperators}.}. To compute their two-point functions, we need the propagators for individual fields
\begin{equation*}
    \wick{\c{(n_i\cdot \Phi(x_1)_{ab})}\ \c{(m_j\cdot \Phi(x_2)_{cd})}} = \frac{n_i\cdot m_j}{(x_1-x_2)^2} \frac{\delta_{bc}\delta_{ad}}N\,.
\end{equation*}
We then construct a six matrix model with the matrix of propagators being\footnote{Here we choose the polarization vectors such that $n_i\cdot n_j=m_i\cdot m_j=0$.} $p_{i(j+3)} = \frac{n_i\cdot m_j}{(x_1-x_2)^2}$. In particular, for the diagonal case where $m_i = n_i^*$, we have
\begin{equation*}
    \langle O_1(\vec{n},x_1)O_2(\vec{n}^*,x_2)\rangle = \int d^2X\,d^2Y\,d^2Z\ e^{-N(x_1-x_2)^2\,\tr\left[X\bar X+Y\bar Y+Z\bar Z\right]} O_1(X,Y,Z) O_2(\bar X,\bar Y, \bar Z)\,,
\end{equation*}
where we assumed that the null vectors are normalized.
\subsection{Three-point functions}
\label{sec:threePointMatrixIntegrals}
Unlike the two-point functions of BPS operators, not all their three-point functions are protected.  The three-point functions $\langle O_{\frac12}O_{\frac12}O_{\frac12}\rangle$ and $\langle O_{\frac12}O_{\frac12}O_{\frac14}\rangle$ have been shown to be protected non-perturbatively, see \cite{Intriligator:1999ff, Baggio:2012rr} and references within. In \cite{DHoker:2001jzy} it was shown that $\langle O_{\frac12}O_{\frac12}O_{\frac18}\rangle$ does not receive corrections at 1-loop. See \cite{Bissi:2021hjk} for a nice summary of 1-loop protected three-point functions.

Let us consider again the simplest case of three $\frac12$-BPS operators. Following the same logic as above, we can reduce the three-point function to a multi-matrix integral. In this case, it is a Hermitian three-matrix integral with the following kinetic term
\begin{equation}
    -S[M_1,M_2,M_3] = \tr\left[\begin{pmatrix}M_1\ M_2\ M_3\end{pmatrix}
\begin{pmatrix}
0 & p_{12} & p_{13} \\
p_{12} & 0 & p_{23} \\
p_{13} & p_{23} & 0
\end{pmatrix}^{-1}
\begin{pmatrix}
M_1 \\ M_2 \\ M_3
\end{pmatrix}\right]\,,
\end{equation}
where $p_{ij}=\frac{n_i\cdot n_j}{(x_1-x_2)^2}$ is the propagator between operators $O_i$ and $O_j$. Here, one can absorb the propagators by rescaling the matrices as
\begin{equation}
  M_i \;\longrightarrow\; M_i\ \underbrace{\frac{\left(\prod_{j<k} p_{jk}\right)^{\frac12}}{p_{\,i+1,i+2}}}_{\kappa_i}\,,\qquad i=1,2,3\ (\bmod\ 3)\,. \label{rescaleThreePt}
\end{equation}
After this rescaling, the three-point function is given by~\cite{Kazakov:2024ald}
\begin{equation}
    \langle O_1(n_1;x_1)O_2(n_2;x_2)O_3(n_3;x_3)\rangle = \int \prod_{i=1}^3 dM_i\ e^{N\, \tr\left[-\sum_i \frac{M_i^2}2+\sum_{i<j} M_iM_j\right]} \prod_i \left(O_i(\kappa_i\,M_i)\ \kappa_i^{N^2}\right)\label{hermitian3Matrix}\,,
\end{equation}
where the factors $\kappa_i$ are defined in \eqref{rescaleThreePt}. Continuing in a similar fashion, one can engineer matrix models for different protected correlators -- for instance $\langle O_{\frac12}O_{\frac12}O_{\frac14}\rangle$ can be cast as a four matrix integral (two matrices for the $\frac12$-BPS operators and two for the $\frac14$-BPS)\footnote{There is one subtlety for $\langle O_{\frac12}O_{\frac12}O_{\frac18}\rangle$ that we are glossing over -- the naive 5 matrix model that we would write down has a zero mode which can be cured by coupling that zero mode to an auxiliary sixth matrix.}.

\subsubsection*{Complex Matrix Integrals for $\frac12$-BPS correlators}

Before closing this section, let us rewrite the correlator $\langle O_{\frac12}O_{\frac12}O_{\frac12}\rangle$ as a complex matrix integral as opposed to the Hermitian integral in \eqref{hermitian3Matrix}. As we will see, this version is more natural for two-point functions and also for the light probes in the background of two huge operators. The parameters -- complex matrix eignevalues distribution -- can be directly and very generally related to the parameters of LLM metric~\cite{Lin:2004nb}. The connection between complex and normal matrix models and $\frac12$-BPS states in $\mathcal{N}=4$ SYM has been noted before in the context of extremal correlators \cite{Corley:2001zk, Jevicki:2006tr}, but not for the general three-point functions. Consider the action in \eqref{hermitian3Matrix}
\begin{equation}
    -S[Z,\bar Z,M_3] = \text{tr}\left[-\frac{Z^2}2-\frac{\bar Z^2}2-\frac{M_3^2}2+Z \bar Z+ZM_3+\bar ZM_3\right] = 2\ \tr Z\bar Z -\frac12\tr(M_3-(Z+\bar Z))^2\,,
\end{equation}
where we suggestively relabeled $M_1\rightarrow Z$ and $M_2\rightarrow \bar Z$. After rescaling the $Z$'s and rotating the contours, we end up with the complex matrix integral 
\begin{equation}
    \langle O_1O_2O_3\rangle = \frac{1}{\mathsf{N}_{123}}\int d^2Z\ e^{-N\tr Z\bar Z} \tilde O_1\left(Z\right)\tilde O_2\left(\bar Z\right)\textcolor{NavyBlue}{\int dM\ e^{-\frac N2\tr (M+i\frac{Z+\bar Z}{\sqrt 2})^2}\tilde O_3\left(i\sqrt2M\right)}\label{probeEquationHHL}\,,
\end{equation}
where $\tilde O_j(Z)=\kappa_j^{N^2}O_j\left(\frac{iZ\, \kappa_j}{\sqrt 2}\right)$ are rescaled versions of the operators. For primary operators, we can pull out these factors and for coherent states and exponentials, we can redefine the parameters to absorb them. For simplicity of exposition, we will drop the tildes and $\kappa_i$'s in the rest of this text.

Now, let $O_1$ and $O_2$ be two huge $\frac12$-BPS operators scaling with $N^2$ and $O_3$ be a probe. In this case, we can evaluate the $Z$ integral using the saddle point method. The density of eigenvalues $\rho(z,\bar z)$ will depend only on the huge operators -- the light insertion does not affect the saddle point equations. Note here that even operators that scale as $N$ (like $\det Z$) are light probes that do not backreact. The Huge-Huge-Light (HHL) correlator is then given by evaluating the probe in the background eigenvalue density created by two-point function of the huge operators
\begin{equation}
    \langle O_1O_2O_3\rangle = \Bigl\langle\textcolor{NavyBlue}{\texttt{probe}(Z+\bar Z)}\Bigr\rangle_{\text{background sourced by }\langle O_1 O_2\rangle},\label{naive-HHL}
\end{equation}
where $\textcolor{NavyBlue}{\texttt{probe}(Z+\bar Z)}$ is the $M$-integral highlighted in blue in \eqref{probeEquationHHL}, which one should understand as a normal ordering of $O_3$ (see appendix \ref{QtransformApp}).

Let us now address one subtlety in the above discussion -- the term $\textcolor{NavyBlue}{\texttt{probe}(Z+\bar Z)}$ depends on the eigenvalues of $Z+\bar Z$, which is not fully determined by the eigenvalues of $Z$ and $\bar Z$ because generically $[Z,\bar Z]\neq 0$. To treat this more carefully, let us Schur decompose the matrices $Z$ and $\bar Z$, which one can do even when they are not simultaneously diagonalizable. We have
\begin{equation}\label{SchurDec}
     Z = U(z+T)U^\dagger \quad \text{ and } \quad \bar Z = U(\bar z+\bar T)U^\dagger\,,
\end{equation}
where $U$ is a unitary matrix, $z$ is a diagonal matrix of eigenvalues and $T,\bar T$ are strictly upper and strictly lower triangular matrices, respectively. The integration measure decomposes as
\begin{equation}
    dZ d\bar{Z}= dU\prod_{i>j} dT_{ij} dT^*_{ji} \; \prod_{i=1}^N dz_i d\bar{z}_i \, |\Delta(z)|^2.
\end{equation}
The operators $O_1$ and $O_2$ only depend on the eigenvalues $z$ and $\bar z$. Also, the unitary integral factorizes completely which leaves us with
\begin{equation}
    \langle O_1O_2O_3\rangle = \int \prod_i d^2z_i\, |\Delta(z)|^2 e^{-N\sum_i |z_i|^2}O_1(z)O_2(\bar z) \int d^2 T\ e^{-N\,\tr T\bar T} \,\textcolor{NavyBlue}{\texttt{probe}(z+\bar z+T+\bar T)}.\label{finiteNintermediatestep}
\end{equation}
Plugging in the definition of \textcolor{NavyBlue}{\texttt{probe}}, we get
\begin{equation*}
\begin{aligned}
    \int d^2T e^{-N\ \tr T\bar T}\textcolor{NavyBlue}{\texttt{probe}}&= \int dM\, e^{-\frac N2\ \tr(M+i\frac{z+\bar z}{\sqrt2})^2 -\frac N2\ \tr M^2+\frac N2\sum_i M_{ii}^2}\ O_3(i\sqrt2M)\,,
\end{aligned}
\end{equation*}
where we performed the Gaussian integral over $T$ by changing the variables $T\rightarrow T-i\sqrt2M^+$ and $\bar T\rightarrow \bar T-i\sqrt2 M^-$ where $M^\pm$ are the strict upper and lower triangular parts of $M$. Now, the term $\sum_i M_{ii}^2$ in the resulting expression can be written as an integral over auxiliary variables $\sigma_i$
\beq
    e^{\frac N2\sum_i M_{ii}^2} \propto \int \prod_i d\sigma_{i}\, e^{-\frac N2\sum_i (\sigma_i^2 - 2\sigma_iM_{ii})}\,.
\eeq
Plugging this in and defining a diagonal matrix $D = \text{diag}(\sigma_1,\ldots \sigma_N)$ we get
\begin{equation}
\begin{aligned}
    \langle\textcolor{NavyBlue}{\texttt{probe}}\rangle_{T} &= \frac1{\mathcal Z_D}\int dM\, dD\ e^{-\frac N2\,\tr\left[(M-D)^2+\left(M+i\frac{z+\bar z}{\sqrt2}\right)^2\right]} O_3(i\sqrt2M)\\
    &=\frac{2^{-N(N-1)/2}}{\mathcal Z_D}\int dM\,dD\ e^{-\frac N2\ \tr M^2-\frac N2\ \tr D^2}\ O_3\left(iM+z+\bar z+iD\right).\\
\end{aligned}\label{probeTeq}
\end{equation}
Finally, the three point function is given by
\begin{equation}\label{probe_T}
    \langle O_1O_2O_3\rangle =\frac{1}{\mathsf{N}_{123}} \int \prod_{i=1}^N dz_i d\bar{z}_i \, \ e^{-N\ |z_i|^2} \, |\Delta(z)|^2\, O_1(z)O_2(\bar{z})\langle\textcolor{NavyBlue}{\texttt{probe}}\rangle_T\,.
\end{equation}

This is an exact equation, valid at finite $N$.  Now, there are a couple of simplifications that occur at large $N$ for $O_{1,2}$ being $\mathcal{O}(N^2)$and $O_3$ being light. Firstly, the diagonal matrix $D$ in \eqref{probeTeq} can be set to zero because the measure does not have a Vandermonde repulsion to fight the quadratic potential. Secondly, the three-point function now manifestly depends only on the eigenvalues of $z$ and therefore we can evaluate it by first computing the density $\rho(z,\bar z)$ sourced by the two point function $\langle O_1 O_2\rangle$. 

It is convenient to define a matrix integral transform of an operator that we call the ``Q-transform" as follows
\begin{equation}
    \mathbb{Q}_f\left(X\right) = \frac{1}{\mathcal{Z}_M}\int dM\ e^{-\frac {N}2\,\tr M^2} f\left(X+iM\right).\label{qtransformDef}
\end{equation}

The Q-transform should be understood as a normal ordering of the operator. This is because by construction, there are no Wick contractions between matrices $Z$ and $\bar Z$ belonging to $\mathbb{Q}(Z+\bar Z)$ . See Appendix \ref{QtransformApp} for examples of the Q-transform for various cases of interest. Using this definition, the HHL correlator takes the following simple form:
\beq
\begin{aligned}
    \langle O_1O_2O_3\rangle &= \frac{1}{\mathsf{N}_{123}} \int \prod_{i=1}^N dz_i d\bar{z}_i \, \ e^{-N\ |z_i|^2} \, |\Delta(z)|^2\, O_1(z)O_2(\bar{z})\,\mathbb{Q}_{O_3}(z+\bar z)\\
    &=\int d^2z\ \rho(z,\bar z)\;\mathbb{Q}_{O_3}(z+\bar z)\label{complex3point}\,. 
\end{aligned}
\eeq

\section{Two-Point Functions of Huge $\frac12$-BPS operators}
\label{Two Point Functions Half BPS}
In this section, we discuss two-point functions of Huge $\frac12$-BPS operators at both finite and large $N$. First we review two-point functions for characters and exponential operators which are well studied in the literature \cite{Corley:2001zk, Zabrodin:2002up, Zabrodin:2004cc, Teodorescu:2004qm}. We rederive  the form of the support of the eigenvalues and identify it with the LLM droplet (details of this identification, along with the details of the LLM background, can be found in the Appendix \ref{App-densities-LLM}). For coherent state operators, much less is known in the literature, and we derive the support of the eigenvalues (which again gives the boundary conditions that determine LLM background). To the best of our knowledge, this is an original result. 

\subsection{Schur polynomials}
Let us concentrate on the case of the two-point function ($R_3=\varnothing$). 
It will allow us to determine the distribution of $Z$ eigenvalues in the large $N$ limit which will be useful for computing certain three-point functions. The two-point function is given by
\begin{equation}
    \mathsf{N}_{R R'}= \frac{1}{\mathcal{Z}_{\bullet}}\int dZ d\bar{Z}\; e^{-N\,\tr [Z\bar{Z}]} \, \chi_{R}(Z) \chi_{R'}(\bar{Z}).
\end{equation}
Here $\mathcal{Z}_\bullet$ denotes the normalization of the complex Gaussian integral\footnote{Below we will derive the support of eigenvalues for this matrix integral and show that the Schur polynomials create concentric ring distributions. The normalization integral $\mathcal{Z}_\bullet$ then will have disk distribution.}.  This problem is conceptually equivalent to the computation of the partition function of a Ginibre ensemble. Since $[Z,\bar{Z}]\neq0$ we cannot simultaneously diagonalize the matrices using unitary transformations, but instead we can perform the Schur decomposition:
\begin{equation}
    Z= U( z+ T)U^\dagger,\la{schurDecomp}
\end{equation}
where $z$ is a diagonal matrix that agrees with the eigenvalues of $Z$ and $T$ is a strictly upper triangular matrix. One important fact is that $\chi_R(Z)$ only depends on the eigenvalues of $Z$.
In this case the integration over $T$ completely decouples from the integrals over the eigenvalues, which leaves us with\footnote{We will also often use the shifted highest weight $h_i=R_{N-j+1}+j-1$.}
\begin{equation}
\begin{aligned}
  \mathsf{N}_{R R'}&= \frac{1}{\mathcal{Z}_{\bullet}}\int \; \prod_{i=1}^N dz_i d\bar{z}_i \, |\Delta(z)|^2\, e^{-N |z_i|^2} \frac{\det\left(z_i^{N-j+\lambda_j}\right)}{\Delta(z)}\frac{\det\left(\bar{z}_i^{N-j+\lambda_j'}\right)}{\Delta(\bar{z})}\\
  &= \frac{1}{\mathcal{Z}_{\bullet}}\int \; \prod_{i=1}^N dz_i d\bar{z}_i \, e^{-N |z_i|^2} \det\left(z_i^{h_j}\right)\det\left(\bar{z}_i^{h_j'}\right),
\end{aligned}
\end{equation}
which clearly vanishes unless $R=R'$. There are various ways of evaluating this expression. After expanding the determinants and performing a series of relabelings of the eigenvalues we get an expression of the form
\begin{equation}\label{2pointSchur}
     \mathsf{N}_{R R'}=  \frac{\det_{i,j}\left[ \int_{\mathbb{C}}dzd\bar{z} e^{-N |z|^2} z^{h_i} \bar{z}^{h'_j} \right]}{\det_{i,j}\left[ \int_{\mathbb{C}}dzd\bar{z} e^{-N |z|^2} z^{N-i} \bar{z}^{N-j} \right]}= \delta_{R,R'} N^{N(N-1)/2}\prod_i\frac{\Gamma(h_i+1)}{N^{h_i} \Gamma(i)}= \delta_{R,R'} \prod_i\frac{h_i!}{N^{|R|} (N-i)!}.
\end{equation}

However, it is more illustrative to analyze the saddle-point calculation of this matrix integral rather than writing down the exact answer. We will see shortly that the eigenvalues $z_i$ in the large $N$ limit will be uniformly distributed in some region of a two-dimensional plane, which is identified with the LLM droplet. 

Using polar coordinates $z_j= r_j e^{i\phi_j}$, the angular integrals can be done by expanding the determinants as sums over permutations. The only terms that contribute are those for which $h_i= h'_j$. This means that $R=R'$ as expected from orthogonality. When $h_i$ is of order $N$ the remaining integration can be done by saddle point:
\begin{equation}\label{YTrings}
    \mathsf{N}_{R R'}=  \delta_{RR'}\frac{\texttt{Vol}[U(N)]N!}{\mathcal{Z}_{\bullet}}\int \; \prod_{i=1}^N dr_i r_i \, e^{-N r_i^2+ h_i \log r_i^2};
\end{equation}
the saddle points are given by 
\begin{equation}\label{hrrings}
    r_i^2= \frac{h_i}{N}= \frac{N-i+\lambda_i}{N}.
\end{equation}
One can see that the insertion of Schur polynomial operators with large dimensions creates \textit{circular} LLM droplets with a uniform density. 
For Young diagrams with large rectangular blocks this implies that the distribution of $z$ is uniform and has support on a collection of concentric rings, in accordance with the LLM bubbling picture~\cite{Lin:2004nb}:
\begin{figure}
    \centering
    \includegraphics[width=0.8\linewidth]{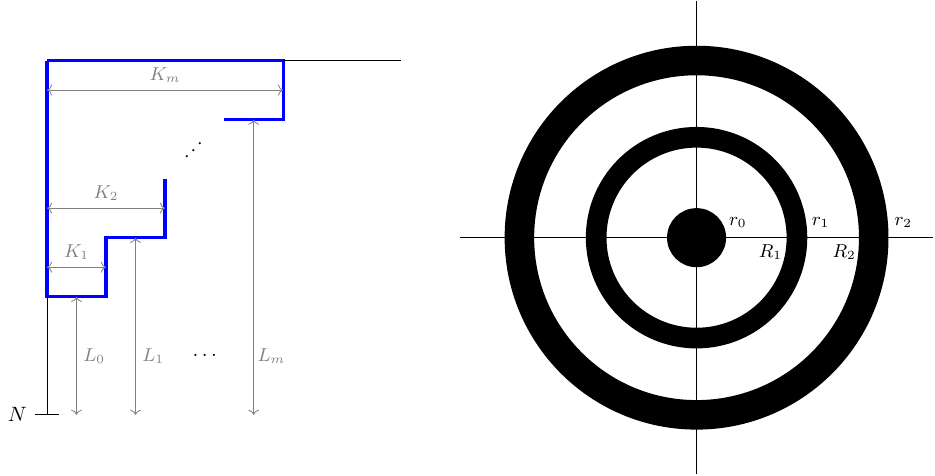}
    \caption{\textbf{Left:} Young tableau consisting of $m$ large rectangular blocks. We parametrize it by the heights of the blocks $L_i$ and by their lenghts $K_i$. \textbf{Right:} An LLM droplet corresponding to this Young tableau (we take the case of 2 rectangular blocks for simplicity). We parametrize the annuli by their inner radii $R_i$ and their outer radii $r_i$. See the paragraph below on the general YT for the expression for $r_i$ and $R_i$ in terms of $K_i$ and $L_i$. See Appendix \ref{App-densities-LLM} for explicit formulas for $r_i$ and $R_i$ in terms of $L_i,K_i$.} 
    \label{fig_YT-Discs}
\end{figure}
See Appendix \ref{App-densities-LLM} for the details of the matching between Young diagrams with large rectangular blocks with the LLM droplets with concentric rings, as well as derivation of the densities and resolvents for the equivalent Hermitian matrix model.

As one can see from \eqref{complex3point}, the latter is important to examine the probes in the background of huge operators, since the probe operators will effectively depend only on $Z+\bar{Z}$. As an example, let us consider the case of a rectangular Young tableau with $N$ rows and $K$ columns. In this case, the corresponding density of eigenvalues will take the shape of an annulus with inner radius $\sqrt{K/N}$ and outer radius $\sqrt{K/N+1}$. Projecting this distribution onto the $x$-axis, we will obtain the following density in the x-space (see Appendix \ref{App-densities-LLM} for the details of derivation):
\begin{align*}
\rho(x)=\frac{2}{\pi} \bigg( \sqrt{r_2^2 -x^2 }- \sqrt{r_1^2-x^2} \, \Theta(r_1^2-x^2) \bigg),
\end{align*}
where
\begin{equation}
    r_1^2 = \frac{K}{N}, \qquad r_2^2 = \frac{K+N}{N}.
\end{equation}

\begin{figure}[H]
\begin{subfigure}{.5\textwidth}
  \centering
  \includegraphics[width=.7\linewidth]{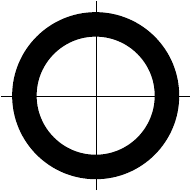}
  \caption{Two dimensional density $\rho_{\mathcal{D}}(z,\bar{z})$.}
  \label{fig:llm_annulus_disc}
\end{subfigure}%
\begin{subfigure}{.5\textwidth}
  \centering
  \includegraphics[width=1.0\linewidth]{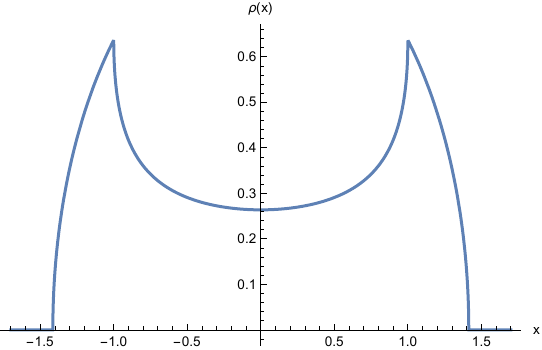}
  \caption{Projection onto the $x$-space.}
  \label{fig:rho(x)-annulus-disc}
\end{subfigure}
\caption{Two-dimensional density of eigenvalues $\rho_{\mathcal{D}}(z,\bar{z})$, corresponding to the Schur polynomial operator with rectangular Young tableau with $N$ rows and $N$ columns, and it's projection onto the $x$-axis. }
\label{fig:annulus}
\end{figure}

Note, that this $x$-distribution agrees with the one obtained in \cite{Kazakov:2024ald}.


\subsection{Exponential operators}
\label{sec:exp-op}

The two-point correlator of exponential operators~\eqref{exponentialOpsDef} takes the form of the partition function of the complex matrix model\footnote{We assume that the contour of integration is chosen such that the integral converges, see for instance \cite{Bleher_Kuijlaars_2012}. This choice of contour should be taken as part of the definition of the operator itself.}
\begin{align}\label{Zcurl}
\langle O_{V}\bar{O}_{V}\rangle&= \int dZ d\bar{Z}\; e^{-N\,\tr \left[Z\bar{Z}-V(Z)-\bar{V}(\bar{Z})\right]} \, \notag\\
&= \int \prod_k d\mu(\bar{z}_{k},z_{k}) \,|\Delta(z)|^{2}\,,\qquad d\mu(\bar{z},z)=dzd\bar{z}\,e^{-N\, \left[z\bar{z}-V(z)-\bar{V}(\bar{z})\right]}\,.
\end{align}
where \( V(z) = \sum_{k=1}^{\infty} t_k z^k \)  and in the second line we used the Schur decomposition \eqref{schurDecomp}.\footnote{The partition function for this matrix model has equivalent eigenvalue representation to the normal matrix model where $[Z,\bar{Z}]=0$ (but not for any other physical quantity!). It has a rich literature and applications list, such as for the problems of Laplacian growth~\cite{Zabrodin:2002up,Zabrodin:2004cc,Kazakov:2002yh}, as well as for the description of time-dependent backgrounds of $c=1$ non-critical string theory~\cite{Alexandrov_2003,Betzios:2025eev}}
At large \(N\) this correlator can be computed via the saddle point. The saddle point equation (SPE) for density of e.v.'s takes the form
\begin{align*}
\int dz'd\bar{z}'\rho(z',\bar{z}')\log|z-z'|^{2}\equiv\pi(\p_{\bar{z}}\p_{z})^{-1}\rho(z,\bar{z})
=|z|^{2}-V(z)-\bar{V}(\bar{z}).
\end{align*}
Acting on both parts by Laplacian \(\p_{\bar{z}}\p_{z}\) we conclude that the density is constant in a domain \(\mathcal{D}\):\begin{align*}
\rho(z,\bar{z})=\begin{cases} 1/\pi, & z\in \mathcal{D}\ \\
0, & z\notin \mathcal{D}\,. 
\end{cases}
\end{align*}  
\vol{Refer to LLM appendix}
This constant density is to be identified with the LLM droplet, which uniquely defines the $10d$ $\frac12$-BPS background (see the details of the identification in the Appendix \ref{App-densities-LLM}). 
Acting on SPE only by \(\p_{\bar{z}}\), integrating over the domain \(\cal{D}\) and applying the Stokes theorem we express the couplings of the potential through the harmonic  moments of the domain: 
\begin{equation} \label{harmom}
    t_k = \frac{1}{2 \pi i k} \oint_{\partial \mathcal{D}} {dz\,\bar{z}(z)}  z^{-k},
\end{equation}
 and similarly for \(\bar{t}_{k}\).

The problem can be solved  for the general set of couplings in terms of the algebraic curve, as was suggested in~\cite{Kazakov:2002yh,Krichever_2004}. That  general construction admits  a multiply connected domain \(\mathcal{D}\) and a generic set of filling fractions of individual sub-domains.  We will concentrate here on the single domain solutions where the algebraic curve can be parameterized as the  following conformal map from the exterior of the unit disc to the exterior of the domain $\mathcal D$
\begin{align}\label{confparam}
z(w) = rw+\sum_{k=0}^{n-1} \frac{a_{k}}{w^{k}} ,
\end{align}
for a polynomial potential \(V(z)\) of degree \(n\). Here \(r\) is the so-called conformal radius. Such an algebraic curve defines a \(2d\) manifold \(\mathbb{C}P^{1}\) of  spherical topology. 
 The shape of the domain in real coordinates is given by the parameterization  
\begin{align*}
\p\mathcal{D}=(x(\theta),p(\theta)):\begin{cases}
    x(\theta)&=\frac12 (z(e^{i\theta})+\bar{z}(e^{-i\theta}))\\
    p(\theta)&=\frac1{2i} (z(e^{i\theta})-\bar{z}(e^{-i\theta}))
\end{cases}
\,,\qquad \quad \theta\in [0,2\pi). 
\end{align*}
Plugging \eqref{confparam} into \eqref{harmom} we obtain  
\begin{equation} 
    t_k =-     \underset{w=\infty}{\text{Res}}\frac{z'(w) \bar{z}(w^{-1})}{k[z(w)]^{k}},\qquad k=0,1,2,\dots\,.
\end{equation}
From here we can express the area of the domain as follows
\begin{align*}
 t_{0} = r^2 - \sum_{j=1}^{n-1} j |a_j|^2.
\end{align*}

This allows to express the data of the curve \(\{r,a_{0},a_{1},a_{2},\dots\}\) through the couplings of the potential \(\{t_{0},t_{1},t_{2},\dots\}\), where the number of the eigenvalues \(N\) is proportional to the area \(t_{0}\). We will keep the area as a free variable. It can be always introduced as a coefficient in front of the first term in the matrix model action:  \(\tr [Z\bar{Z}]\to\frac{1}{t} \tr [Z\bar{Z}]\), so that we can fix one of the couplings, for example \(t_{n}=\frac{1}{n}\), without a loss of generality. 

To compute this correlator in the saddle point approximation we have to compute the free energy \(\mathcal{F}=\frac{1}{N^{2}}\log \mathcal{Z}\). The best is to use an elegant formula from~\cite{Wiegmann:2003xf} for its 2nd derivative in the area:   
\begin{align}\label{FreeE}
\p^{2}_{t_{0}}\mathcal{F}=\log r^{2}\,.
\end{align}

One can find in the literature the construction of algebraic curve and computation of the are for  a few particular  examples of potentials~\cite{Teodorescu:2004qm}: Gaussian \(V(Z)=t_{2}Z^{2}\) (ellipse domain \(\mathcal{D}\)),  cubic \(V(Z)=t_{3}Z^{3}\) (hypotrochoid domain), and the potential with a pole at finite position: \(V(Z)=-\alpha\log(1-Z/\beta)-\gamma Z\) (wing of aircraft domain). 

\subsubsection{Critical regimes}

For particular values of coupling constants the algebraic curve can develop singularities leading to the formation of cusps (``beaks") in the shape of the curve \cite{Teodorescu_2005,Bettelheim_2005}. It happens when a singularity  of the curve, generically positioned somewhere in the complex plane, touches the boundary of the droplet. Tuning further the parameters of the potential we can also make \(m\) singularities touching the boundary of the droplet at the same point \footnote{Multi-critical points of that kind have been first introduced in~\cite{Kazakov:1989bc} for the one-matrix model and in~\cite{Daul:1993bg} for the 2-matrix model, in the context of their applications to the 2d quantum gravity with various matter fields.}.  At such multi-critical point the curve satisfies the criticality condition 
\begin{align*}
z^{(k)}(w_{c})=0,\,\,\,(k=1,2,\dots,m-1),\quad \text{for }\,\,|w_{c}|=1 .
\end{align*}
At these values of parameters, the area \(t(r)\) develops a higher order zero in the radius:  \(t^{(m)}(r_{c})=0 \), so that  \(t_{c}-t=C \left(r-r_{c}\right)^{m}+O\left(\left(r-r_{c}\right)^{m+1}\right)\), which gives for the inverse \(\left(r-r_{c}\right)=C'(t_{c}-t)^{\frac{1}{m}}+O\left(\left(t_{c}-t\right)^{1+\frac{1}{m}}\right)\). Plugging it into \eqref{FreeE} we find the critical behavior of free energy (up to an irrelevant constant term):
\begin{align*}
\mathcal{F}=\text{const}\cdot(t_{c}-t)^{2+\frac{1}{m}}+O\left(\left(t_{c}-t\right)^{3+\frac{1}{m}}\right).
\end{align*}

This critical behavior is typical for the matrix models describing the \(2d\) quantum gravity partition function  coupled to various rational matter fields with central charge  \(c_{matter}=1-6\frac{(p-q)^2}{pq}\,,\,\,\,(p>q=1,2,3,\dots)\) ( for the  spherical topology of the \(2d\) ``universe"). The case \(m=2\) describes the pure \(2d\)  gravity ($p=3,q=2$). The case \(m=3\)  can be either Ising matter (free fermions)  \(p=4,\, q=3\) or non-unitary Yang-Lee point \(p=5,\, q=2\), etc (see e.g. the review~\cite{DiFrancesco:1993cyw}).

These results can be generalized to all orders of \(1/N^{2}\) expansion, giving at each order the universal critical behavior, via the double scaling procedure~\cite{Brezin:1990rb,Douglas:1989dd,Gross:1989vs}. Namely, the free energy reads in this regime
\begin{align*}
\p^{2}_{t}\mathcal{F}_{m}\,\sim\, N^{-m-1/2}f_{m}(x),
\end{align*}
where the parameter \(x\sim N^{m/m+1/2)}(t_{c}-t)\)is kept finite in the double-scaling limit \(t\to t_{c},\,\,N\to\infty\). 

The universal function $f_{m}(x)$  is described in the double-scaling limit via  specific reductions of generalized KdV equations~\cite{Douglas:1989dd}. For example, for $m=2$ 
 \(f_{2}(x)\)  satisfies the Painlevé-I equation: \(f''+f^{2}=x\).   For \(m=3\)  the double scaled free energy   satisfies the equation \(f^{3}-ff''-\frac{1}{2}(f')^{2}+\alpha \,f''''=x\), where for the Yang-Lee matter \(\alpha=1/10\) whereas for Ising matter
\(\alpha=2/27\)~\cite{Brezin:1990rb}. 

All these results hint on important opportunity: to study analytically the \(10d\) quantum supergravity effects arising in the corresponding   critical regimes of LLM gravity. It would be interesting to reproduce these results  computing directly the quantum corrections to the LLM gravity.~\footnote{It must be a topological sector of quantum gravity, in the spirit of the B-model and Kodaira-Spencer theory~\cite{Dijkgraaf:2002fc,Bershadsky:1993cx}. }

\subsubsection{Quartic potential example} 
Let us consider now the example of a general quartic potential and study its critical behaviour. In appendix \ref{cubicApp}, we study the cubic potential and its criticality. Consider the potential \(V(z)=\frac{\tau}{2}  z^2+\frac{  1}{4}z^4\). It easy to establish using~\eqref{harmom} that the curve takes the form
\begin{align*}
z(w)=r w-\frac{r \tau }{\left(3 r^2-1\right) w}+\frac{r^3}{w^3},
\end{align*}
and the area \(t\equiv t_{0}\) is
\begin{align*}
t=r^2 \left(-3 r^4-\frac{\tau ^2}{\left(1-3 r^2\right)^2}+1\right)\,.
\end{align*}

{\it Criticality:} The curve becomes singular when the singularity touches the curve. In our case, due to the symmetry \(z\to -z\) it generically happens at \(w=1\). So the formation of the beak happens when 
\begin{align}\label{critz}
z'(1)=\frac{-9 r^5+6 r^3+r \tau -r}{3 r^2-1}=0,
\end{align}
or at (positive) radii 
\begin{align*}
r_{c}=\frac{\sqrt{1\pm\sqrt{\tau }}}{\sqrt{3}}\,.
\end{align*}
Say, around \(r_{c}=\frac{\sqrt{1-\sqrt{\tau }}}{\sqrt{3}}\) the area has the expansion
\begin{align*}
t_{c}-t=C \left(r-r_{c}\right)^2+O\left(\left(r-r_{c}\right)^3\right)
\end{align*}
where \(t_{c}=\frac{2}{9} \left(1-\sqrt{\tau }\right) \left(\sqrt{\tau }+1-2 \tau \right)\) and \(C=8 \left(\tau -3 \sqrt{\tau }+2\right)\).  Solving it for \(r(t)\) and plugging  into the formula for free energy \eqref{FreeE} we integrate twice in \(t\) and obtain the value for free energy near criticality
\begin{align*}
\mathcal{F}_{m=2}=\text{reg}(t)+\frac{8 }{15\sqrt{C}   r_{c}} (t_{c}-t)^{5/2}+O\left(\left(r-r_{c}\right)^{7/2}\right).
\end{align*} 
The part reg\((t)     \) can be replaced by reg\((t_{c})\): it is non-universal and depends on normalization of the correlator. 

\begin{figure}[htb]
 \begin{center}
\includegraphics[scale=0.7]{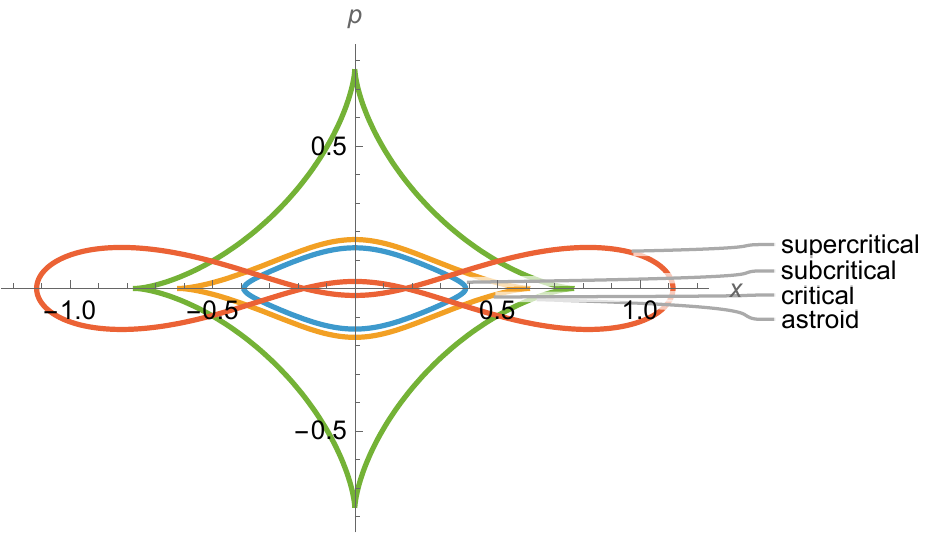} 
\end{center}
 \caption{Shapes of domains \(\mathcal{D}\) (droplets) for different regimes in the complex matrix model: a) Blue: subcritical \(r= (r_{c}-0.1),\,\,\tau=0.4\); b) Orange: critical \(r= r_{c},\,\,\tau=0.4\); c) Red: supercritical (forming intersections) \(r= (r_{c}+0.118),\,\,\tau=0.4\); Green: ``astroid" \(r= r_{c},\,\,\tau=0\).}
 \label{Fig:quartic_curves}
 \end{figure}

On the Fig.\ref{Fig:quartic_curves} we presented the shapes of the droplets for various regimes (choices of couplings): subcritical, critical, supercritical and purely cubic critical potential. 

{\it Double criticality:}
When we impose, in addition to
\eqref{critz}, the double-criticality condition
\begin{align}\label{critz2}
z''(1)=-48 (r^2 + 3 r^4)=0,
\end{align}
we obtain an \(m=2\) critical point with parameters \(r_{*}=\pm\frac{i}{\sqrt{3}}\,,\,\,\,\tau_{*}=4\,,\,\,\, t_{*}=\frac{10}{9}\).
 We have
 \begin{align*}
t_{*}-t=C \left(r-r_{*}\right)^{3}+O\left(\left(r-r_{c}\right)^4\right),
\end{align*}
which leads to 
\begin{align*}
\mathcal{F}_{m=3}=\text{reg}(t)+C' (t_{*}-t)^{7/3}+O\left(\left(r-r_{*}\right)^{10/3}\right),
\end{align*}
where the constants $C,C'$ can be easily computed. The algebraic curve is fixed at this criticality to
\begin{align}\label{doublec}
z_{*}(w)=\frac{1}{3 \sqrt{3} w^3}-\frac{w}{\sqrt{3}}+\frac{2}{\sqrt{3} w}.
\end{align}
We replaced here \(w\to i\,w\) which simply rotates the picture by $\pi/2$. 

We cannot immediately conclude from here whether this singularity is of Yang-Lee   or Ising type. For that one needs a deeper analysis of the  that matrix model, for example deriving the double scaled equation  via the orthogonal polynomial approach, but it is not so important for our current purposes.           
\begin{figure}[htb]
 \begin{center}
\includegraphics[scale=0.3]{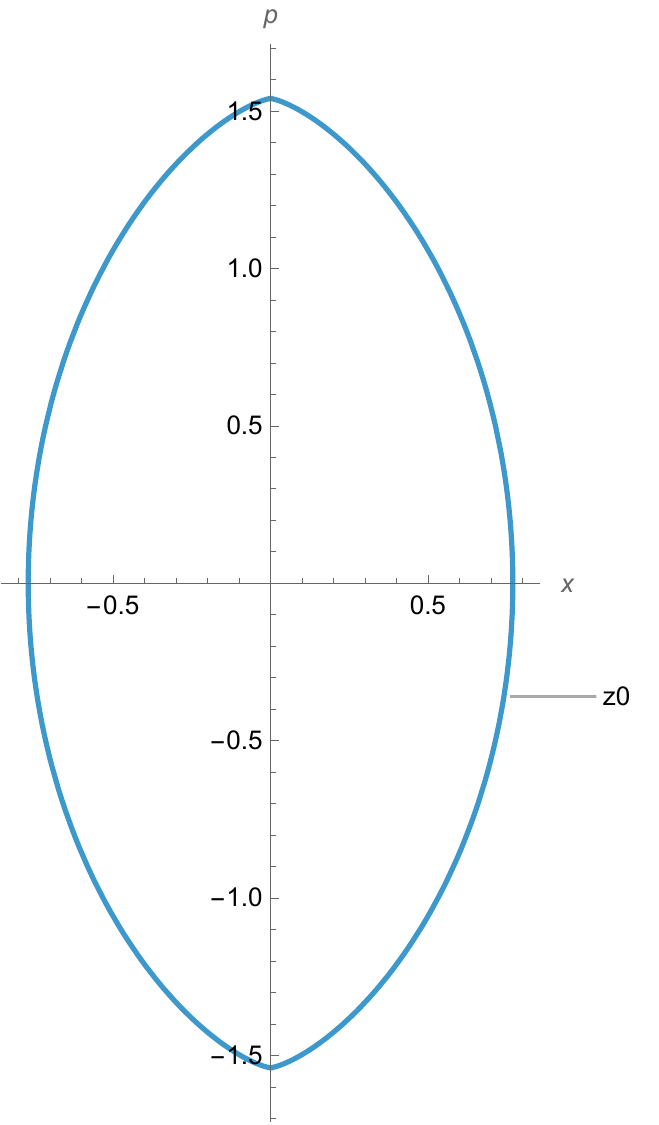} 
\end{center}
 \caption{Shapes of domain \(\mathcal{D}\) (droplet) for the double-critical regime \(m=3\) given by the curve~\eqref{doublec}. Notice that the beaks  (on the top and bottom) are less pronounced than in the \(m=2\) case.        }
 \label{Fig:quartic_curves2}
 \end{figure}

\subsection{Coherent states}
\label{coherentStatesSection}

Another interesting class of half-BPS states are those generated via coherent states \cite{Berenstein:2022srd,Lin:2022wdr, Holguin:2022drf}. This class of states is useful since we can control the placement of the eigenvalues. For the matrix model this amounts to the introduction of a background field $\Lambda=\text{diag}(\lambda_1,\ldots\lambda_N)$
\begin{equation*}
    e^{V(Z)}= \int dU e^{N\,\Tr[UZU^\dagger \bar{\Lambda}]}.
\end{equation*}
The two-point function of these coherent state operators is given by
\begin{equation}
    \mathcal{Z}_{CS} = \int d^2Z\,dU\,dV\, e^{N\,\tr\left[-Z\bar Z + \bar{\Lambda}UZU^{\dagger} + \Lambda V\bar Z V^{\dagger}\right]}\,.\la{12bpsCoh}
\end{equation}
Note that the action is quadratic in the $Z$'s and the integral can be computed by shifting $Z \rightarrow Z+U^\dagger \Lambda U$ and $\bar Z\rightarrow \bar Z + V^\dagger \bar\Lambda V$
\begin{align*}
    \mathcal{Z}_{CS} &\propto \int dU e^{N\,\tr U\Lambda U^\dagger \bar\Lambda}
    = \frac{\det_{ij}e^{N\,\lambda_i \bar\lambda_j}}{\Delta(\lambda)\Delta(\bar\lambda)}\,,
\end{align*}
where we used the HCIZ formula in the last line. While the final result is simple, we are typically interested in the eigenvalue distribution of $Z$ -- this is what encodes information about light probes in the new background. Note that coherent state simplifies the computation of holomorphic moments in the coherent state background:
\begin{equation}
\begin{aligned}
    \langle \tr Z^n\rangle &= \frac1{\mathcal{Z}} \int dU e^{N\,\tr U\Lambda U^\dagger \bar\Lambda} \int d^2Z e^{-N\,\tr Z\bar Z}\ \tr(Z+U^\dagger\Lambda U)^n\\
    &= \tr \Lambda^n. \la{momentsZCS}
\end{aligned}
\end{equation}
Similarly, we have $\langle\tr\bar Z^n\rangle = \tr \bar\Lambda^n$. The same is true for all holomorphic multi-trace operators. The mixed traces, involving both $Z$ and $\bar Z$ are harder to compute due to the residual unitary integrals \cite{Holguin:2023orq}. Note that the problem of determining the CS parameter $\Lambda$ that gives rise to a particular droplet shape is not well defined -- many $\Lambda$'s can lead to the same shape. A trivial example is the following: the moments $\tr \Lambda^n$ are the same when eigenvalues of $\Lambda$ are uniformly distributed inside a disc of \textit{any} radius. Below, we study the inverse problem of determining the shape given harmonic moments $\langle\tr Z^n\rangle$. In general the harmonic moments are insufficient to recover the shape of a $\mathcal{D}$. For some special cases, like when the moments obey quadrature identities, the shape can be determined explicitly. Let us look at some examples.

\subsubsection*{Example: $\rho(z,\bar{z}$) and Quadrature Domains}
Since the coherent state operators are described by a holomorphic (albeit very complicated) potential, the saddle point analysis still applies. We now assume that there exist dominant saddle point configuration, and we denote $\lambda_i$ the eigenvalues of $\Lambda$. Then the coherent state condition guarantees that the expectation value of any holomorphic function is determined by the $\lambda_i$, that is
\beq
    \langle \Tr \, f(Z)\rangle= \frac1\pi\int_{\mathcal{D}} d^2 z\ f(z) =\sum_{k=1}^{n} a_k f(z_k).\label{quadratureDomainDefn}
\eeq
Domains satisfying this condition are known as (pure point) \textit{quadrature domains} \cite{Aharonov_Shapiro_1976}. In our case $f(z)$  can be taken to be any analytic function on $\mathcal{D}$, $a_k$ are some constants and the points $z_k\in \mathcal{D}$ are called the nodes. Note that we can rewrite the LHS using Stokes' theorem as
\begin{equation*}
    \frac{1}{2\pi i}\oint_{\partial \mathcal D} dz\ S(z) f(z)\,,
\end{equation*}
where we introduced the Schwarz function $\bar z=S(z)$ \cite{Davis1974} that defines the boundary  $\partial\mathcal{D}$. Now, one can also characterize quadrature domains as shapes whose Schwarz function is
\beq
S(z) = \sum_{k=1}^n \frac{a_k}{z-z_k} + (\texttt{analytic inside }\mathcal{D})\,.\la{schwarzQuad}
\eeq

The simplest example is a single node, which corresponds to $\Lambda = z_0 I$. In this case, the Schwarz function is simply $S(z) = \frac1{z-z_0}$, which corresponds to a unit disc centered at $z_0$. The unit disc corresponds to (shifted, for $z_0\neq 0$) $AdS_5\times S^5$ geometry \cite{Lin:2004nb}.

Now, consider the case where the eigenvalues of $\Lambda$ are chosen in the following way
\beq
\lambda_i = \begin{cases}
    \alpha_1 \qquad \text{for } 1\le i\le p\\
    \alpha_2 \qquad \text{for } p+1\le i\le N\\
\end{cases}\la{twoEvCS}
\eeq
The moments of $Z$ are then given by
\begin{equation*}
    \langle \Tr Z^n\rangle = \frac pN \alpha_1^n + \left(1-\frac{p}N\right) \alpha_2^n\,.
\end{equation*}
This corresponds to a quadrature domain with nodes at $\alpha_{1,2}$ with weights $\frac pN$ and $\left(1-\frac pN\right)$. The domain for this case is the image of the unit disc under the following conformal map 
\begin{equation*}
    \Phi(w) = \mathsf{N}\left(\frac{w}{\xi_1 w-1} + \frac{\gamma w}{\xi_2 w-1}\right)
\end{equation*}
for an appropriate choice of constants $\xi_{1,2}$, $\gamma$ and $\mathsf{N}$. It is easy to show that the Schwarz function for this domain is of the form \eqref{schwarzQuad}, with
\begin{eqnarray*}
    a_1 &=& \mathsf{N}^2\left(\frac{\gamma }{\left(\xi _1 \xi _2-1\right){}^2}+\frac{1}{\left(\xi _1^2-1\right){}^2}\right)\ , \qquad \qquad z_1 = \mathsf{N}\xi _1 \left(\frac{\gamma }{\xi _1 \xi _2-1}+\frac{1}{\xi _1^2-1}\right)\\
    a_2 &=& \mathsf{N}^2\left(\frac{\gamma^2}{\left(\xi _2^2-1\right){}^2}+\frac{\gamma}{\left(\xi _1 \xi _2-1\right){}^2}\right)\ ,\qquad \qquad    z_2 = \mathsf{N}\xi _2 \left(\frac{\gamma }{\xi _2^2-1}+\frac{1}{\xi _1 \xi _2-1}\right)
\end{eqnarray*}
We can now solve for the parameters of the map by requiring that $\frac{a_1}{a_2} = \frac{p}{N-p}$ and $z_{1,2} = \alpha_{1,2}$. The normalization constant $\mathsf{N}$ 
is fixed by imposing $a_1 + a_2=1$, which follows from \eqref{quadratureDomainDefn}.
\begin{figure}
    \centering
    \includegraphics[width=0.8\linewidth]{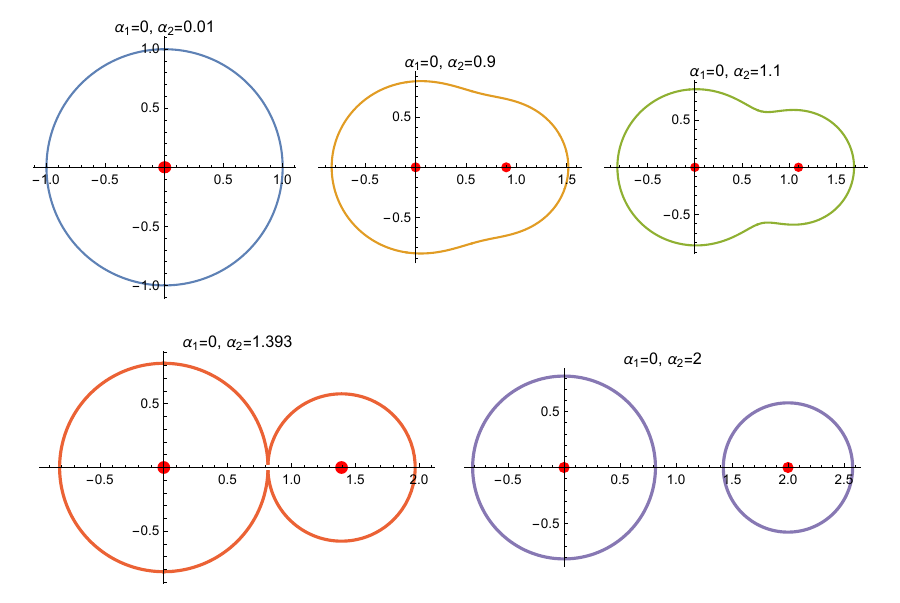}
    \caption{The support of eigenvalue distribution for the coherent state \eqref{twoEvCS} with $\frac pN = \frac23$ for various centers. When $\alpha_2=\frac{1+\sqrt 2}{\sqrt 3}\approx 1.394$, the droplet degenerates into two discs.}
    \label{fig:unequalSpots}
\end{figure}

The case $\gamma=1$ corresponds to $p = \frac N2$. The quadrature domain in this case is called the ``Neumann oval". This is the same as the Ripplon deformation considered in \cite{Skenderis:2007yb} for $n=2$. In figure \ref{fig:unequalSpots}, we plot the boundary of the domain when $p = \frac{2N}3$. Note that the shape degenerates into two discs at the critical value of separation
\beq
(\alpha_2 - \alpha_1)_* = \sqrt{\frac pN} + \sqrt{1-\frac pN}\nn\,.
\eeq
For larger separation, the map $\Phi(u)$ fails but the domain $\mathcal{D}$ is simply two discs centered at $\alpha_1$ and $\alpha_2$. This generalizes in a straightforward way to domains with higher numbers of nodes. For a pure point quadrature domain with $n$ nodes that is simply connected the conformal map from the unit disk to the domain satisfies the relation 
\begin{equation}
    \Phi_n(u)= \sum_{i=k}^n \frac{\bar{a}_k}{\overline{w}_k}\frac{u}{1-u \bar{\lambda}_k}\,,
\end{equation}
where $\lambda_i= \Phi_n^{-1}(z_i)$ and $w_i = \Phi_n'(\lambda_i)$, see for instance \cite{Ameur_2021} and the references therein for a proof. One can go beyond quadrature identities with simple poles by taking superpositions of coherent states. Since we can choose these to be approximately orthogonal in the large $N$ limit by picking for sufficiently different distributions, the expectation values of holomorphic single trace functions can satisfy more general quadrature identities with arbitrary complex weights $a_i$, and possibly with higher poles.

\begin{figure}
    \centering
    \includegraphics[width=0.8\linewidth]{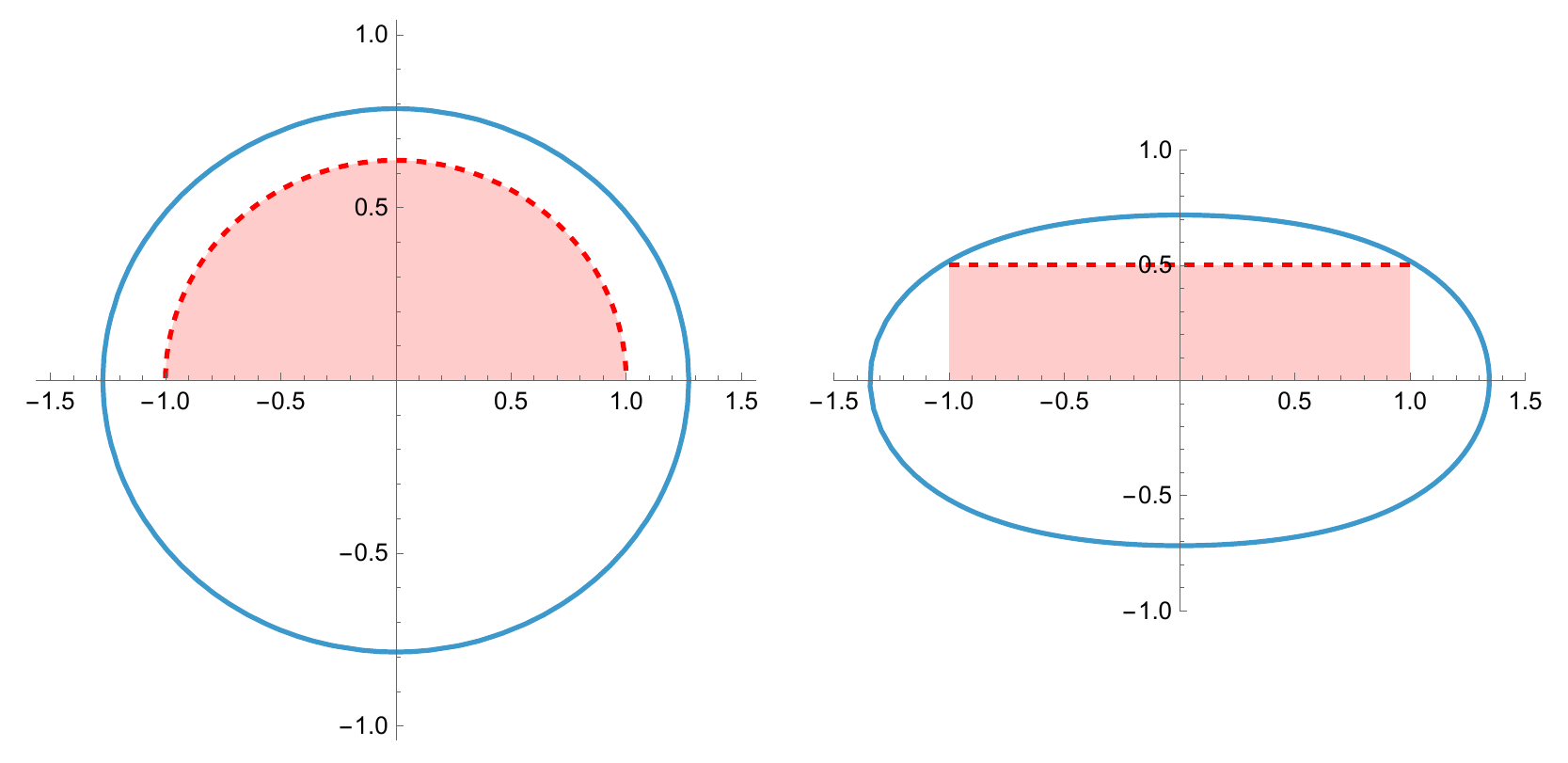}
    \caption{Support of the eigenvalues (shown in blue) for two coherent states, whose $\Lambda$ eigenvalue densities are shown in red. In the figure on the left, the density of $\lambda$ is a semi-circle distribution and the spectral curve is \eqref{spectralCurveSC}. On the right, we have uniform density for $\lambda$ and the spectral curve is \eqref{spectralCurveUnif}.}
    \label{fig:coherentLinear}
\end{figure}

We could also consider generalizations of quadrature domains such that the Schwarz function has branch cuts, and not just poles inside $\mathcal{D}$. For instance, say we distribute the eigenvalues of $\Lambda$ according to the semi-circle law
\beq
    \rho(\lambda) = \frac{2\sqrt{c^2-\lambda^2}}{\pi c^2}\,.
\eeq
In this case, the Schwarz function should have a square root branch cut between $-c$ and $c$. The following function fits the requirement
\beq
    S(z) = \left(\frac{2a^2}{c^2}-1\right)z-\frac2{c^2}\sqrt{z^2-c^2}\la{spectralCurveSC}\,,
\eeq
where $a$ obeys $c^2 = a^2-a^{-2}$. Notice that this is Schwarz function of an ellipse of unit area with focii at $\pm c$, as shown in figure \ref{fig:coherentLinear}.

Let us consider one final example for the coherent states, where the eigenvalues of $\Lambda$ are chosen to be uniformly distributed on a line
\beq
    \lambda_j = \alpha\frac{j-1}{N}\nn\,,
\eeq
for $\alpha > 0$. The Schwarz function has a logarithmic cut from $0$ to $\alpha$. In this case, it is harder to guess the regular part which leads to a closed contour. However, we can use the saddle point equation to get to the result. Note that the coherent state is given by
\[
    \int dU e^{N\,\tr[UZU^\dagger \Bar\Lambda]} \propto \frac{\Delta(e^{\alpha z})}{\Delta(z)}\,,
\]
where we omitted the Vandermode factor $\frac{1}{\Delta(\lambda)}$ which does not affect the saddle analysis. The SPE is
\[
    -\bar z + \alpha  e^{\alpha z} \int \frac{d^2 w}{\pi} \frac{1}{e^{\alpha z} - e^{\alpha w}} = 0\,,
\]
where we have already used that $\rho(w,\bar w) = \frac1\pi$ for $w\in \mathcal{D}$. The integral in the second term can be evaluated using the known moments of $Z$~\eqref{momentsZCS}. Plugging that in, we obtain the following equation for the spectral curve
\beq
    \bar z  = S(z) = \alpha + \frac1\alpha \log\left(\frac{e^{\alpha z}-1}{e^{\alpha z}-e^{\alpha^2}}\right)\,.\la{spectralCurveUnif}
\eeq
It's easy to show that this is the equation of a circle in the variable $w=e^{\alpha z}$. We can also parameterize the curve as follows
\beq
    z(\theta) = \frac1\alpha \log\left(e^{\alpha^2} + e^{i\theta} \sqrt{e^{2\alpha^2} -e^{\alpha^2}}\right)\,.\nn
\eeq
We can easily generalize this to any uniform line distribution of $\lambda$'s by noting that translations and rotations in $\Lambda$ can be absorbed into translations and rotations of the $Z$ integral. 

Quadrature domains are known to be dense in the space of plane domains \cite{Gustafsson_1983}. In the large $N$ limit the number of nodes can become arbitrarily large so we can in principle approximate any shape with a coherent state operator. This is not surprising because coherent states form an over complete basis of the half-BPS Hilbert space.

\subsubsection{Simply connected domains}

To address more general distributions that do not follow a quadrature rule, we need to first solve the problem of simply connected domains. We can then take the direct sum of these simply connected domains to obtain the general case -- this is of course possible because the harmonic moments are linear in the density $\rho(z,\bar z)$. When the droplet is simply connected, we can use numerical methods to reconstruct the shape. Here, we briefly sketch how this can be done. The idea is that simply connected domains can be approximated by polygon with large number of vertices. Furthermore, for an $n$-sided polygon, knowing the first $2n-3$ moments is enough to reconstruct the shape \cite{shapeMoments}. First, we define two Hankel matrices
\begin{equation}
    \begin{aligned}
    [H_0]_{ij} &= \tau_{i+j-2}\ ,\\
    [H_1]_{ij} &= \tau_{i+j-1}\ ,
\end{aligned}
\end{equation}
where $\tau_k = k(k-1) \langle \tr Z^{k-2} \rangle$ and $\tau_0=\tau_1=0$. As shown in \cite{shapeMoments}, the vertices of the polygon $z_j$ are solutions to the following generalized eigenvalue problem
\beq
    H_1 \ket{v_i} = z_i H_0 \ket{v_i}\,.
\eeq
Therefore, if the domain $\mathcal{D}$ can be obtained as the limit of a sequence of polygons, we can reconstruct it by knowing the moments. In order to implement this in practice, we would need to make the problem well conditioned by introducing a regulator, say Tikhonov regularization.
When $\mathcal{D}$ is not simply connected, there exist examples where the domain is not uniquely determined by its harmonic moments \cite{uniquenessShape}. In such cases, we would have to evaluate the action to determine the leading saddle. 
\subsection{Multi-trace operators}
Perhaps the simplest huge operator one could consider is $O(N^2)$ copies of $\Tr Z^k$ where $k=O(1)$. When $k=2$, the distribution is an ellipse as we will see shortly. When $k>2$, the situation is quite different. As pointed out in \cite{Guerrieri:2025ytx}, the leading saddle Abelianizes -- one eigenvalue runs off to infinity\footnote{In general, we can have $O(1)$ number of eigenvalues run off to infinity. All these saddles are degenerate in the large $N$ limit.} and the other eigenvalues are distributed uniformly inside the unit disc as in the vacuum.

First, let us study the case of $k=2$, i.e. $O_2=(\Tr Z^2)^{\frac{\alpha N^2}{2}}$. The SPEs for the two points function $\langle O_2\bar O_2\rangle $ are
\begin{equation}
    -\bar z_i + \frac 1N \sum_{j\neq i}\frac{1}{z_i-z_j} + \alpha N \frac{z_i}{\tr Z^2} = 0\,.\nn
\end{equation}
Now, we can solve this using the mean field method where we say $\langle \Tr Z^2\rangle=c(\alpha)$ which is some constant and then self-consistently solve for $c(\alpha)$. Note that the SPEs above are precisely what we would get if we introduce a potential operator of the form $\exp(\frac{N\alpha}{2c(\alpha)}\, \tr Z^2)$. Proceeding as we did in section \ref{sec:exp-op}, we obtain the following conformal map for an elliptic droplet
\begin{equation}
    z(w) = \frac{c(\alpha)}{\sqrt{c(\alpha)^2-\alpha^2}}\left(w+\frac{\alpha}{c(\alpha)w}\right)\nn\,.
\end{equation}
We now need to impose $c(\alpha)=\langle\Tr Z^2\rangle$, which gives
\begin{equation}
    c(\alpha)^2 = \alpha(1+\alpha)\,.\nn
\end{equation}
Now, let us turn to huge operators of the type $O_k = (\Tr Z^k)^{\frac{\alpha N^2}k}$ for $k>2$. The SPEs are again identical to those of an exponential operator with an effective coupling 
\begin{equation}
    -\bar z_i + \frac1N \sum_{j\neq i} \frac1{z_i-z_j} +  \frac{\alpha N}{\tr Z^k}\,z_i^{k-1}=0\,.\label{speOkAbel}
\end{equation}
As before, we can make an ansatz for the conformal map and self-consistently solve for $c(\alpha)=\langle\Tr Z^k\rangle$. For instance at $k=3$, the distribution is a hypotrochoid. In addition to this saddle, we have a new Abelianized saddle where say $z_1\gg z_{2,3\ldots}$. In this case, in the equation for $z_1$ in \eqref{speOkAbel}, we can drop the Vandermonde repulsion and we have $\tr Z^k \approx z_1^k$. Therefore,
\begin{equation}
    |z_1| = \sqrt{\alpha N}\nn\,.
\end{equation}
In the SPEs for $z_2,\ldots z_N$, we can effectively drop the huge operator term which scales as $N^{1-\frac k2}$ and the problem reduces to the Gaussian model whose eigenvalue density is uniform in the unit disc. The free energy of this saddle at leading order is 
\begin{equation}
    \mathcal F = \mathcal F_{\texttt{disc}} - \frac1N|z_1|^2 + \frac{\alpha}k \log \langle\Tr Z^k\rangle =\mathcal F_{\texttt{disc}} - \alpha\left(1-\frac12\log\alpha\right) + \alpha\left(\frac 12-\frac1k\right)\log N\,.\nn
\end{equation}
For $k>2$, this is the leading saddle whose free energy scales with $\log N$ and it dominates the mean field saddle discussed above where the free energy is $O(1)$. This result can be interpreted as follows: we can expand $O_k$ as a sum over characters, $O_k=\sum_R c_R \chi_R(Z)$ and the two point function $\langle O_k\bar O_k\rangle$ is dominated by characters with $O(1)$ rows containing $O(N^2)$ boxes.  

\section{BPS Probes of LLM backgrounds}
\label{Light probes half BPS}
In this section we will consider expectation values of light half-BPS operators ($\Delta= O(1)$) in the background of two huge half-BPS operators and compare them with the results of gravity calculations~\cite{Skenderis:2007yb}. For this it is important to organize the correlation functions according to the symmetries that are left unbroken by the presence of the huge background. Before giving explicit formulas for these kinds of correlators we will first discuss the constraints imposed by the unbroken supersymmetry algebra on the three-point functions. A general background is created by a linear combination of huge $\frac{1}{2}$-BPS operators of different dimensions -- BPS primaries -- and as such the conformal dimension and $U(1)_R$ charge $J$ are no longer conserved. In the presence of these huge operators the superconformal symmetry is broken down to a  $PSU(2|2)^2\rtimes \mathbb{R}$ subalgebra, where the central generator corresponds to $\Delta-J_R$. Only in the case that the background has a fixed dimension is $J_R$ conserved on its own.
The bosonic symmetry of the background will in general be $SO(4)\times SO(4)_R$, where the first $SO(4)$ is the residual spacetime symmetry. The Ward identities for these symmetries imply that the net $SO(4)\times SO(4)_R$ charges in a three-point correlator must vanish which means that the only operators with non-zero expectation value are those which are singlets under this $SO(4)\times SO(4)_R$. These are for instance scalar operators with vanishing $SO(4)_R$ quantum numbers. Focusing on half-BPS operators, the branching from $SO(6)_R\rightarrow SO(4)_R\times SO(2)_R$ is given by 
\begin{equation}
    [0,\Delta,0]= \bigoplus_{h=0}^{\Delta} \;\;\bigoplus_{j\,=-\left(\frac{\Delta}{2}-h\right)}^{\left(\frac{\Delta}{2}-h\right)}\left[\frac{h}{2}, \frac{h}{2}\right]_{2j},
\end{equation}
where $\left[\frac{h}{2}, \frac{h}{2}\right]_{2j}$ denote the spin $\left[\frac{h}{2}, \frac{h}{2}\right]$ representations of $SO(4)_R=SU(2)\times SU(2)$ with $SO(2)_R$ charge $2j$. The only components of this $SO(6)$ multiplet that contribute are those with vanishing $R$-symmetry spins $h=0$, of which there are $\Delta+1$ such states with distinct charges $2j= -\Delta, -\Delta+2, \dots, \Delta-2, \Delta $.
Since generic third $\frac{1}{2}$-BPS operator is described by a projective null six-vector $n\cdot n=0$ which we can split as $(n_{A}, n_{5,6}) $ with $A=1,2,3,4$.  The residual $SO(4)_R$ symmetry can then be used to set $n_A\propto (i,0,0,0)$ and the remaining degree of freedom is $n_5+ i n_6= e^{i\psi}= q$. This means that we can work in coordinates where the huge background is made out of $Z$'s and the general form of a probe operator will be 
\begin{equation}
    O_\Delta(n\cdot \Phi)\equiv O_\Delta(qZ+ \bar{q}\bar{Z}+ 2i \Phi_1)= \sum_{j\,=-\frac{\Delta}{2}}^{\frac{\Delta}{2}} \;q^{2j}\,O_{\Delta, 2j}+\;\;\texttt{non-singlet parts}.
\end{equation}
The operators $O_{\Delta, 2j}$ are $SO(4)_R$ invariant half-BPS operators with $SO(2)_R$ charge $2j$. The expectation value of this operator in a general half-BPS background $\ket{\mathcal{B}}$ then takes the form
\begin{equation}
\begin{aligned}
 \bra{\mathcal{B}} O_\Delta(n\cdot \Phi)\ket{\mathcal{B}}&= \sum_{j\,=-\frac{\Delta}{2}}^{\frac{\Delta}{2}} \;q^{2j}\; \mathcal{C}^{(2j)}_{\mathcal{B} \,\mathcal{B}^\dagger O_{\Delta}}\\
 \ket{\mathcal{B}}&= \sum_{O_B} \alpha_{O_{B}}O_{B}\ket{0}.
\end{aligned}
\end{equation}
The sum over states in $\ket{\mathcal{B}}$ is schematically taken over a complete set of primaries. The coefficients $\mathcal{C}^{(2j)}_{\mathcal{B} \,\mathcal{B}^\dagger O_{\Delta}}$ are generally linear combinations of OPE coefficients: for instance those with $j=0$ correspond to sums of diagonal structure constants,
\begin{equation}
\mathcal{C}^{(0)}_{\mathcal{B} \,\mathcal{B}^\dagger O_{\Delta}}= \sum_{O_B} |\alpha_{O_B}|^2 \; C_{O_B \bar{O}_B O_{\Delta}},
\end{equation}
while those with $j\neq0$ correspond to sums of off-diagonal structure constants. The case $j=\pm\Delta$ corresponds to extremal correlators. In general it is difficult to extract specific OPE coefficients from these sums, but the sums are enough to compare with holographic computations. One exception is the case where the background is itself is a primary operator, in which case the $SO(2)_R$ symmetry is restored. Then the only non-vanishing contributions are those with $j=0$ which is only possible for even $\Delta$.

\subsection{Matrix Model for HHL Correlators}
\label{sec:HHL}

In this section we will demonstrate how to compute expectation values of light BPS probes in terms of a matrix model. The class of probes that we will consider are the $SO(4)_R$ invariant single trace primaries \cite{Drukker:2008wr}
\begin{equation}
    O_{\Delta, k}=  c^{\Delta, k}_{i_1 \dots i_l}\tr\left[\Phi_{i_1}\dots\Phi_{i_L} \right].
\end{equation}
Their coefficients are given by those of the $SO(4)$ invariant spherical harmonics on $S^5$, $c^{\Delta, k}_{i_1 \dots i_l} x_{i_1}\dots x_{i_L}$. These operators are labeled by their dimension $\Delta$ and $U(1)_R$ charge $k$. The operators with non-zero correlators are those with  $k= -\Delta, -\Delta+2, \,\dots, \Delta-2, \Delta$. A couple of examples are \cite{Choi:2024ktc}:
\begin{equation}
\begin{aligned}
      O_{2, 0} &= \sqrt{\frac{2}{3}}\tr[Z \bar{Z}-\frac{1}{2}\sum_{i=1}^4 \Phi_i^2]\\
       O_{2, 2} &= \frac{1}{\sqrt{2}}\tr[Z^2]\\
      O_{3, 1} &= \frac{1}{\sqrt{2}} \tr[Z  (Z\bar{Z}- \sum_{i=1}^4 \Phi_i^2)],\\
\end{aligned}
\end{equation}
where $Z=\Phi_{5}+i\Phi_{6}$. For higher values of $\Delta$ it is straightforward but tediuous to find the form the coefficients, but their form can be determined by acting with $SO(6)_R$ generators \cite{Turton:2025svk}. In practice this procedure is very cumbersome since we need to integrate rather complicated polynomials over $\Phi_i$ and the triangular parts of the matrices $Z, \bar{Z}$.

Another way of determining them is to use the residual symmetry to fix the $R$-symmetry frame as discussed before and then to expand in the parameter $q$ (see for instance the discussion in \cite{Choi:2024ktc, 
Holguin:2025bfe})
\begin{equation}
\begin{aligned}
 \tr[(n\cdot \Phi)^2]&= \tr[q^2Z^2+ \bar{q}^2\bar{Z}^2 -4 \Phi_1^2 + 2 Z \bar{Z} + 4i \Phi_1(Z+\bar{Z)}]\\
 &= q^2 \tr[Z^2]+\bar{q}^2 \tr[\bar{Z}^2]+ \tr[2Z\bar{Z}-4 \Phi_1^2]+ \texttt{non-singlet}.
\end{aligned}
\end{equation}
The first two terms are clearly the contributions of $O_{2, \pm 2}$, while the third corresponds to $O_{2,0}$. This is because the insertion of $4\,\tr[\Phi_1^2]$ is the same as that of $\sum_{i=1}^4 \tr [\Phi_i^2]$ in the correlation function by $SO(4)_R$ symmetry. The remainder terms, i.e. the \texttt{non-singlet} parts, vanish because they involve odd numbers of $\Phi_1$ which means that they have uncontracted $SO(4)_R$ indices.

This procedure works generally for all values of $\Delta$, although it does not determine the normalization of the operator. 
Starting at dimension 4 the ordering of the fields matter and the operators are more complicated, so instead we consider the correlation function of the operator with fixed polarization $n$
\begin{equation}  
   O_\Delta(n\cdot \Phi)=  \frac{1}{\sqrt{\Delta}}\tr[(q Z + \bar{q}\bar{Z} +2 i \Phi_1)^\Delta],
\end{equation}
from which we can read off the $U(1)_R$ charge of each contribution after integrating out $\Phi_1$ and the triangular matrices $T, \bar{T}$ coming from Schur decomposition~\eqref{SchurDec}. The reduction to eigenvalues of these operators is carried over in Appendix \ref{QtransformApp}, both by explicit integration and using a generating function. We denote by 
\begin{equation}
  \tau_{\Delta}(q)=  \mathbb{Q}\left[ O_{\Delta, q}\right],
\end{equation}
where the  $\mathbb{Q}$-transform is defined in \eqref{qtransformDef} and the first few  of these operators in the large $N$ limit are
\begin{equation}
\begin{aligned}
\tau_{1}(q)&= \tr m= \tr \text{He}_1(m)\\
\tau_{2}(q)&= \tr \left[m^2- \mathbb{I}\right]=\tr \text{He}_2(m)\\
\tau_{3}(q)&= \tr \left[m^3- 3m\right]=\tr \text{He}_3(m)\\
\tau_{4}(q)&= \tr \left[m^4- 4m^2+2\right]= \tr \text{He}_4(m)+2\, \tr\text{He}_2(m) + \tr\text{He}_0(m)\\
\tau_{5}(q)&= \tr \left[m^5- 5m^3+10 m\right]=\tr \text{He}_{5}(m)+ 5\, \tr \text{He}_3(m)+ 10\, \tr\text{He}_1(m),\\
\end{aligned}
\end{equation}
where $m= q z + \bar{q}\bar{z}$ and $z$ is are the eigenvalues of $Z$. In general we expect the operator $\tau_\Delta(q)$ to be a linear combination of Hermite polynomials in $x$ with positive integer coefficients.\footnote{For example the coefficients for $\Delta=4,5$ are the third and sixth row of Pascal's triangle!} The $SO(4)_R$ invariant operators are then obtained by grouping the coeffients of $q^k$ and then normalizing this polynomial. The normalization can be found for instance by comparing with spherical harmonics build as homogeneous coordinates\cite{Drukker:2008wr} or by performing the integration or Wick contractions explicitly. To compute the three point correlator of these operators we simply insert into \eqref{complex3point}.

\subsection{Comparison with Holographic vevs}

\vol{Refer to new appendix on LLM}
Let us now discuss the corresponding situation in gravity~\cite{Skenderis:2007yb}. The holographic dual description of a huge half-BPS operator is a bubbling geometry specified by the $Z$-eigenvalue density $\rho$  through the potential function \cite{Lin:2004nb}
\begin{equation}
    \Phi(z,\bar{z}, y) = \frac{1}{\pi} \int dz' d\bar{z}' \frac{\rho(z', \bar{z}')-\frac{1}{2}}{(|z-z'|^2+y^2)^2}.
\end{equation}
This is a harmonic function on $\mathbb{R}^2\times \mathbb{R}^4$ with $SO(4)$ symmetry and the values of the LLM metric and five-form are determined by it\footnote{To fully determine the metric and five-form field strength, one also needs the vector $V_i(z,\bar{z},y)$ (see Eq. (2.15) in \cite{Lin:2004nb}).}. Inserting a third small half-BPS operator measures the holographic one-point function of the corresponding supergravity mode which is sourced by the operator, which is read off from the asymptotic expansion of the fields near the boundary \cite{Skenderis:2006uy}. Since the supergravity solution breaks the $SO(6)_R$ symmetry into $SO(4)_R$ only operators with vanishing $SO(4)_R$ quantum numbers have non-vanishing vevs. Additionally the $SO(2)_R$ charge  is only conserved for a subset of the solutions corresponding to rotationally symmetric distributions, so the fields are indexed by both their dimension and charge in general LLM backgrounds. This analysis was carried out in \cite{Skenderis:2007yb} for scalar operators of dimension $\Delta \leq 4$ in addition to the R-charge currents and energy-momentum tensor. The idea is that the asymptotics of the fields can be determined in terms of the multi-pole expansion of the harmonic form $\Phi$ 
\begin{equation}
    \Phi(z,\bar{z},y)= \Phi^0(z,\bar{z},y)+ \sum_{ L,k}  \frac{(\Delta\Phi)_{L,k} Y_{L,k}(\phi, \theta)}{R^{L+4}},
\end{equation}
where $R= \sqrt{|z|^2+y^2}$, and $Y_{L,k}$ are $SO(4)_R$ invariant spherical harmonics.  

This series encodes the deviations of an arbitrary LLM geometry from the reference $AdS_5\times S^5$ vacuum $\Phi^0(z,\bar z,y)$ and it can be related to the Fefferman-Graham expansion of the metric and other background fields after a change of coordinates.  $\Delta\Phi_{L,k}$ are sourced by perturbations around the disk distribution; these are precisely single trace operators. This is clear from the droplet picture since the insertion of a single trace operator excites a collective mode of the eigenvalues which slightly deforms the Fermi surface of the droplet configuration and condensation of these modes corresponds to large deformations of the eigenvalue configuration. Similarly from the gravity description these chiral primaries correspond to deformations of the metric and five-form field which are consistent with the symmetries of a general LLM geometry. Since all of the fields are specified in terms of the density $\rho(z, \bar{z})$, the chiral primaries with non-zero one-point functions source a deformation $\delta\rho$ and accordingly a mode $\Delta \Phi_{L,k}$.  For extremal correlators the dictionary is simple, and the modes $\Delta\Phi_{k,\pm k}$ correspond to (anti-)holomorphic insertions in the matrix model:
\begin{equation}
  \langle\mathcal{B}|\, \tr Z^k\,|\mathcal{B}\rangle_{\text{SYM}}=   \langle \tau_{k, k}\rangle_{MM}\leftrightarrow \Delta\Phi_{k,k}.
\end{equation}
This is because the mode $\Delta \Phi_{k,k}$ represents a small ripple with angular momentum $k$ around the boundary of the droplet, which is precisely the collective mode $\tr Z^k$ in the matrix model.
For general half-BPS single-trace operators one needs to consider a non-linear Kaluza-Klein reduction of the type IIB supergravity fields with $AdS_5\times S^5$ asymptotics. The starting point of this analysis is to consider spectrum of fluctuations around $AdS_5\times S^5$ \cite{Kim:1985ez}. Working in harmonic gauge on the sphere the relevant modes  needed to determine the one-point functions of chiral primaries are the trace of the $S^5$ metric $h^{\alpha}_\alpha$ and the five-form field strength on the sphere $f_{abcde}$\footnote{It could happen that the asymptotic expansion of fields in the LLM solution are not in de Donder gauge. One would then need to form gauge invariant combinations of fields as done in \cite{Skenderis:2007yb}, or change coordinates such that the expansion is in de Donder gauge order by order as in \cite{Turton:2025svk}.}
\begin{equation}
\begin{aligned}
    &h^{a}_{a}(\mathsf{x},\mathsf{y})= \sum_{I} \pi_{I}(\mathsf{x}) Y_{I}(\mathsf{y}),\\
   &f_{abcde}= \epsilon_{abcde}\sum_I b_I(\mathsf{x}) \Lambda^I   Y_{I}(\mathsf{y}).
\end{aligned}
\end{equation}
Here the index $I$ labels the $SO(6)$ quantum numbers, $Y_I(\mathsf{y})$ are scalar spherical harmonics on $S^5$, $\Lambda^I=-L(L+4)$ is the eigenvalue of the Laplacian, and $\mathsf{x}$ and $\mathsf{y}$ denote the coordinates of the asymptotic $AdS_5$ and $S^5$ metrics respectively. All other field perturbations carry vector indices along the $S^5$ or $AdS_5$ directions and they correspond to descendants of chiral primaries. At the linearized level the modes $\pi_I$ and $b_I$ mix and the linear combination modes dual to chiral primaries are given by
\begin{equation}
    s_I= \frac{1}{20(\Delta+1)}\left( \pi_I -10(\Delta+4) b_I \right).
\end{equation}
These enter the linearized 5d action as scalar fields which satisfying the Klein-Gordon on $AdS_5$ with conformal dimension $\Delta=L$ \cite{Kim:1985ez}. One should then define 5d fields $S_I= s_I + D_{IJK} s_J s_K +\dots$ for which the  equations of motion reduced from 10d to 5d come from a purely 5d effective action. The non-linear terms in this effective action are needed to consistently reproduce extremal couplings \cite{Skenderis:2006uy}.
The expectation value of the $SO(4)_R$ invariant chiral primaries $O_{L,k}$ with $L\leq 4$ are given by~\cite{Skenderis:2007yb}   
\begin{equation}
\begin{aligned}
  \langle O_{2,k}\rangle &=   \frac{N^2}{2\pi^2} \frac{2\sqrt{8}}{2} [s_{2,k}]_2,\\
   \langle O_{3,k}\rangle &=   \frac{N^2}{2\pi^2} \sqrt{3} [s_{3,k}]_3,\\
   \langle O_{4,k}\rangle &=   \frac{N^2}{2\pi^2} \frac{4\sqrt{3}}{5} [2 s_{4,k}+\frac{2}{3 \mathsf{n}_4} a^{(4k)(2 l)(2 m)} \,s_{2,l} \,s_{2,m}]_4,
\end{aligned}
\end{equation}
where $1/\mathsf{n}_L=2^{L-1}(L+1)(L+2)$ is the normalization for the spherical harmonics chosen in \cite{Skenderis:2007yb}, and $a^{IJK}$ are defined by a triple integration over spherical harmonics.  The fields $s_{\Delta,k}$ can be expressed in terms of the moments $\Delta \Phi_{\Delta, k}$\footnote{Here $\phi_{(s)}$ is the scalar mode of the metric perturbations on $S^5$. This mode is pure gauge but it is needed here to make gauge invariant combinations of the fields $\pi$ and $b$ because the LLM solutions are not in harmonic gauge. In \cite{Skenderis:2007yb} the gauge invariant combinations are hatted $\hat{s}_{L,k}$. }
The overall normalization is chosen to match the normalization of the effective type IIB action and it differs from the field theory normalization \cite{Skenderis:2006uy}. The brackets $[f]_{k}$ mean the $k$-th term in the Fefferman-Graham expansion of $f$. The non-linear terms for $O_{4,k}$ arise due to extremal couplings between chiral primaries. 

For the LLM solutions these take the following forms \cite{Skenderis:2007yb}:
\begin{equation}
\begin{aligned}
 \langle O_{\Delta,\Delta}\rangle &=   \frac{N^2}{\sqrt{\Delta}\pi^2} (\Delta-2)\sqrt{\Delta-1}\int dzd\bar{z}\rho(z,\bar{z})z^\Delta, \\
  \langle O_{2,0}\rangle &=   \frac{N^2}{\pi^2} \frac{\sqrt{2}}{\sqrt{3}} \int dzd\bar{z} \rho(z,\bar{z})\left(z\bar{z}-\frac{1}{2}\right) ,\\
   \langle O_{3,1}\rangle &=   \frac{N^2}{\pi^2} \int dzd\bar{z} \rho(z,\bar{z}) z^2\bar{z},\\
   \langle O_{4,0}\rangle &=   \frac{N^2}{\pi^2} \frac{\sqrt{3}}{\sqrt{5}}\int dzd\bar{z} \rho(z,\bar{z}) \left(3 (z\bar{z})^2 -4 z\bar{z}+1\right),\\
    \langle O_{4,2}\rangle &=   \frac{N^2}{\pi^2} \frac{4\sqrt{3}}{\sqrt{10}}\int dzd\bar{z} \rho(z,\bar{z})\, z^2\left(z\bar{z}-1\right).
\end{aligned}\label{skenderisTs}
\end{equation}
The remaining vevs up to dimensions 4 can be determined by complex conjugation. We should compare these expressions with the complex matrix model correlators $\tau_{\Delta, k}$. 
For  dimensions $\leq 4$ they are given by
\begin{equation}
\begin{aligned}
\tau_{\Delta, \Delta}&= \frac{1}{\sqrt{\Delta}}\tr[z^\Delta]\rightarrow \textcolor{NavyBlue}{\frac{N}{\sqrt{\Delta}}\int dzd\bar{z} \,\rho(z,\bar{z})\,z^\Delta},\\
\tau_{2,0}&= \mathsf{N}_{2,0}\tr\left[z \bar{z} -\frac{1}{2}\left(1+\frac{1}{N}\right)\mathbb{I}\right]\rightarrow  \textcolor{NavyBlue}{\sqrt{\frac{2}{3}}N\int dz d\bar{z} \rho(z,\bar{z})\left(z \bar{z} -\frac{1}{2}\right)},\\
\tau_{3,1}&= \mathsf{N}_{3,1}\tr\left[z^2 \bar{z}- \left(1+\frac{1}{N}\right) z\right]\rightarrow \textcolor{NavyBlue}{\frac{N}{\sqrt{2}}\int dz d\bar{z} \rho(z,\bar{z})\left(z^2 \bar{z}-  z\right)},\\
       \tau_{4,0}&= \mathsf{N}_{4,0}\tr\left[3(z\bar{z})^2 -4\left(1+\frac{2}{3N}\right)z\bar{z} + \left(1+\frac{1}{N}\right)\left(1+\frac{3}{N}\right)\mathbb{I}\right]-\frac{4}{N} (\tr z)(\tr\bar{z}),\\
       &\rightarrow \textcolor{NavyBlue}{\frac{N}{2\sqrt{5}}\int dz d\bar{z} \,\rho(z,\bar{z})\left(3(z\bar{z})^2 -4z\bar{z} + 1\right)}\\
 \tau_{4,2}&= \mathsf{N}_{4,2}\tr\left[z^3\bar{z} -\left(1+\frac{2}{3N}\right)z^2 \right]-\frac{2}{N} (\tr z)^2\rightarrow  \textcolor{NavyBlue}{\sqrt{\frac{2}{5}}N\int dz d\bar{z} \,\rho(z,\bar{z})z^2\left(z\bar{z} -1 \right)},
\end{aligned}
\end{equation}
see appendix \ref{QtransformApp} for details. In the large $N$ limit we can replace the traces by integrals and drop subleading terms.
 To match with the calculations of \cite{Skenderis:2007yb} we needed to set the center of mass of the distribution at the origin, meaning that $\langle \tau_{1, \pm}\rangle=0$. The normalization of the maximally charged operators  used by Skenderis and Taylor is
\begin{equation}
\begin{aligned}
 \langle O_{\Delta,k}\rangle_{\texttt{gravity}} &= \textcolor{Red}{\mathcal{N}^{\texttt{sugra}}_k}\times \textcolor{NavyBlue}{\langle \tau_{\Delta,k}\rangle_{MM}}\\
  \mathcal{N}^{\texttt{sugra}}_k&= \frac{N}{\pi^2}(k-2+\delta_{k,2}) \sqrt{k-1}
\end{aligned}
\end{equation}
The normalization factors $\mathsf{N}$ of the charged chiral primary operator for $\Delta=4$ were determined by matching with the leading coefficients from \cite{Choi:2024ktc}\footnote{Our conventions differ from theirs by $\Phi_{\texttt{theirs}}= Z_{\texttt{ours}}$ and $(\Phi_i)_{\texttt{theirs}}= \sqrt{2}(\Phi_i)_{\texttt{ours}}$ which changes the normalizations of the propagators by factors of 2.}. These normalization coefficients can also be determined by making sure that the $\tau_{\Delta,k}$ have unit normalized two point correlators in the large $N$ limit. 

These results are in exact agreement, and in fact the matrix model allows us to compute beyond $\Delta=4$. Since we have finite formulas in principle we can compute leading order corrections in $1/N$ to these vevs in the single trace basis. One important comment is that the intricacy of the single trace contributions to the vevs boils down to expanding in terms of fields with definite $SO(2)_R$ charge. If instead the fields had been grouped with respect to a fixed $SO(6)_R$ polarization one would have noticed that these were given by integer combinations of Hermite polynomials. We take this as a sign of the integrability of the  half-BPS subsector of type IIB supergravity \footnote{This integrability should not be confused with the integrability of the $\mathcal{N}=4$ SYM spin chain.}. Afterall the quantization of this subsector leads to a relatively simple quantum theory, that of a chiral boson \cite{Grant:2005qc, Maoz:2005nk}. This explains why the modes $\tau_\Delta(q)$ are given by harmonic oscillator wavefunctions.  One should note that this model is identical to Kodaira-Spencer theory on non-compact Calabi-Yau threefolds \cite{Bershadsky:1993cx}. There the chiral boson governs deformations to the complex structure at infinity, much like in the type IIB theory where the chiral boson describes the deformations $\Delta\Phi_{L,k}$. In both cases the geometry is specified by a spectral curve which describes the positions of the branes creating the background. The B-model in this context is also known to be intimately related to integrable hierarchies \cite{Aganagic:2003qj}, and we should understand these topological models as being embedded in corners of the half-BPS sector of the full type IIB theory, although this relation is not well understood beyond the planar limit \cite{Costello:2018zrm}. This is to say that one would expect that there exists a reorganization of the type IIB fields in the half-BPS sector that manifests this integrable structure.

\section{Giant Probe Correlators}
\label{Giant Probes}
We turn to study of three-point functions such that two of the operators are huge, meaning that they have dimensions of order $N^2$, and the third operator is giant, i.e. it has dimension of order $N$. In this regime the correlator is dominated by the saddle point of the huge operators which defines the density for the eigenvalues $\rho(z,\bar{z})$ that we discussed in the previous section. More precisely, we will study correlation functions in the presence of half-BPS states that create a smooth eigenvalue profile in the matrix model, but do not necessarily have a fixed conformal dimension. For the third operator we consider two examples in detail, namely Schur polynomials corresponding to single row or single column.

\subsection{Sphere Giant graviton}
 For Young diagrams with a single column with $\Delta$ boxes the transformed operators \eqref{qtransformDef} are
\begin{align}\label{Q-determinant}
\mathbb{Q}[\chi_{\texttt{asym},\Delta}](M)= \frac{(\sqrt{N})^{N-\Delta}}{(N-\Delta)!}\frac{1}{\mathcal{Z}_\sigma}\int d\sigma \;e^{-\frac{N}{2}\sigma^2}\, \text{He}_{N-\Delta}(\sqrt{N}\sigma)\;\det(m+\sigma\,\mathbb{I}),
\end{align}
where we chose $M= qZ+ \bar{q}\bar{Z}$ and $m$ is defined as $m=qz+ \bar{q}\bar{z}$ (note that these are not the eigenvalues of $M$). For the cases we will calculate the background state is created by a character so it will not be necessary to keep track of the phase $q$ since only the $U(1)_R$ neutral contribution to the correlator is non-vanishing, meaning that we can set $q=1$. 
First we will consider the case when the background huge state is created by two characters for the representations $R$ and $R'$. A general expression for the HHG structure constant is obtained by plugging in \eqref{Q-determinant} into \eqref{probe_T}:

\begin{equation}
\begin{aligned}
C_{R\,R'\,\Delta}=&\,\frac{1}{\mathsf{N}_{RR' \Delta}}\int d\sigma\, e^{-\frac{N}{2}\sigma^2 } \, \text{He}_{N-\Delta}(\sqrt{N}\sigma)\,\int \prod_k |dz_k|^2  \,e^{-N z_k\bar{z}_k}\left(z_k+\bar{z}_k+\sigma\right)\det\left(z_j^{h_l}\right)\det\left(\bar{z}_n^{h'_m}\right),\\
\mathsf{N}_{RR' \Delta}&=\sqrt{2 \pi N} \times \left[\frac{(N!)^3\; (N-\Delta)! \prod_i h_i! \,h'_i!\; \pi^{2N}}{  N^{|R|+|R'|+ 3N}}\right]^{\frac{1}{2}}.
\end{aligned}
\end{equation}

\subsubsection{Exact Solution and Large $N$}
For these simple correlators the integration over the eigenvalues can be carried out exactly. For the exact integration it turns out to be better to work with the rescaled variables $\zeta_k= \sqrt{N} z_k$ and $x= \sqrt{N}\sigma$ in which the integration becomes 
\begin{multline}
\int \prod_k |d\zeta_k|^2  \,e^{-\zeta_k \bar{\zeta}_k}\left(\zeta_k+\bar{\zeta}_k+x\right)\det\left(\zeta_j^{h_l}\right)\det\left(\bar{\zeta}_n^{h'_m}\right)\\
   = N!\,\pi^{N}\prod_j^N h_j!\times \det_{j,k}\left(x  \,\delta_{h_j, h'_k}+ (h_j+1)\delta_{h_j, h_k'-1}+  \delta_{h_j, h_k'+1}\right).
\end{multline}
The overall common factor of $\pi^{N}\prod_j^N  h_j!$ can be recognized as the normalization of two-point function of the characters. For simplicity let us assume that the huge states correspond to the same Young diagram, meaning that $h_j=h'_j$. 
Suppose that the Young diagram is made out of large rectangular blocks. Then it is not hard to see that the matrix inside the determinant becomes block diagonal with each block corresponding to a section of the Young diagram where the number of boxes per row is constant. For each of these sections $h_i$ decreases linearly in unit steps, so if the number of rows as a function of $i$ changes, the off-diagonal parts of the matrix vanish. Therefore the determinant factorizes into a product where each term corresponds to a section of the Young diagram where row length is constant. Each of these terms is schematically the same what we would obtain for a single large rectangle.

For a moment let us concentrate on the lowest most part of the Young diagram which we take to be made out of $M$ rows of constant length $L$. 
To that part of the Young diagram we can associate a $M\times M$ matrix which we call $\mathcal{M}_M^L(x)$:
\begin{equation}
\begin{aligned}
\mathcal{M}_M^L(x)&=\begin{pmatrix}
        x& h_1+1&0&0&\dots&0\\
        1& x& h_2+1&0&\dots&0\\
        0&1&x &h_3+1&\dots&0\\
       0& 0&1&x&\dots&0\\ \vdots&\vdots&\vdots& \vdots& \ddots& h_{M}+1\\
       0&0&0&0&1& x
    \end{pmatrix}  .
\end{aligned}
\end{equation}
From here we can either upper triangularize each block, or compute the determinant exploiting the fact that $\mathcal{M}_M^L(x)$ is tridiagonal. The upper triangularization turns out to be better suited for taking the large $N$ limit, while working with the tridiagonal form is better for finite $N$ calculations.
For  this block we will have that $h_i= L+i-1$ as long as $i$ is less than $M$. 


It will turn out to be useful to think of the determinants $\{\det\mathcal{M}^L_n(x)\}_{n=0}^\infty=\{F^L_n(x)\}_{n=0}^\infty$ as a sequence of orthogonal polynomials in the variable $x$. Since $\mathcal{M}_M^L(x)$ is tridiagonal its determinant $F_M^L(x)$ can be expressed through $F_{M-1}^L(x)$ and $F_{M-2}^L(x)$ via its cofactor expansion along the last row and column
\begin{equation}
    F_M^L(x)= x F_{M-1}^L(x)- (h_{M-1}+1) F_{M-2}^L(x).\label{threeTermRecursionGeneral}
\end{equation}
This gives a three-term recursion relation for $F_M^L(x)$ which is enough to guarantee that these functions form a sequence of orthogonal polynomials with respect to a positive weight.  With this result we have enough information to compute the determinant for arbitrary representations $R$.
 
To do this we split the Young diagram into rectangular blocks of increasing lengths $R_k> R_{k-1}$ whose heights are $M_k$. Then we define $N_k= \sum_{l< k} M_l$ which is the vertical position on the diagram  at which the length changes from $R_{k-1}$ to $R_k$ when read from bottom to top and $L_{k}= R_k+ N_k$.

\begin{figure}[ht]
  \centering
  \scalebox{0.85}{
  \input{tikz/YDblocks}}
  \caption{Correspondence between the block matrices $\mathcal{M}_{M_i}^{L_i}(x)$ and a Young diagram. In this example $M_1=2$, $M_2=3$ and $M_3=4$. The shifts $L_i$ are related to the row lengths by $L_1=3$, $L_2=6+M_1=8$, $L_3=9+M_2+M_1=14$ which are the lengths of the shifted Young diagram.}
  \label{fig:TDblocksDeterminant}
\end{figure}
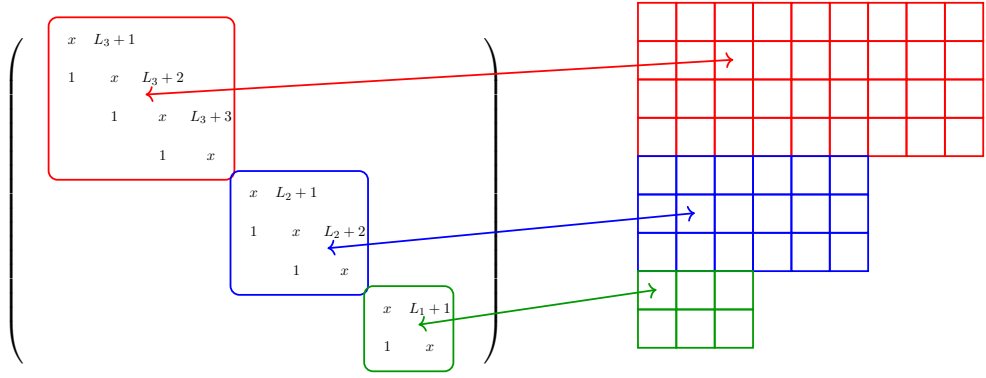

For each of the blocks we get a factor of $F_{M_{k}}^{L_k}(\sigma)$ so in general we obtain 
\begin{equation}
\frac{1}{\pi^{N}\prod_j h_j!}\int \prod_k |d\zeta_k|^2  \,e^{-\zeta_k \bar{\zeta}_k}\left(\zeta_k+\bar{\zeta}_k+x\right)\det\left(\zeta_j^{h_l}\right)\det\left(\bar{\zeta}_n^{h'_m}\right)= N!\prod_{k=1}^{\texttt{\# of blocks}} F_{M_k}^{L_k}(x).
\end{equation}
Substituting this into the formula for the structure constant we get
\begin{align}
C_{R\,R\,\Delta}=&\sqrt{\frac{(N-\Delta)!}{N!}}\times \frac{1}{(N-\Delta)!}\frac{1}{\mathcal{Z}_x}\int dx\, e^{-\frac{1}{2}x^2 } \, \text{He}_{N-\Delta}(x) \prod_{k=1}^{\texttt{\# of blocks}} F_{M_k}^{L_k}(x).
\end{align}
\subsubsection{Some examples}
Let us now consider some simple examples. As a sanity check we can take two of the operators to be the identity $R=\varnothing$ so that the correlator vanishes trivially. This checks that the intermediate steps in the integration are self-consistent. In this case the determinant can be evaluated explicitly. For $R=\varnothing$ the three term recursion relation \eqref{threeTermRecursionGeneral} is given by 
\begin{equation}
    F^{\varnothing}_{n+1}= x F^{\varnothing}_{n}-n \,F^{\varnothing}_{n-1},
\end{equation}
which can be recognized as the recursion relation satisfied by the Hermite polynomials $\text{He}_n(x)$. Substituting this in gives
\begin{equation}\label{AdSGG}
    C_{\varnothing, \varnothing, k} \sim \int dx e^{-\frac{1}{2}x ^2} \text{He}_{N-k}(x) \det\begin{pmatrix}
        x&1&0&0&\dots\\
        1&x&2&0&\dots\\
        0&1&x&3&\dots\\
        \vdots&\vdots&\vdots& \vdots&\ddots
    \end{pmatrix}= \int dx\, e^{-\frac{1}{2}x ^2} \,\text{He}_{N-k}(x )\text{He}_N(x )=0.
\end{equation}
Clearly the only non-zero possibility is $k=0$ which just picks up the constant factor of the generating function, meaning that all the operators are taken to be the identity. This is of course  trivial for any finite $N$, but in the large $N$ limit it is not easy to see this orthogonality.

The next simplest example is when the Young diagram associated to $R$ is a rectangle of $K$ columns and $N$ rows. In that case the shifted highest weights are $h_{K,j}= K+j-1$ which gives the following recursion relation for the determinant in the correlator
\begin{equation}
  F^{K}_{n+1}= x F^{K}_{n}- (K+n) F^{K}_{n-1},
\end{equation}
which is the recursion relation for the \textit{associated Hermite polynomials} $F^{K}_{n}(x)=\text{He}_{n}(x;K)$ \cite{Askey_Wimp_1984}. These polynomials are orthogonal with respect to a positive measure $d\mu^K(x)= |D_{-K}(ix)|^{-2}$ where $D_{\nu}$ are parabolic cylinder functions. The associated Hermite polynomials are related to the usual Hermite polynomials by the formula \cite{Askey_Wimp_1984, Drake_2009}:
\begin{equation}
    \text{He}_{n}(x;K)= \sum_{j=0}^{[n/2]}(-1)^j \frac{\Gamma(K+j)}{\Gamma(K)} \binom{n-j}{j} \text{He}_{n-2j}(x)\,.
\end{equation}
We will call the corresponding structure constant $C^{\,\texttt{asym}}_{K\,K\, \Delta}$ and it is given by 
\footnote{For the correct normalization, one should divide the expression for $C_{K \, K\, \Delta}$ by the square roots of the two-point functions, as well as by the matrix integral $\int dZ d\bar{Z} e^{-\tr Z\bar{Z}}$ and the gaussian integral $\int d\sigma\,  e^{-\sigma^2/2}$.} 
\begin{equation}\label{ExplicitC}
    C^{\,\texttt{asym}}_{K\,K\, \Delta}= (-1)^{\Delta/2}\sqrt{\frac{(N-\Delta)!}{N!}}\times\frac{(K+\Delta/2-1)!}{(K-1)!}\frac{(N-\Delta/2)!}{(N-\Delta)!(\Delta/2)!}  .
\end{equation}
The first factor is is nothing more than the normalization of the subdeterminant operator, while the other two terms carry all the dependence on the background.  This formula is exact and it is valid for all values of $K$ and $N$ as long as $\Delta$ is even so that the integral is non-zero. 

Now we can take the large $N$ limit of this expression. We are interested in the scaling limit $K=N\gamma$ and $\Delta=N \delta$, for finite $\gamma$ and $\delta$.
The full large $N$ asymptotics of \eqref{ExplicitC} is
\begin{align}\label{CKKDeltaAss}
    C_{KK\Delta}^{\,\texttt{asym}}\sim(-1)^\frac{\Delta}{2} \left[  \frac{\left(1-\frac{\delta}{2}\right)^{1-\frac{\delta }{2}} \left(\gamma +\frac{\delta }{2} \right)^{\gamma +\frac{\delta }{2}}}{\left(1-\delta\right)^{\frac{1-\delta}{2}}\gamma ^{\gamma }\left(\frac{\delta}{2}\right) ^{\frac{\delta }{2}}}\right]^N.
\end{align}
This could also be rewritten in terms of more geometric bulk variables as follows. The distribution of eigenvalues created by the operator $\left(\det Z\right)^K$ is given by an annulus whose inner and outer radii are given by $N a^2= K$ and $N b^2= N+K$. This gives that $b^2-a^2=1$ and $\gamma= a^2$. The parameter $\delta$ is related to the energy of the wavefunction of the hole on the Fermi sea by $\varepsilon=1-\delta$.

Another simple example is the case of a trapezium Young diagram $\boldsymbol{T}$. In that case $h_j=K+2j-2$ so integral over the $\zeta_k$ and $\bar{\zeta}_k$ results only in the diagonal terms. In that case the structure constant evaluates to 
\begin{equation}
 C^{\,\texttt{asym}}_{\boldsymbol{T}, \boldsymbol{T}, \Delta} = \frac{1+ (-1)^\Delta}{2} \times\sqrt{\frac{N!}{(N-\Delta)!}} \frac{1}{2^{\Delta/2} (\frac{\Delta}{2})!}\sim \left[(1-\delta)^{1-\delta} \, \delta^\delta\right]^{-\frac{N}{2}}.
\end{equation}
Note that this is independent of the details of the Young diagram ($K$-independent) in contrast to the previous cases. In fact for any Young diagram that leads to distributions where the local density of eigenvalues is less than one will have the same result since the off-diagonal $\zeta_k$ integrals vanish. This is consistent with the fact that the triangular and trapezoidal Young diagrams are generic states in the superstar ensemble which serves as an analog of the thermal state (albeit at infinite temperature)\cite{Balasubramanian:2005mg}. 

\subsubsection{Saddle point analysis}

Now we turn to the saddle point analysis of the correlator. At large $N$, the three point function of a giant graviton with two generic Huge operators is
\begin{equation}
\begin{aligned}
    C^{\,\texttt{asym}}_{\mathcal{B} \mathcal{B} \Delta} &= \biggl\langle\mathbb{Q}[\chi_{\texttt{asym},\Delta}](z+\bar z)\biggr\rangle_{\mathcal B}\\
    &\sim\frac{1}{\sqrt{\mathsf{N}_\Delta}}\frac{1}{\mathcal{Z}_\sigma} \oint \frac{d\alpha}{2\pi i \,\alpha^{N-\Delta+1}} \int_{\Gamma} d\sigma\,\, e^{-\frac{N}{2}(\sigma-\alpha)^2}\;\underbrace{e^{N \,\int dzd\bar{z}\,\rho_\mathcal{B}(z, \bar{z})\, \log(z+ \bar{z}+\sigma)}\vphantom{\Big|}}_{\equiv \Psi_{\mathcal B}(\sigma)},
\end{aligned}
\label{BBD-integrand-saddle}
\end{equation}
where the first line follows from \eqref{complex3point} and in the second line, we plug in the normal ordering \eqref{Q-determinant}. We also introduced above
\[\Psi_{\mathcal B}(\sigma) = \biggl\langle\det(z+\bar z+\sigma)\biggr\rangle,\]
which can be interpreted as the wavefunction for a probe eigenvalue in the background created by the huge operators. This integral is reminiscent of eigenvalue instanton contributions to partition functions in matrix models. It would be very nice to reproduce this form from a bulk gravity computation. 
 
Formally at large $N$, the wavefunction can be written in a WKB form
\begin{equation}
    \Psi_{\mathcal B}(\sigma)= \exp\left[N \,\int dzd\bar{z}\,\rho_{\mathcal B}(z, \bar{z})\, \log(z+ \bar{z}+\sigma)\right]\sim \exp\left[N \int^\sigma d\sigma' \; \mathcal{P}(\sigma'; {\mathcal B})\right],\label{wavefunctionWKB}
\end{equation}
where $\mathcal{P}(\sigma'; {\mathcal B})$ is the momentum conjugate to $x$, which is essentially the resolvent for the distribution of $z$ projected into the $x=\frac{z+\bar{z}}{2}$ axis\footnote{For general backgrounds, there is no radial symmetry in the $z$ plane and thus the choice of the axis of projection matters.}. The precise relation between the WKB momentum and the resolvent is 
\begin{equation}
    \mathcal{P}(\sigma; {\mathcal B})= \frac{1}{2} G_{{\mathcal B}}(\sigma/2).
\end{equation}
As is well known from WKB analysis, the function $\mathcal{P}(\sigma; R)$ is not singled valued on the complex plane since it is a solution to a quadratic equation so we must ensure that the asymptotic behavior matches that of the expectation value of the determinant we are estimating. At large $\sigma$ this means that $\int^\sigma d\sigma'\,\mathcal{P}(\sigma';\mathcal B)\sim \log(\sigma)$. That is why the asymptotics of $\mathcal{P}(\sigma;R)$ are the same as the asymptotics of the resolvent on the first sheet. The behaviour of the wavefunction depends on whether $\sigma$ sits inside or outside the classically allowed region -- for large values of $|\sigma|$ it decays exponentially, while for small $|\sigma|$ it is highly oscillatory.

As a check let us take the case of a single rectangle with $N$ rows for which we know that the integral over the $z_k$ variables evaluates to an associated Hermite polynomial. Plugging in the resolvent of the annulus background \eqref{annulusResolvent}, the wavefunction is given by
\begin{align}
   \Phi_K(\sigma) 
    =e^{-\frac{N}{2}}\frac{\left(\frac{\sigma }{2}\pm\sqrt{\frac{\sigma
   ^2}{4}-(\gamma+1)}\right)^{(\gamma+1) N}}{\left(\frac{\sigma }{2}\pm\sqrt{\frac{\sigma ^2}{4}-\gamma}\right)^{\gamma N}}\,\exp \left[\pm N \left(\frac{\sigma  }{2} \left(\sqrt{\frac{\sigma
   ^2}{4}-\gamma}-\sqrt{\frac{\sigma ^2}{4}-(\gamma+1)}\right) \right)\right],\label{assHermAsymptotics}
\end{align}
where the signs are determined by whether $\sigma$ is to the right or the left of the cut. These choices are the two asymptotics on either side of the classically allowed region. The same wavefunction can be computed using the asymptotics of the associated Hermite polynomials using the following generating function
\begin{equation}
    \text{He}_{N}(\sqrt{N} \sigma;K)= \binom{K+N-1}{N}\frac{K}{N^{N/2}}\int_0^1 dv\; v^{K-1}\;\oint \frac{d\alpha}{\alpha^{N+1}}e^{-\frac{N}{2}\alpha^2(1-v^2)+ N \sigma \alpha(1-v)}.
\end{equation}
The integrals over $v$ and $\alpha$  can be evaluated using the saddle point method at large $N$ and large $K$; there are four saddle points, and the leading saddles yield the correct asymptotic behavior \eqref{assHermAsymptotics}\footnote{Note that these asymptotics are only valid when we are away from the branch points.}. 

For generic Young diagrams with large edges the finite $N$ form of $\Psi_{\mathcal B}(\sigma)$ factorizes into terms associated to each large rectangle, which means that we simply need to replace the function inside the exponential by a sum of terms associated to each block. This gives the resolvent for an arbitrary distribution of concentric rings as expected (see Appendix \ref{App-densities-LLM}), which is consistent with taking products of asymptotics of associated Hermite polynomials.

Let us now evaluate the asymptotic form of the HHG correlator. The structure constant is given by the integral
\beq\label{KKD-integrand-saddle}
C^{\,\texttt{asym}}_{KK \Delta}\sim\frac{1}{\sqrt{\mathsf{N}_\Delta}}\frac{1}{\mathcal{Z}_\sigma}\int_{\Gamma} d\sigma\, e^{-\frac{N}{2}\sigma^2} \oint \frac{d\alpha}{2\pi i \,\alpha^{N-\Delta+1}} \; e^{-\frac{N}{2}\alpha^2+ N \alpha \sigma}\;\Phi_K(\sigma).
\eeq
This integral can be evaluated using stationary phase approximation. The saddle point equations for $\sigma$ and $\alpha$ are 
\begin{equation}
\begin{aligned}
    \alpha-\sigma + \sqrt{\left(\frac{\sigma}{2}\right)^2- \gamma}-\sqrt{\left(\frac{\sigma}{2}\right)^2- (\gamma+1)}&=0,\\
    \sigma-\left( \alpha+\frac{\varepsilon}{\alpha}\right)&=0,
\end{aligned}
\end{equation}
where the physical parameter associated to the wavefunction of the determinant is $\varepsilon=1-\delta$. 
Because of the form of the saddle point equations most of the terms in the exponent of \eqref{KKD-integrand-saddle} cancel, which simplifies the final result greatly. Evaluating the integrand at the solutions of the SPEs gives:
\begin{equation}
\begin{aligned}
    C_{K\,K\, \Delta}&\sim  \,
   \varepsilon^{-\frac{N}{2}\varepsilon}\;\left[\frac{(\varepsilon+1)^{\varepsilon+1} (\varepsilon-1)^{\varepsilon-1} (2\gamma+1-\varepsilon)^{2\gamma+1-\varepsilon}}{2^{2\gamma+1 +\varepsilon}\,\gamma^{2\gamma}}\right]^{\frac{N}{2}}.
\end{aligned}
\end{equation}

This exactly reproduces the large $N$ asymptotics \eqref{CKKDeltaAss} of the finite $N$ correlator including the sign (and without doing any combinatorics!). For generic values of the parameters the correlator is exponentially large. Note that in this notation some of the terms inside the bracket are negative but this form will be useful for comparing with the saddle point result for the symmetric representations. 
\subsection{AdS Giant Graviton}
We now comment on the analogous calculation for Young diagrams with a single row instead of column. In the large $N$ limit the normal ordered version of these tensors is given by 
\begin{equation}
    \mathbb{Q}[\chi_{\texttt{sym}, \Delta}]\simeq  \oint ds\; \text{He}_{N+ \Delta-1}(\sqrt{N} s) \;\frac{1}{\det\left(z+\bar{z}+s\right)}+ O(1/N).
\end{equation}
The contour of integration for $s$ encloses all poles in integrant. For general characters we do not have a simple way of evaluating the expectation value of this operator  at finite $N$, although they should be related to the correlation functions of determinants by Hilbert transform \cite{Kimura:2021hph}. However in the large $N$ limit we expect that the correlator is given by a saddle point computation.

For the ring distribution, the calculation turns out to be identical to the one for the determinant operator with the replacement $\varepsilon\rightarrow\frac{N+ \Delta-1}{N}$ and changing the sign of $N$ in $\Phi_R(\sigma)$. The saddle points are the same as before with the appropriate replacements of the energy $\varepsilon$ as a function of the dimension $\Delta$. This replacement is consistent with particle-hole duality. For the ring distribution this is given by

\beq
C^{\texttt{symmetric}}_{KK \Delta}\sim\frac{(N+\Delta-1)}{N^{\frac{1}{2}\left(N+\Delta-1\right)}\sqrt{\mathsf{N}^{\text{s}}_{\Delta}}}\int_{\Gamma} d\sigma\, \oint \frac{d\alpha\, e^{-\frac{N}{2}\alpha^2+ N \alpha \sigma}}{2\pi i \,\alpha^{(N+\Delta-1)+1}} \; \; \exp\left[-N \,\int dzd\bar{z}\,\rho_K(z, \bar{z})\, \log(z+ \bar{z}+\sigma)\right]. \label{sym_HHGG}
\eeq
Let us consider the case when $4 \gamma \geq (\varepsilon+3)(\varepsilon-1)= \delta(\delta+4)$. As we will see for these choices of parameters we safely can reuse the expression for $\Phi_K(\sigma)$ without worrying about picking branches for the logarithm.  In those cases the saddle point is real and takes the form
\begin{equation}
\begin{aligned}
    \sigma^*&=\frac{2\varepsilon(2\gamma+1) -\varepsilon^2-1}{\sqrt{2(\varepsilon^2-1)(2\gamma+1-\varepsilon)}},\\
    \alpha^*&=\sqrt{
    \frac{\varepsilon^2-1}{2(2\gamma+1-\varepsilon)}}.
\end{aligned}
\end{equation}
The saddle points are clearly analytic continuations of the saddle points for the antisymmetric representations and in the relevant set of parameter space the saddle points lie on the positive real axis. The asymptotics of the structure constant are then given by 
\begin{equation}
  C^{\texttt{symmetric}}_{KK \Delta}\sim \varepsilon^{\frac{N}{2}\varepsilon} \left[\frac{(\varepsilon+1)^{\varepsilon+1}(\varepsilon-1)^{\varepsilon-1}(2\gamma+1-\varepsilon)^{2\gamma+1 -\varepsilon}}{2^{2\gamma+1 +\varepsilon}\,\gamma^{2\gamma}}\right]^{-\frac{N}{2}}.
\end{equation}

One remarkable property is that the term in the brackets is precisely the same as the term appearing for the antisymmetric representations, the only change being the replacement $N\leftrightarrow-N$ and $\varepsilon=\frac{N+\Delta-1}{N}$. One important point is that whenever $4 \gamma = (\varepsilon+3)(\varepsilon-1)= \delta(\delta+4)$ the position of $\sigma^*$ lies at the edge of the eigenvalue distribution, so the expression for $\Phi_K(\sigma^*)$ becomes discontinuous for smaller values of $\gamma$. For the choices of parameters that we are considering we can check that $\sigma^*$ larger than the position of the edge of the distribution $2\sqrt{\gamma+1}$. This means that we can interpret the insertion of a symmetric representation character as tunneling an eigenvalue to a position $\sigma^*$ outside the cut.  

Although we only explicitly computed the asymptotics for backgrounds created by character type operators, the formulae \eqref{BBD-integrand-saddle}, \eqref{sym_HHGG} are valid for general backgrounds such as those created by exponential and coherent state operators.

 \section{HHH Correlators}
 \label{HHH Correlators}
Here we consider some examples of correlators where all three operators are huge, i.e. they are $\mathcal{O}(N^2)$. In particular, we consider the case of exponential operators with cubic  potential and the case of rectangular characters, both of which are analytically tractable. 

For the HHH case, it is convenient to treat all three operators on equal footing. Therefore, we work with the Hermitian three-matrix integral representation~\eqref{hermitian3Matrix}. Another useful representation is obtained by decoupling the integrals in $M_1,M_2,M_3$ in~\eqref{hermitian3Matrix} by introducing a Gaussian integral $\int DX\ e^{-\frac N2\,\tr(X-\sum_i M_i)^2}$. Upto some irrelavant normalization factors, this then results in
\beq
    \mathcal{Z} = \int DX\, e^{-\frac N2\, \tr X^2} \prod\limits_{i=1}^3 \mathbb{Q}_{O_i}(X)\,,\label{disentanglingX}
\eeq
where $\mathbb Q$ is defined in \eqref{qtransformDef}.


\subsection{Three Rectangular Young Tableaux}
\begin{figure}
    \centering
    \includegraphics[width=\linewidth]{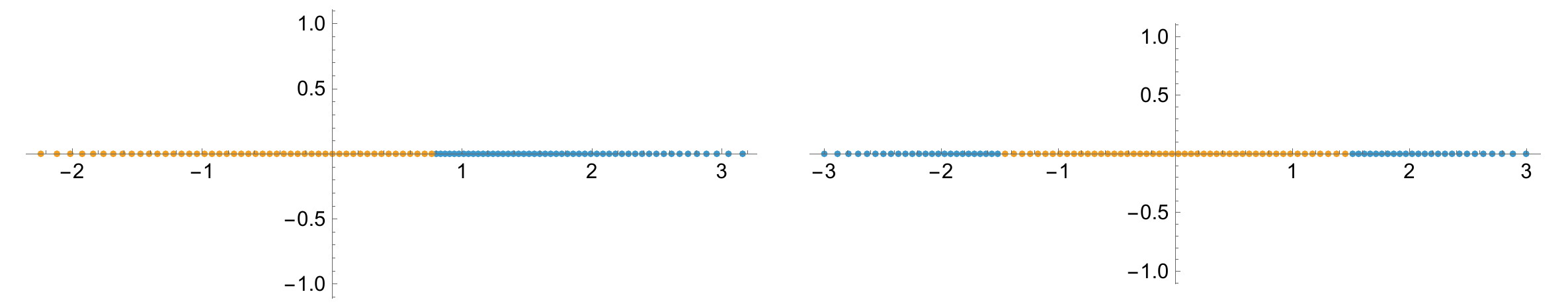}
    \caption{Two solutions of the discretized SPEs for $N=50$ and $K=50$. The positions of $x$'s are shown in blue and those of $y$'s are shown in orange. On the left, we have a single cut for $x$ and a single cut for $y$. On the right, we have one cut for $y$ and two symmetric cuts for $x$. The two-cut saddle is the dominant one.}
    \label{fig:discreteSPESols}
\end{figure}

\begin{figure}
    \centering
    \includegraphics[width=\linewidth,trim=0cm 21cm 0cm 0cm,clip]{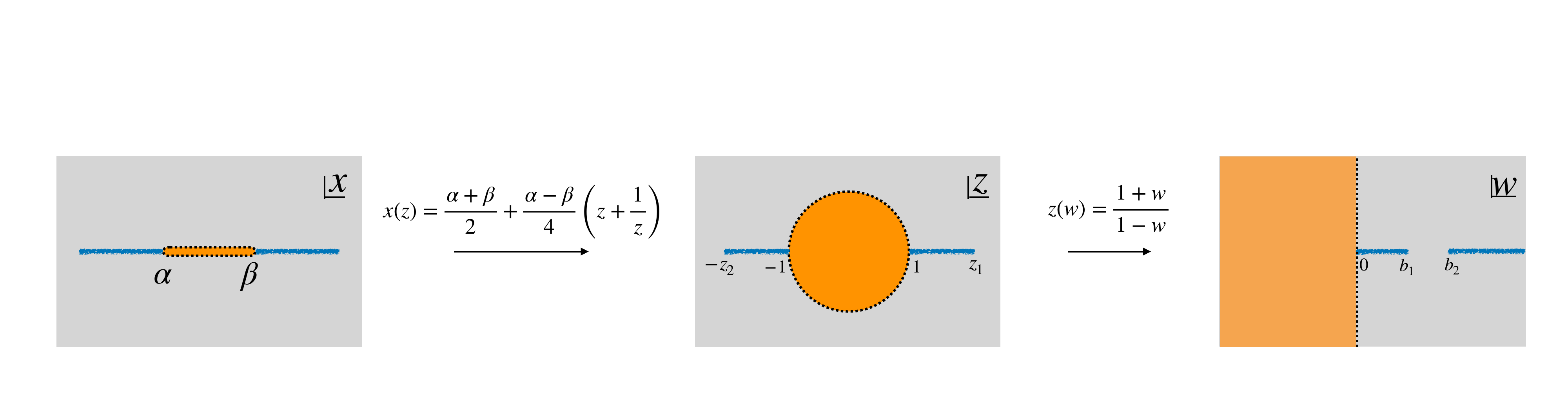}
    \caption{A series of conformal maps $x\rightarrow z\rightarrow w$ that brings the $x$-SPE \eqref{OGspeRect} to the nice form \eqref{speRectLikeOn} in $w$. The cuts in the resolvent are denoted in blue.}
    \label{fig:changeOfVariables}
\end{figure}

The case of three equal rectangular Young Tableaux was considered in \cite{Kazakov:2024ald}
\beq
    \mathcal Z = \int DX\, e^{-\frac N2 \tr X^2} Q_K(X)^3\,, \label{rectHHHdef}
\eeq
where $Q_K(X)$ is the $\mathbb Q$-transform of a rectangular Young Tableau with $N$ rows and $K$ columns. In \cite{Kazakov:2024ald}, the saddle point equations were simplified, but were not solved. In this section, we complete the solution. As shown in Appendix \ref{QtransformApp}, $Q_K(X)$ is given by
\beq
    Q_K(X) = \int \prod_{i=1}^K dy_i\ e^{-\frac N2 \sum y_i^2} \prod_{\substack{1\le i\le N\\1\le j\le K}} (x_i-y_j)\,.\label{rectQtransform}
\eeq
We can now plug this into \eqref{rectHHHdef} and then evaluate the integral using the saddle point method at large $N$. The integral is over $N$ $x$-variables and $K$ $y$-variables -- let their densities be $\rho(x)$ and $\sigma(y)$ respectively. Let us also define the corresponding resolvents $G(x)$ and $g(y)$ which obey the following saddle point equations in the continuum limit
\begin{equation}
    \begin{aligned}
        -y + \frac2\gamma \text{Re}\,g(y) + G(y) &= 0\,,\\
        -x + 2\, \text{Re}\,G(x)+\frac3\gamma g(x)&=0\,,\label{OGspeRect}
    \end{aligned}
\end{equation}
where $\gamma\defeq\frac NK$ is an $\mathcal O(1)$ number. Note that in addition to the usual Vandermonde repulsions, we have repulsive forces between the $x$'s and $y$'s coming from the $(x_i-y_j)$ factors in \eqref{rectQtransform}. In order to gain some intuition for the equilibrium configurations, we can discretize these saddle point equations and solve them numerically. In figure \ref{fig:discreteSPESols}, we plot two solutions of the discrete SPEs for $N=50$ and $\gamma=\frac NK=1$. An interesting feature is that the cuts of $x$ and $y$ touch each other -- this is natural because the logarithmic repulsions lead to separations of $\mathcal{O}(1/N)$ between them. Furthermore, we have a large number of solutions of the SPEs with different cut topologies. In figure \ref{fig:discreteSPESols} we consider the cases with a single cut for $y$'s and one and two cuts for $x$'s. With the discrete SPE numerics, one can check that the dominant saddle is the two-cut one.
Now, following \cite{Kazakov:2024ald}, we can solve the first SPE in \eqref{OGspeRect} and plug it into the second one for $x$. We review the details of this step in Appendix \ref{ThreeRectanglesAppendix}. It is convenient to perform a series of change of variables $x \rightarrow x(z) \rightarrow x(z(w))$ (see figure \ref{fig:changeOfVariables}) which are defined as
\begin{equation}
        x(z) = \frac{\alpha+\beta}2 + \frac{\alpha-\beta}4\left(z+\frac1z\right), \qquad z(w) = \frac{1+w}{1-w}\,.
    \label{variableChangeRect}
\end{equation}
Here, $z$ is a Joukowsky variable that uniformizes the cut in $y$ and $z(w)$ is a Möbius transformation. We also define a new function $f(w)$ which is the resolvent of $x$ up to some rational terms. Explicitly, it is related to the density $\rho(x)$ via
\begin{equation}
    f(w) = z(w) \left(\int d\xi \frac{\rho(x(\xi))x'(\xi)}{z(w)-\xi}\right) + \frac1\gamma -\frac{(\alpha-\beta)^2}{16 z(w)^2} + \frac{\beta^2-\alpha^2}{8z(w)}\,,\nn
\end{equation}
where $x(\xi)$ and $z(w)$ are defined in \eqref{variableChangeRect}. With these definitions, the SPE for $x$ is now converted into the following simple equation
\beq
    f(w+i\epsilon) + f(w-i\epsilon) + f(-w) = 0\,, \label{speRectLikeOn}
\eeq
for $w$ being on the cut (see Appendix \ref{ThreeRectanglesAppendix} for details). This is the same saddle point equation as for the $O(n)$ model on random graphs\cite{Eynard:1992cn}. In particular, this corresponds to $n=1$,
i.e. the Ising model on random graphs~\cite{Kazakov:1986hu,Boulatov:1986sb}. While the equation is the same, $f(w)$ has very different analytic properties which changes the solution. Let us summarize them:
\begin{itemize}
    \item Zeros: Requiring that $g(y)=\frac1y+\mathcal{O}(\frac1{y^2})$ imposes that $f(w)$ has the following zeros
    \beq
        f(0) = f(\infty)=0\,.\nonumber
    \eeq
    \item Asymptotics: Expanding near $w=1$, which corresponds to $x\rightarrow\infty$, we get
    \beq
        f(w) = 1+\frac1\gamma + \mathcal{O}(w-1)\,.\nonumber
    \eeq
    \item Poles: By construction, $f(w)$ has a double pole at $w=-1$
    \beq
        f(w) = -\frac{(\alpha -\beta )^2}{4 (w+1)^2}+\frac{\beta  (\beta -\alpha )}{2 (w+1)} + \left(\frac{1}{16} (\alpha -\beta ) (\alpha +3 \beta )+\frac{1}{\gamma }\right) +\mathcal{O}(w+1)\,.\label{doublePolefw}
    \eeq
    \item Analyticity: Let us focus on the two cases considered in figure \ref{fig:discreteSPESols}. As discussed above, the cuts in $x$ and $y$ touch, which means that the branch points are at $w=0$ and/or $w=\infty$. Explicitly, we have the following cuts
    \begin{equation}
        \begin{cases}
            \texttt{one-cut: } w\in (0,b)\\
            \texttt{two-cut: } w\in (0,b_1)\cup(b_2,\infty)
        \end{cases}\nonumber
    \end{equation}
\end{itemize}

Now, let us solve \eqref{speRectLikeOn}, taking inspiration from the  solution for $O(1)$ model on random graphs \cite{Eynard:1992cn,Eynard:1995nv,Eynard:1995zv,Kostov:2006ry}. We define the following auxiliary functions
\beq
    f_\pm(w) = \pm \frac{1}{i\sqrt3}\left(e^{\mp \frac{2\pi i}3} f(w) - e^{\pm \frac{2\pi i}3}f(-w)\right)\,,\nonumber
\eeq
such that by construction, $f_-(w)=f_+(-w)$. In terms of these auxiliary function, the SPE takes the form
\begin{equation}
    f_\pm(w-i\epsilon) = e^{\pm \frac{2\pi i}3}f_\mp(w+i\epsilon)\,.\label{SPEfpfm}
\end{equation}
Next, we define two more auxiliary functions $r(w)$ and $s(w)$ as follows
\begin{align*}
    r(w) &= f_+(w)f_-(w) = f(w)^2 + f(-w)^2 + f(w)f(-w)\,,\\
    s(w) &= \frac12 (f_+(w)^3+f_-(w)^3)\,.
\end{align*}

We can easily see from the SPEs that the discontinuities of $r(w)$ and $s(w)$ for $w$ on the cut of $f(w)$ vanish. Also, these functions are symmetric under $w\rightarrow-w$. Therefore, $r(w)$ and $s(w)$ are meromorphic even functions. We can also invert the above equations to write $f_\pm(w)$ in terms of $r(w)$ and $s(w)$ as follows
\beq
    f_\pm(w)^3 = s(w) \pm \sqrt{s(w)^2 - r(w)^3}\,.\label{fpmsrdef}
\eeq

Now, we know from \eqref{doublePolefw} that $f_\pm$ must have double poles at $w=\pm 1$. Therefore, $r(w)$ has a fourth order pole at $w=\pm 1$. Naively, $s(w)$ has a sixth order pole at $w=\pm 1$, but the leading order cancels so that $s(w)$ also has a fourth order pole. We parameterize these even meromorphic functions as
\beq
    r(w) = \frac{c_2 w^2+c_4 w^4+c_6 w^6}{(w^2-1)^4}\ ,\qquad s(w) = \frac{d_2 w^2+d_4 w^4+d_6 w^6}{(w^2-1)^4}\,.
\eeq
Recall that we needed $f(0)=f(\infty)=0$, which is why there are no $w^0$ and $w^8$ terms in the numerators. With these definitions, the spectral curve $\mathcal S$ of the rectangle three point function takes the following simple form
\beq
    \mathcal{S}(f,w) = f^3 -3r(w)f -2 s(w) = 0\,.
\eeq
We still need to solve for the moduli $c_i$ and $d_i$ of the curve by imposing the conditions that must be obeyed by $f(w)$ that were listed above. In particular, we must impose that the we have only two (four) branch points on the first sheet for the one (two) cut case. Let us explicitly solve for the moduli for these cases.

\subsubsection{Two-cut Solution -- Dominant Saddle}
\begin{figure}
    \centering
    \begin{subfigure}[t]{0.47\textwidth}
        \centering
        \includegraphics[width=\linewidth]{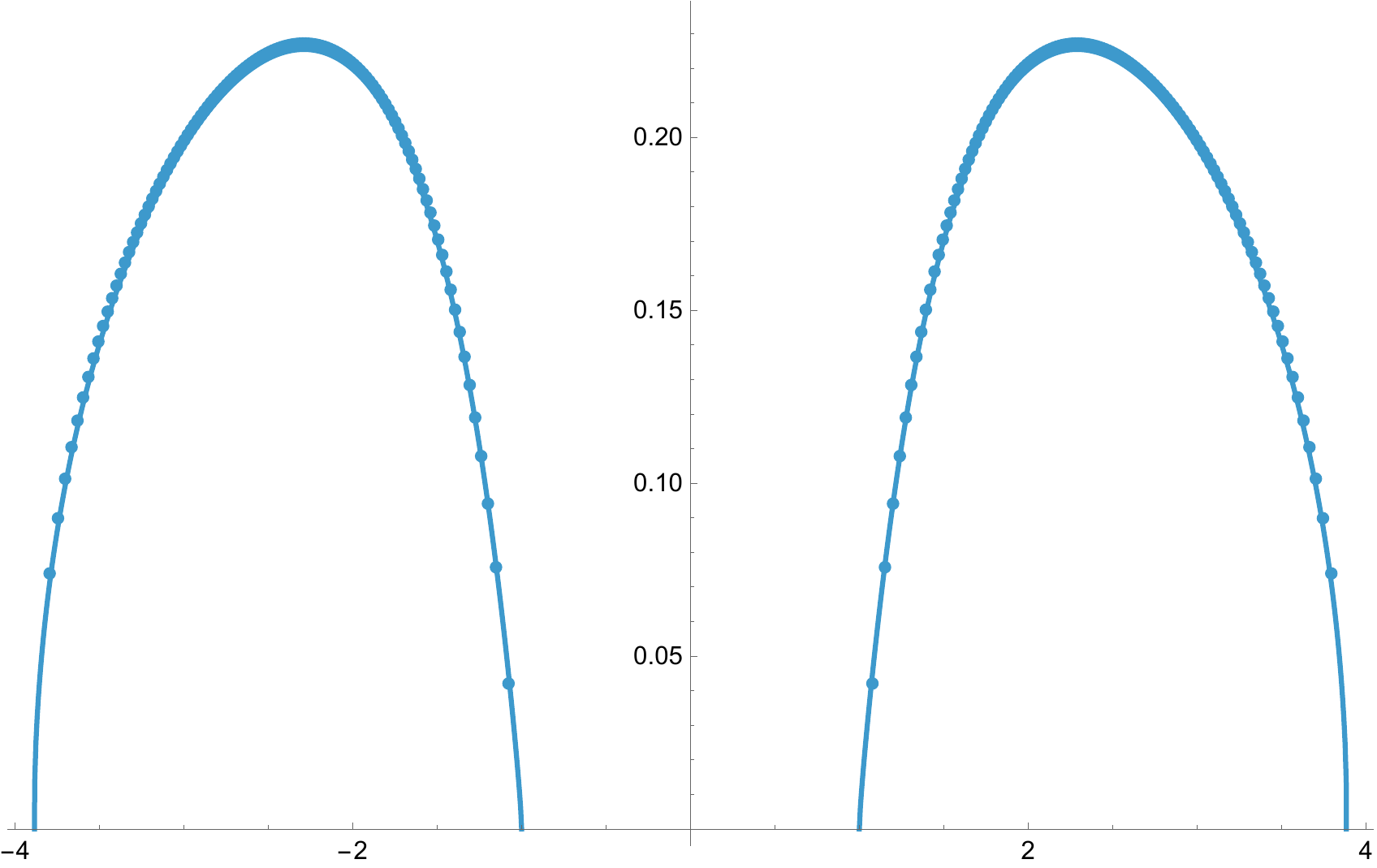}
        \caption{Two cuts}
    \end{subfigure}\ \ \ \ \ 
    \begin{subfigure}[t]{0.47\textwidth}
        \centering
        \includegraphics[width=\linewidth]{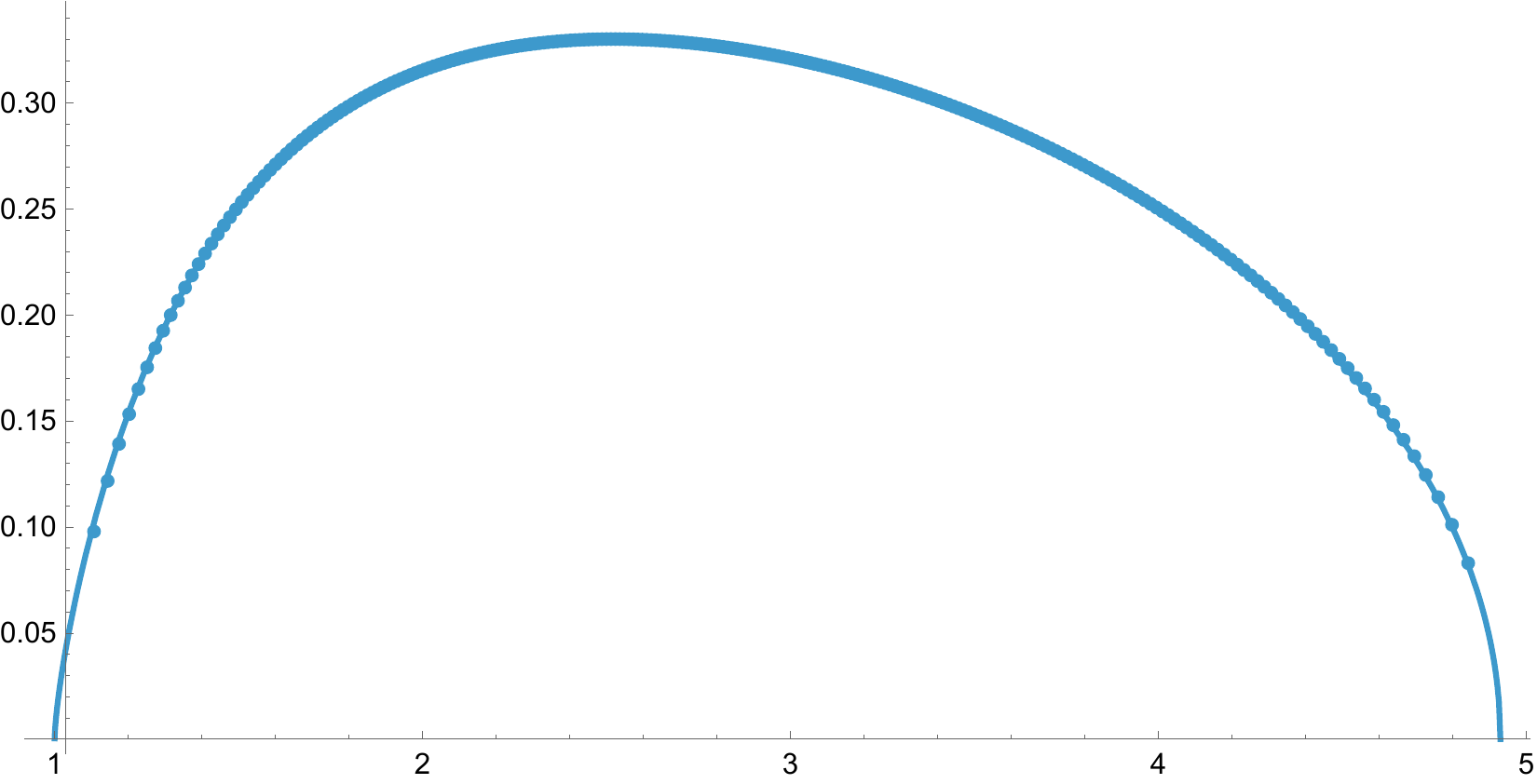}
        \caption{One cut}
    \end{subfigure}
    \caption{Eigenvalue densities of $F(z)$ for $\frac NK=1$. The blue dots are from the solution of the discretized saddle point equations with $N=K=250$ and the solid lines are analytic densities derived above}
    \label{fig:densityAnalyticNumeric}
\end{figure}

For the symmetric two-cut case, the branch points in $y$ are at $\alpha$ and $\beta=-\alpha$. The cuts in $x$ are also symmetric. This implies that $f(w)$ has the symmetry,
\beq
    f(w)=f\left(\frac1w\right)\,,\nonumber
\eeq
and the cuts are on $(0,b)\cup(\frac1b,\infty)$. The $w\leftrightarrow\frac1w$ symmetry imposes that $s(w)$ and $r(w)$ take the form
\beq
    r(w) = \frac{c_2(w^2+\frac1{w^2}) + c_4}{(w-\frac1w)^4} \ ,\qquad s(w) = \frac{d_2(w^2+\frac1{w^2}) + d_4}{(w-\frac1w)^4}\,.
\eeq
Imposing the correct behavior at $w=\pm 1$ gives us
\begin{equation}
    \begin{aligned}
        c_2 &= \frac{4\alpha ^2}{3} \left(\alpha ^2-1\right)-\frac{4 \alpha ^2}{\gamma }\,,\\
        c_4 &= \frac{8\alpha^2}{3} \left(\alpha ^2+1\right)+\frac{8 \alpha ^2}{\gamma }\,,\\
        d_4 &= -2d_2-\frac{8\alpha^4(1+\gamma)}\gamma\,.
    \end{aligned}
\end{equation}
We are now left with two parameters $\alpha$ and $d_2$. In order to fix these, we need to impose that the only branch points in \eqref{fpmsrdef} are at $w=0,b \text{ and }\frac1b$. We must have
\beq
    s(w)^2-r(w)^3 = \frac{w^4 (w^2-b^2)(1-b^2 w^2) P(w^2)^2}{(w^2-1)^{12}}\,,\nn
\eeq
where $P(x)$ is a cubic. Using the symmetry $w\leftrightarrow\frac1w$ we can bring it to the following form
\beq
    s(w)^2-r(w)^3 = \frac{(w^2-b^2)\left(\frac1{w^2}-b^2\right)\left(q_1\left(w+\frac1w\right) + q_3\left(w^3+\frac1{w^3}\right)\right)^2}{(w-\frac1w)^{12}}\,.\nn
\eeq
We can now solve this equation for $b, d_2,\alpha,q_1$ and $q_3$. For instance, for $\gamma=1$, we get
\begin{eqnarray}
    \alpha= 1.5143, \quad d_2&=&0.46865, \quad d_4 = -84.601\nn\\
    b=0.59054, \quad c_2&=&-5.2188,\quad c_4=38.482\nn\\
    q_1=113.22, &~& q_3=0.79360\nn
\end{eqnarray}
In figure \ref{fig:densityAnalyticNumeric}, we plot the density of $x$ eigenvalues in the Joukowsky variable ($z$) and there is perfect agreement with numerics.
\subsubsection{One-cut Solution}
For the one-cut case, we could follow the same steps as above but we will have many more equations due to the lack of $w\leftrightarrow\frac1w$ symmetry. So, motivated by the critical Ising model solution \cite{Eynard:1992cn}, let us make the following ansatz
\beq
    f_\pm(w) = \left(\sqrt{1-\frac{b^2}{w^2}} \mp \frac{ib}w\right)^{\frac13}\left(A(w)\sqrt{1-\frac{b^2}{w^2}} \pm B(w)\frac{ib}w\right)\,,
\eeq
where $A(w)$ and $B(w)$ are even meromorphic functions with double poles at $w=\pm1$.
\begin{eqnarray*}
    A(w) &=& \frac{c_2 w^2}{(w^2-1)^2}\,,\\
    B(w) &=& \frac{d_2 w^2+d_4 w^4}{(w^2-1)^2}\,.
\end{eqnarray*}
There are 6 unknowns -- $(\alpha,\beta,b,c_2,d_2,d_4)$ and 6 equations, from the expansions around $w=\pm 1$. To illustrate the solution, let's consider a particular value of $\gamma=\frac NK=1$. The solution for this case is
\begin{eqnarray*}
\alpha=0.798989, \quad \beta&=& -2.39611, \quad b=0.662703\\
c_2=5.67715, \quad d_2&=&2.48655, \quad d_4=3.67324
\end{eqnarray*}
Once again, as seen in figure \ref{fig:densityAnalyticNumeric}, we have excellent agreement with the density of solution of discrete SPEs.

\subsection{Three-point functions for exponential operators}
\label{sec:3point-Potts}

Let us consider three-point functions for the exponential operators. We take all three operators to have the form $e^{\tr V_{\boldsymbol{t}}(Z)}$, and we restrict our attention to the potential
\begin{equation}
    V_{\boldsymbol{t}}(Z) = t_2 \,  Z^2 + t_3 \,  Z^3.
\end{equation}
As we see shortly, the matrix integral for the three-point function in this case will reduce to the partition function for the 3-state Potts model on random graphs \cite{Kazakov:1987qg,KOSTOV1989295,Daul:1994qy}. This model was extensively studied in the literature, with the main focus on finding the critical points, critical exponents, and algebraic curve \cite{Zinn-Justin:1999qww,Kulanthaivelu:2019zia,Kulanthaivelu:2019atg,Atkin:2015ksy,Eynard:1999gp}. However, to the best of our knowledge, there was no explicit calculation of the partition function, which is of primary interest to us.

As shown in section \ref{sec:threePointMatrixIntegrals}, the three-point function for these operators can be written as:
\begin{equation}
    C\,[t_2,t_3] = \frac{1}{\mathcal{Z}} \int \prod_{a=1}^3 dM_a\, e^{N\,\tr\left[-\sum_a \left(\frac{M_a^2}2-V_{\boldsymbol{t}}\left(\kappa_a M_a\right)\right)+\sum_{a<b}M_aM_b\right]},
\end{equation}
where $\kappa_a$, defined in \eqref{rescaleThreePt}, account for the spacetime dependence of the correlators. Here, we will consider an equilateral configuration such that $\kappa_{a=1,2,3}=\kappa$. We can disentangle the $M_aM_b$ interactions by introducing an auxiliary matrix $X$ as in \eqref{disentanglingX} which leads to
\begin{equation}
    C\,[t_2,t_3] = \frac{1}{\mathcal{Z}}\int dX \, e^{-\frac{N}{2} \, \tr X^2}  I(X)^3\,, \qquad I(X) \equiv \int dM \,  e^{N \, \tr \left(-\frac{1}{2\alpha^2} M^2 -\frac{{g}}{3 \alpha^3} M^3+XM\right)},
\end{equation}
where we have introduced the couplings
\begin{equation}
    \frac{1}{\alpha^2}=1-2t_2\kappa^2\,,\qquad  g = -3 \alpha^3 \,t_3\kappa^3.
\end{equation}

The integral $I(X)$ may be written as a function of the density $\rho(x)$ of eigenvalues of the matrix $X$ using the loop equations \cite{Gross:1991aj,Daul:1994qy}. We are interested in calculating the free energy $\mathcal{W}$ for the correlator $ C\, [t_2,t_3] = e^{N^2 \mathcal{W}[\rho(x)]}$.
Discretizing the problem, we can solve the saddle-point equation numerically and we find that only the one-cut solution exists. We will assume this in the following.

Instead of solving the saddle-point equations for $x$, it is convenient to reformulate the problem in terms of the variable $z = \sqrt{\alpha x+c}$ \cite{Daul:1994qy}. Defining the resolvent 
\begin{equation}
    G_0(z) \equiv \int^B_A dp \frac{\pi(p)}{z-p}, \qquad z = \sqrt{\alpha x+c},
\end{equation}
where we have defined the new density $\pi(z)$ as
\begin{equation}
    \pi(z) \equiv \rho(x(z)),
\end{equation}
it turns out that the saddle-point equation for the $G_0(z)$ is the same as for the $O(n)$ model \cite{Eynard:1992cn,Eynard:1995nv,Eynard:1995zv,Eynard:1999gp} with $n=-1$.
From the analytic structure of the resolvent, we see that on the first (physical) sheet it has a single cut on the line $[A,B]$. Then, passing onto the next sheet we find that there is a reflected cut on the negative axis. 

From the saddle-point equation, we can write down cubic algebraic equation for the resolvent
\begin{equation}
    G^3(z) - 3 G(z) r(z)-\frac{2}{3\sqrt{3}\alpha^3} R(z)=0,
\end{equation}
where $r(z)$ is an even polynomial of order $4$ and $R(z)$ is an odd polynomial of order $5$ \cite{Eynard:1992cn}. 
\begin{equation} \label{non_crit_sol}
\begin{aligned}
    G(z) &= - ( e^{\frac{i \pi}{6}} G_+(z) + e^{-\frac{i \pi}{6}} G_-(z)), \\
    G_{\pm}(z) &= \frac{1}{\sqrt{3}\alpha} \big( \sqrt{(z^2-A^2)(z^2-B^2)} S(z) \pm i R(z)   \big)^{\frac{1}{3}},
\end{aligned}
\end{equation}
where $S(z)$ is even monic polynomial of order 4.

 Here the endpoint of the cut $A,B$ and the parameters of the polynomials can be found as functions of $\alpha$ and $g$ via SPEs, which boil down to a system of $10$ algebraic equations for 10 unknwons that impose the correct asymptotics and analyticity properties of $G(z)$, see Appendix \ref{app-3pt-exp} for the details.
We find that the free energy, evaluated on the solution of SPEs, can be rewritten as follows:
\beq\label{freeW}
\begin{aligned}
    \mathcal{W}[\pi(z)] &= - \frac{1}{2\alpha^3}G^{(5)}+\frac{2 }{\alpha g^{1/2}} G^{(4)} + \bigg( \frac{c}{\alpha^3} -\frac{3 }{2\alpha g} \bigg) \, G^{(3)}  + \frac{3}{\alpha g^{1/2}} \bigg( \frac{1}{4g}-c \bigg) G^{(2)}  \\
    &\qquad+\frac{1}{\alpha} \int dp \,\pi (p)\, p \log p+ \bigg(\frac{3c}{4g}-\frac{c^2}{4\alpha^2} \bigg)\bigg(1 + \frac{1}{\alpha}\bigg)-\frac{3}{4}\log g - \log \alpha - \frac{1}{4g^2},
\end{aligned}
\eeq
where we have denoted
\begin{equation}
    G^{(n)} \equiv \int dz \, \pi (z) z^n.
\end{equation}
These integrals can be found from the expansion of \eqref{non_crit_sol} at $z \to \infty$.
The density $\pi(z)$ is written as
\begin{equation}
    \pi(z) = \frac{1}{2\pi \alpha } \bigg[ \bigg(  R(z) + \sqrt{(B^2-z^2)(z^2-A^2)}  \bigg)^{1/3}  - \bigg(  R(z) - \sqrt{(B^2-z^2)(z^2-A^2)}  \bigg)^{1/3}  \bigg].
\end{equation}

\begin{figure}[H]
    \centering
    \includegraphics[width=0.7\linewidth]{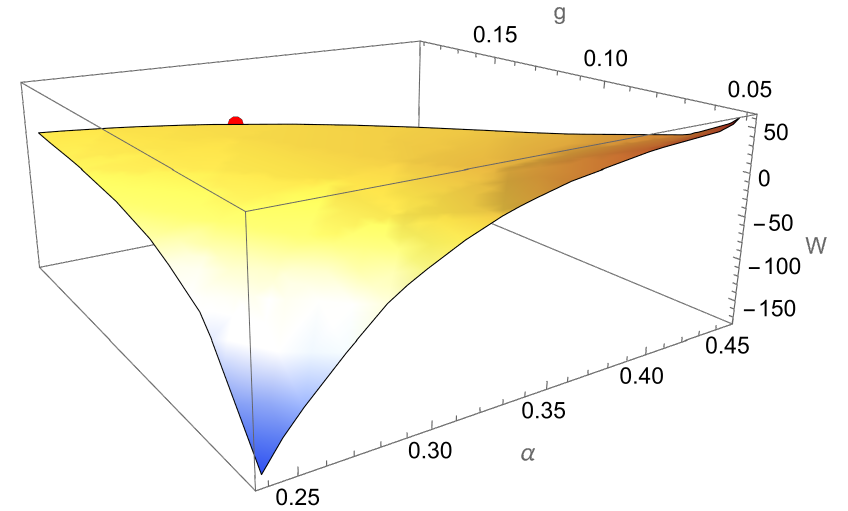}
    \caption{Behavior of the free energy $\mathcal{W}$ as a function of $\alpha$ and $g$. The red dot here is the position of the critical point. The value of the correlator for given $\alpha$ and $g$ is given by $C = e^{N^2 \mathcal{W}}$.}
    \label{fig:3pt-exp-free-energy}
\end{figure}

Then, solving the SPEs numerically, we find that the free energy behaves as 
depicted on Fig.\ref{fig:3pt-exp-free-energy}.

For a fixed $\alpha$, increasing $g$ we reach a critical line $g=g_c(\alpha)$, on which $A=0$, i.e. at this line, the two cuts ($[A,B]$ and $[-B,-A]$) of the resolvent in the unphysical Riemann sheets touch. After this line, the physical solution cease to exist. The red dot here denotes the postion of the critical point, which is determined below by examining the critical solution.

Let us consider the critical case, which corresponds to setting $A=0$. Then, we can write down the following solution for the resolvent:
\begin{equation}
    \begin{aligned}
    G(z) &= - \left( e^{\frac{i \pi}{6}} G_+(z) + e^{-\frac{i \pi}{6}} G_-(z)\right)\,, \\
    G_{\pm} (z)&= \left(\sqrt{1-\frac{B^2}{z^2}} \pm \frac{iB}{z}  \right)^{-\frac{2}{3}} \left[ \frac{z^2}{\sqrt{3}\alpha}   \sqrt{1-\frac{B^2}{z^2}} \pm  i B \, b_2 z   \right]\,,
    \end{aligned}
\end{equation}
where $B$ and $b_2$ are some constants. 

\begin{figure}[H]
    \centering
    \includegraphics[width=8cm]{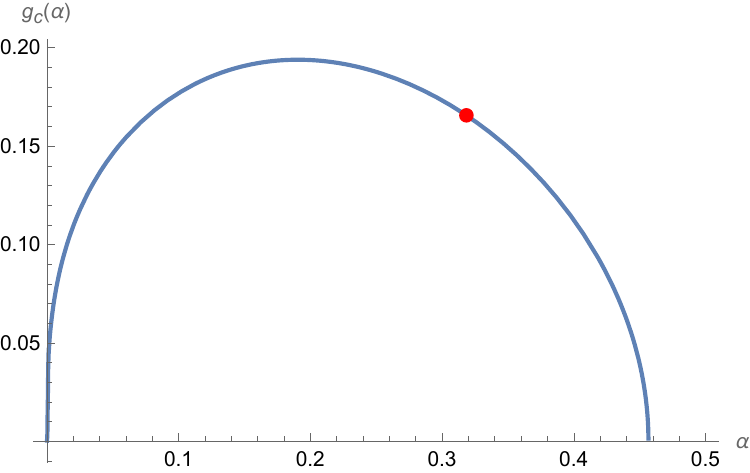}
    \caption{The critical line $g_c(\alpha)$. The red dot is the critical point $(\alpha_c,g_c(\alpha_c))$. }
    \label{exp_3pt_crit_line}
\end{figure}

Now, one needs to determine  $b_2$, $B$, and $c$ by comparing asymptotics of this solution with the required asymptotics for $G(z)$. we again arrive at the system of algebraic equations, solving which we obtain the critical line $g=g_c(\alpha)$. The details of this computation and the explicit expression for $g_c(\alpha)$ can be found in the Appendix \ref{app-3pt-exp}. The obtained critical line is plotted in Fig. \ref{exp_3pt_crit_line}.

We can obtain the critical point by imposing a constraint $\pi(z) \sim z^{\frac{5}{3}}$ as $z \to 0$. Taking the limit $z \to 0$, we find that at the critical point
\begin{equation}
    b_2^{\text{crit}} = \frac{1}{\sqrt{3}\alpha}.
\end{equation}
Combining it with the solution for the critical line, we find the position of the critical point, which agrees with \cite{Daul:1994qy, Zinn-Justin:1999qww, Bonnet:1999nf}:
\begin{equation}
    \alpha_c = \sqrt{\frac{1}{38} \left(\sqrt{47}-3\right)} ,\qquad g_c(\alpha_c) = \frac{1}{76} \sqrt{\frac{105}{19} \left(37 \sqrt{47}-225\right)}.
\end{equation}
The density of eigenvalues $\pi(z)$ behaves as follows outside and at the critical line:
\begin{figure}[H]
\centering
\begin{subfigure}{.5\textwidth}
  \centering
  \includegraphics[width=.8\linewidth]{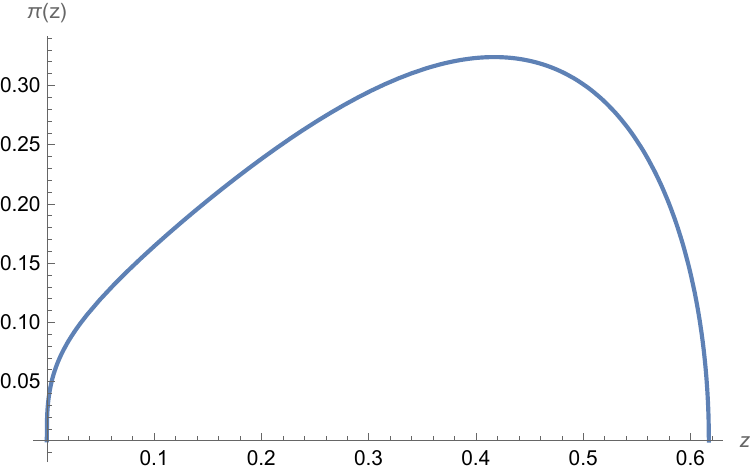}
  \caption{The density $\pi(z)$ on the critical line outside of the critical point. Here $\alpha=0.1$.}
  \label{fig:sub1}
\end{subfigure}%
\begin{subfigure}{.5\textwidth}
  \centering
  \includegraphics[width=.8\linewidth]{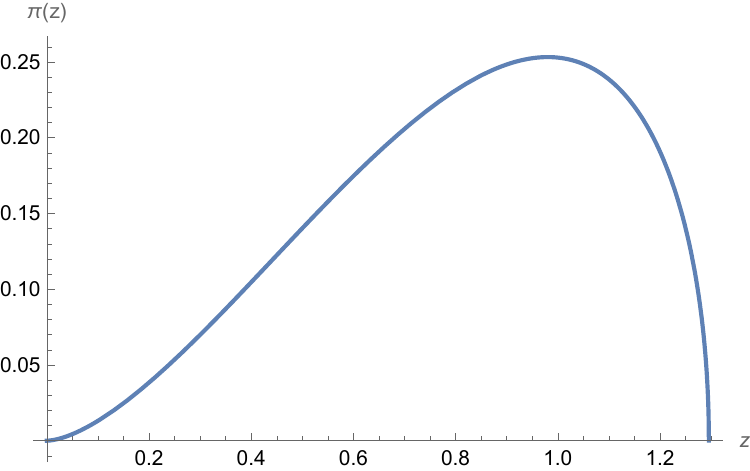}
  \caption{The density $\pi(z)$ at the critical point. Here $\alpha=\alpha_c$.}
  \label{fig:sub2}
\end{subfigure}
\caption{Comparison of the density of states $\pi(z)$ on the critical line outside and at the critical point. }
\label{fig:test}
\end{figure}


From the explicit free energy of the model~\eqref{freeW} we can see that in the critical regime\cite{KOSTOV1989295,Daul:1994qy}~\vol{We promised the solution for $\mathcal{W}$, so may be to put this of formula here, together with density, and refer to its parameters to App?}
\begin{equation}
    \frac{\partial^2}{\partial g^2} \mathcal{W} \sim (g_c-g)^{\frac{1}{5}}.
\end{equation}
 It corresponds to the non-critical string theory in terms of the  Liouville CFT  coupled to a matter  field with the central charge $c_{m}=4/5$ and the string susceptibility $\gamma_{str} = -1/5$~\cite{Kazakov:1987qg,Daul:1994qy}, as it should be for the 3-state Potts model coupled to 2d gravity.

\section{$\frac{1}{4}$- and $\frac{1}{8}$-BPS operators}
\label{Quarter BPS}
In general, the study of the matrix integrals related to the protected correlators  of  $\frac14$-BPS and $\frac18$-BPS operators is a difficult task hardly accessible via the standard large \(N\) matrix model approaches which work well for the $\frac12$-BPS case mostly considered before in this paper, such as diagonalization of matrices, saddle point calculations, orthogonal polynomials, and HCIZ integrals, etc.). The dual gravity description of  $\frac14$-BPS and $\frac18$-BPS operators is also much more intricate~\cite{Donos:2006iy, Chen:2007du} than  the LLM geometry~\cite{Lin:2004nb} for $\frac12$-BPS operators.
 So any solvable cases or new observations and mathematical analogies with the
known matrix models and QFTs, especially any signs of integrability (see next section), would be of a great physical importance. Previous approaches to this problem involve heuristic approximations by commuting matrix models \cite{Berenstein:2005aa, Berenstein:2008eg, Correa:2010zj, Filev:2014qxa} although this naive approximation is not under control; see also \cite{Kimura:2011df} for some evidence of the correspondence between Young diagrams and higher dimensional droplets. 

We will present in this section a few observations related to the correlation functions involving $\frac14$-BPS and $\frac18$-BPS operators which may be useful for the further progress on this very interesting subject. In particular, we will see that the holomorphic moments  of $X,Y,Z$ and the 3-point correlator of such a 1/4-BPS CS operator with two exponential 1/2-BPS operators with gaussian potential can be trivially computed. 
Another potentially promising  observation presented below relates the two-point functions of  $\frac14$-BPS and $\frac18$-BPS CS operators to Eguchi-Kawai quenched reduction of the principal chiral $\sigma$-model (PCM) in 2D and 3D, respectively.  

\subsection{Coherent State Form of $\frac18$-BPS and $\frac14$-BPS Operators}

The $\frac18$-BPS  coherent state operator is defined as a unitary integral~\cite{Berenstein:2022srd}
\begin{align}\label{BPS_CS}
O_{P}&=\int_{U(N)} dU\,e^{N\tr [U^{\dagger}P^{\mu} UX_{\mu}]}\,,\qquad \mu = 1,2,3,
\end{align}
where \(\{X_{1},X_{2},X_{3}\}=\{X,Y,Z\}\) are complex \(N\times N\)  matrices, \(P^{\mu}=\text{diag}\{p^{\mu}_{1},\dots,p^{\mu}_{N}\}\) are  diagonal matrices parameterizing the operator. The notation $p_i^\mu$ is chosen to make contact with the quenched momemtum prescription of PCM in the next section. The $\frac14$-BPS and $\frac12$-BPS cases correspond to \(P_{X}=0\) and \(P_{X}=P_{Y}=0\), respectively. 

The $\frac{1}{4}$-BPS coherent state operator has been shown to be annihilated by the two-loop non-planar dilatation operator in \cite{Lin:2022wdr} and it was conjectured in \cite{Pasukonis:2010rv, Lewis-Brown:2020nmg} that operators that are $\frac{1}{4}$-BPS at one-loop are not renormalized. The argument relies on the fact that the counting of quarter BPS states only changes when the gauge coupling is turned on \cite{Kinney:2005ej}. Here we restate the argument for the coherent state operators. The counting of states at weak coupling is then in one-to-one correspondence with representatives in the chiral ring. For quarter BPS states in the $\mathcal{N}=4$ theory these are given by invariants built from two commuting complex matrices. Classically this associated with the moduli space $(\mathbb{C}^2)^N/S_N$, and the representatives are obtained by quantizing the moduli space at a generic point. The wavefunctions in that case are holomorphic polynomials of the eigenvalues of complex matrices. Clearly these are exactly the same parameters controlling the BPS coherent state, so the number of states generated by the coherent state matches the counting of quarter BPS operators. Since the coherent state operator generates the same number of independent operators at two-loops as there are quarter BPS operators \footnote{This can be checked by expanding the exponential and performing the unitary integrations.}, we can expect that the coherent state remains in the kernel of the dilatation operator at all orders of perturbation theory. This is because at arbitrarily weak coupling the leading contribution to the anomalous dimension comes from the one-loop contribution and higher loop contributions can be taken arbitrarily small, so multiplet recombination cannot occur. Non-perturbatively one has to worry about possible recombinations, but in the non-planar regime, which is non-integrable, the level repulsion should prevent this from occurring. Similar arguments apply to $\frac{1}{8}$-BPS coherent state operator, since the $SU(2|3)$ sector is also closed to all loops. 

The pair correlator is computed by a  Gaussian integral with the matrix action \(S=\tr(\bar X^{\mu}X_{\mu})\), so that we are left with a single group unitary integral:
\begin{align}\label{2pointf}
\langle O_{P}&\bar O_{\bar{P}}\rangle=\int_{SU(N)} dg\,\,e^{\tr [g^{\dagger}\bar{P}^{\mu} g P_{\mu} ]}\,,\qquad g\in SU(N). 
\end{align}
In general this integral cannot be evaluated exactly by any known methods, except the 1/2-BPS case where it is reduced to the IZHC integral, see sec.\eqref{coherentStatesSection} discussing this case. However because these operators are coherent states all (anti-)holomorphic moments can be solved  in terms of the parameters $P_\mu$, i.e.
\begin{equation}
 \langle O_{P} \,\tr\left[f(X,Y,Z)\right]\,\bar O_{\bar{P}}\rangle = \tr\left[f(\bar{P}_X,\bar{P}_Y, \bar{P}_Z)\right]\langle O_{P}\bar O_{\bar{P}}\rangle.
\end{equation}
Notice that the ordering of the letters $X^\mu$ in $f(X,Y,Z)$ do not affect the value of the correlator because the $P_\mu$ commute.  Since these type of three-point correlators are holomorphic they are sums of extremal ones so they are protected in the quarter BPS case $P_X=0$\footnote{For example in this case there are no possible Wick contractions between $O_P$ and $\tr\left[f(X,Y,Z)\right]$.}. For more complicated moments mixing holomorphic and anti-holomorphic fields things become much more complicated. However some observations proposed  below give a hope for a deeper insight into this problem.

\subsubsection{Relation to Eguchi-Kawai reduction of principal chiral model}

Curiously, the correlator~\eqref{2pointf} appears to be equal to the partition function of Eguchi-Kawai (EK)  reduction of PCM with quenched momenta. PCM is a $D$-dimensional field theory with the partition function~\footnote{See \cite{Polyakov:1987hqn} for definition and description of physical properties of PCM.} 
\begin{align}\label{PCM}
Z_{\rm PCM}=\int [\mathcal{D}g]_{_{SU(N)}}\,\,e^{-\frac{1}{2\lambda}\int d^{D}x\,\tr[\p_{\mu}g \p^{\mu}g^{\dagger}]}\,,\qquad g(x)\in SU(N). 
\end{align}
 Formally, the action in this functional integral can be   reduced to~\eqref{2pointf} via the Eguchi-Kawai 0-dimensional reduction~\cite{Gross:1982at,Eguchi:1982nm} 
\begin{align*}
g(x)\to e^{-iP_\mu\, x^\mu}g\,e^{iP_\mu\, x^\mu},\quad \text{or with color indices: }\quad g_{kj}(x)\to e^{-ip_{k}\cdot x}g_{kj}e^{ip_{j}\cdot x}\quad \text{(no summation!)}
\end{align*}
where \(g\in SU(N)\) is already a coordinate independent group matrix. Substituting this prescription into the path integral \eqref{PCM} gives
\begin{align}\label{ZqBPS}
Z=\int dg\,\,e^{\frac{V_{D}}{2\lambda}\,\tr[ P^{\mu},\-g][ P_{\mu},\-g^{\dagger}]}=\int dg\,\,\exp\left(\frac{V_{D}}{2\lambda}\sum_{ k,j=1}^{N}(p^{k}-p^{j})^{2} |g_{kj}|^{2}\right),
\end{align}
where \(V_{D}\) is the space volume. This coincides, up to a trivial factor
 \(e^{-\frac{V_{D}}{\lambda}\tr P^{\mu}  P_{\mu} }\) and after the  rescaling  \(P\to \sqrt{\frac{2\lambda}{V_{D}}}P\), for \(D=1,2,3\) with the $\frac12$-, $\frac14$-, or $\frac{1}{8}$-BPS  correlator  \eqref{2pointf}, respectively.
The  parameters \(\{p^{\mu}_{1},\dots,p^{\mu}_{N}\}\)  (quenched momenta) should now be taken  real and randomly scattered in the momentum box: \(0<p^{\mu}_{j}<\Lambda\).~\footnote{The overall shift of momenta is irrelevant since \eqref{ZqBPS} depends only on their differences. }   They play the role of quenched momenta introduced in~\cite{Gross:1982at}.
The characteristic distance between the neighboring momentum points in the momentum volume \(\Lambda^{D}\) is \(\frac{\Lambda}{N^{1/D}}\), which is related to the space volume $V_{D}$ through the uncertainty relation \(\Lambda^{D}V_{D}\rightarrow(2\pi N)^{D}\).\footnote{Notice that in the reduced model the size of the matrix $N$ is not the same parameter  as in  the original unreduced QFT. It it also contains the information about the spacial size of the original QFT, as foolows from this argument.} 

It was claimed in original papers on EK reduction, such as  \cite{Gross:1982at}, that in the large $N$ 't~Hooft limit a reduced matrix theory, like PCM~\eqref{ZqBPS}, is equivalent to its  full $D$-dimensional prototype QFT.  
In our case, this equivalence  between the PCM QFT~\eqref{PCM} and its EK reduction~\eqref{ZqBPS}  would give a tempting perspective to study the 2D and 3D PCM\footnote{The 2D PCM is an interesting asymptotically free QFT with dynamically generated spectrum of particles.} in the 't~Hooft limit by employing the gravity duals of $\frac{1}{4}$-BPS and $\frac{1}{8}$-BPS operators proposed in~\cite{Donos:2006iy,Chen:2007du,Gava:2006pu}. Alternatively, one could hope to apply the integrability methods developed for 2D PCM~\cite{Polyakov:1983tt,Polyakov:1984et,Wiegmann:1984ec,Wiegmann:1984pcm,Wiegmann:1984ec}, especially powerful in the 't~Hooft limit~\cite{Fateev:1994ai,Fateev:1994dp,Kazakov:2019laa,Kazakov:2023imu}, to the study of $\frac14$-BPS correlator~\eqref{2pointf} at particular distributions of eigenvalues of the parameter matrices $P_\mu$.   Interestingly, the $\frac14$-BPS correlator  even inherits from PCM a certain integrability property at finite \(N\), at least on the classical level (see Appendix~\ref{sec:Integrability1/4BPS}). 

However, it is still not clear, at least to our knowledge, whether the quenched version of EK reduction of PCM~\eqref{ZqBPS}  fails or not to describe the correct asymptotically free behavior of the original PCM. The arguments of~\cite{Bringoltz:2008ek}   (and references therein) point out to the failure of   the quenched EK reduced lattice Yang-Mills theory to describe the right physical behavior, due to a pathological behavior of the e.v.'s of gauge matrices. But these arguments might be not applicable to the non-gauge theories, like PCM. Formally, one can reproduce the asymptotically free behavior (say, in the presence of external fields, as in~\cite{Fateev:1994dp,Fateev:1994ai} by expanding around the group unity: $g\simeq \mathbb{I}+iA-\frac{1}{2}A^{2}-i\frac{1}{6}A^{3}+\dots$ and doing the perturbation theory. However, non-perturbatively the other classical minima apart from unity, such as permutations, can contribute due to their large entropy and disorder the system even in the weak coupling limit. A further study is needed to clarify this question, may be a precision simulation of the quenched model or some deeper analytic arguments.   

\subsubsection{ Coherent state (CS) operators in $\beta$-deformed $\mathcal{N}=4$ SYM}
\label{quenchedPCM}

This subsection will be somewhat speculative.
It is tempting to introduce the analogue of the $\frac14$-BPS  CS operator~\eqref{BPS_CS} in $\beta$-deformation of \(\mathcal{N}=4\) SYM  theory as follows (see Appendix \ref{sec:gammaSYM} recalling the definition of the theory)
\begin{align}\label{PQtwisted}
O_{q}=\int_{SU(N)} dU\,e^{\tr(U^{\dagger}PUX+U^{\dagger}QUZ)},
\end{align}
where the coefficient matrices \(P,Q\) \(q\)-commute:
\begin{align}\label{qcom}
[P,Q]_{q}=0\,.
\end{align}  

The one-loop dilatation operator of this theory in the \(SU(2)\) sector was computed in~\cite{Fokken:2013mza} (see also \cite{McLoughlin:2020siu,McLoughlin:2020zew})\footnote{In planar limit it was computed up to 4 loops in~\cite{Minahan:2011bi}.}. It reads 
\begin{equation}\label{1loopD}
\mathcal{D}_2 = -\frac{2g_{\beta}^{2}}{N} \left( \operatorname{Tr}\Bigl([X,Z]_{q}[\check{X},\check{Z}]_{q}\Bigr) - \frac{\bigl(q - q^{-1}\bigr)^2}{N}\operatorname{Tr}(XZ)\operatorname{Tr}(\check{X}\check{Z}) \right),
\end{equation}
where \(|g_{\beta}|^{2}=\frac{g^{2}}{1-\bigl(q - q^{-1}\bigr)^{2}/N^{2}}\) and  \([X,Z]_{q}\equiv qXZ -q^{-1}ZX\) and  \(\check{X}\equiv \frac{d}{dX}\), etc. The last term comes from the double-trace counter-term added to the action to preserve the conformality.

It is evident that such operator has vanishing one-loop anomalous dimension
\begin{align*}
\mathcal{D}_2 O_{q}=0.
\end{align*}
The first term in \eqref{1loopD} gives zero due to~\eqref{qcom} and the second - due \(\tr(PQ)=0\) following from  the trace of~\eqref{qcom}.

If the operator \eqref{PQtwisted} is indeed protected, we could again use the complex  matrix measure \(dXdZ\exp\{-\tr(\bar{Z}Z+\bar{X}X)\}\) to compute the two point function, which would give for the two-point correlator of operators~\eqref{qcom} 
\begin{align}\label{PQtwistedCorr}
\langle O_{q}\overline{O_{q}}\rangle=\int [dg]_{SU(N)}\,e^{\tr(g^{\dagger}PgP^{\dagger}+g^{\dagger}QgQ^{\dagger})}.
\end{align}
This correlator looks similar  to the partition function of EK reduced {\it twisted} lattice PCM of the papers~\cite{Das:1983jv, Gonzalez-Arroyo:2018aus,Gonzalez-Arroyo:2022irw} 
but unfortunately it is not the same: the former contains also the complex conjugate terms in the exponent. 


 We don't know whether the CS operator \eqref{PQtwisted} is all-loop protected.
Unfortunately, the two- and higher-loop dilatation operator is still not worked out, so we cannot have a direct check.  The one loop argument could be made stronger if we new how to count all protected operators built out of $X$ and $Z$ fields in the $\beta$-deformed SYM and to make it sure that the CS operators \eqref{PQtwisted} provide the full list of them. However, we are unaware of any such counting in the literature.
Another way to see whether \eqref{PQtwisted} is protected is to try to match short operators produced by the expansion in powers of $P,Q$ of~\eqref{BPS_CS} with the protected  correlators  of operators from the chiral ring  computed in a few loops  in~\cite{Mauri:2006uw}.\footnote{We could introduce a similar CS operator in the $\gamma$-deformed $\mathcal{N}=4$  SYM, where the supersymmetry is completely broken, but it would be obviously not protected. } We leave the work on these issues for the future.    



\subsection{Three-point functions of one $\frac{1}{4}$-BPS or $\frac{1}{8}$-BPS correlator  with  two $\frac12$-BPS  operators }

It is established~\cite{Bissi:2021hjk} that the three-point  correlators of two $\frac12$-BPS operators and one $\frac12$-BPS, $\frac14$-BPS or $\frac18$-BPS operator are protected, see also \cite{Heslop:2001gp}. Let us consider the following example of such correlator: two exponential $\frac12$-BPS operators 
\begin{align*}
O_{\rm exp}[t]=e^{N\tr V_{\boldsymbol{t}}(Z)}\,,\qquad  V_{\boldsymbol{t}}(Z)=\sum_{k=1}^{q}t_{k}Z^{k}.
\end{align*}
and one CS $\frac14$-BPS operator
\begin{align*}
O_{A,B}&=\int_{U(N)}dU\,e^{N\tr [U^{\dagger}AU[Z+\bar Z+i(Y+\bar Y)]+U^{\dagger}BU[
i(Z-\bar Z)-(Y-\bar Y)]},
\end{align*}
where $A,B$ are mutually commuting Hermitian matrices.
Such a three point correlator  reduces to the integral
\begin{align*}
K[t,P]\sim\int d^{2N^{2}}Y \int d^{2N^{2}}Z\, e^{-N\tr(\bar ZZ+\bar YY)}O_{\rm exp}[t]\bar O_{\rm exp}[\bar t]O_{A,B}&.
\end{align*}
The integral over \(Y\) is trivial and it gives the unsignificant factor 
\begin{align}\label{PPbar}
e^{-N\tr \bar PP},\qquad P=\frac{1}{2}(A+iB),\quad \bar P=\frac{1}{2}(A-iB).
\end{align}
 Then this correlator is given by the following   integral over the unitary matrix
\begin{align*}
K[t,P]\sim\int [dU]_{_{SU(N)}} \int d^{2N^{2}}Z\, e^{N\tr[-\bar ZZ+ V_{\boldsymbol{t}}(Z)+\bar  V_{\boldsymbol{t}}(\bar Z)+U^{\dagger}PUZ+U^{\dagger}\bar PU\bar Z]}.
\end{align*}
The  \(U\)-dependence of the action can be absorbed into the similarity rotation  of \(Z\), so that we are left with the following complex matrix integral in the presence of a diagonal external field \(P\):
\begin{align}\label{twoPfunctionCS}
K[t,P]\sim \int d^{2N^{2}}Z\, e^{N\tr[-\bar ZZ+ V_{\boldsymbol{t}}(Z)+\bar  V_{\boldsymbol{t}}(\bar Z)+PZ+\bar P\bar Z]},\qquad [\bar P,P]=0.
\end{align}

For  \(\bar P=P\) it becomes the (1/2,1/2,1/2)BPS SC and can be in principal computed in the large $N$ limit. For $P\ne \bar{P}$ the known analytic methods fail to compute the integral over $Z,\bar{Z}$, except the case of quadratic potential considered below.

\subsubsection{ Gaussian case}

Let us take \(V_{\boldsymbol{t}}(Z)=\frac{t }{2}Z^{2} \) which would be characterized by a droplet of elliptic shape, see section~\ref{sec:exp-op}. The three-point function \eqref{twoPfunctionCS} then takes the form
\begin{align}\label{3p14BPS}
C_{P}[t]\sim\int [dU]_{_{SU(N)}} \int d^{2N^{2}}Z\, e^{N\tr[-\bar ZZ+t Z^{2}+\bar  t\bar Z^{2}+U^{\dagger}PUZ+U^{\dagger}\bar PU\bar Z]}\,.
\end{align}
Integrating over \(Z,\bar Z  \)   we obtain for the structure constant
\begin{align}
C_{P}[t]\sim\exp N\tr\Big[\frac{-\bar{t }P^2 -t  \bar{P}^2-2 P \bar{P}}{2 \left(t \bar{t }-1\right)}\Big]\,.
\end{align}
Interestingly, it almost repeats the weight for \(Z,\bar Z\), up to some signs and the overall coefficient in~\eqref{3p14BPS}.

\section{Conclusions and prospects}

We propose in this work a fresh look at an old subject -- the analysis of dynamics of protected states in $\mathcal{N}=4$ SYM in terms of equivalent matrix models \cite{Corley:2001zk, Jevicki:2006tr}.
The matrix model approach was extensively studied in the literature \cite{Berenstein:2004kk,Takayama_2005,Balasubramanian:2001nh,Hashimoto:2000zp} (for a modern treatment see \cite{Kazakov:2024ald,Jiang:2019xdz, Chen:2019gsb, Yang:2021kot}) and the one we propose in this paper allows us to study very general classes of operators and to match them with the corresponding LLM geometry. Furthermore, our approach allows us to systematically study non-extremal three-point functions involving huge operators in the large $N$ limit. 
It is known~\cite{Berenstein:2004kk, Lin:2004nb} that the matrix model approach is the most natural way of relating half-BPS operators  of $\mathcal{N}=4$ SYM to their geometric duals. In that respect, the complex matrix integral representation of protected correlation functions for $\frac12$, $\frac14$, and $\frac18$-BPS operators of $\mathcal{N}=4$ SYM theory which we are using here presents a great opportunity to apply well established matrix model methods for the analysis of the intricate large $N$ BPS dynamics.

For the $\frac12$-BPS case, the complex matrix representation of these correlators gives a direct correspondence of the SYM parameters to those of the LLM metric. Namely, the distribution of complex eigenvalues in the matrix integral for the two-point function of an operator gives directly the shape of the droplet of the corresponding LLM geometry.  Moreover, we propose a complex matrix representation for the the most general protected  three-point correlators and of the related CFT structure constants of three $\frac12$-BPS operators. This representation has certainly advantages to the Hermitian matrix representation of~\cite{Kazakov:2024ald}, due to a more direct analogy with LLM geometry. Using this analogy, we managed to put into exact one-to-one correspondence the results of the SYM computations to those of supergravity~\cite{Skenderis:2006di} for HHL structure constants.
To our knowledge, it is the first direct computation of such correlators in the field theory for arbitrary backgrounds. Previous results rely heavily on the free fermion description in this sector. For example \cite{Skenderis:2007yb} used various three-point functions calculations to determine the form of the non-extremal light operators in the free fermion picture. Instead we give first principle reductions of these operators to the complex matrix model, which are not only more practical but also reveal a hidden integrable structure.

We also emphasize that the three-point functions involving maximally charged operators is the only information needed to reconstruct the geometry dual to a particular $\frac12$-BPS state in gauge theory in the large $N$ limit. For instance the operators of the form $e^{\,\tr V(Z)}$, which we call exponential $\frac{1}{2}$-BPS operators, have as a feature the fact that a given connected distribution of complex eigenvalues of the $Z$ matrix can be reproduced  by manipulating the couplings in the $V(Z)$ ``potential" \cite{Vazquez:2006id}. In the large $N$ limit the two-point correlator of exponential operators is described by a complex two-matrix model closely related to the beautiful subject of Laplacian growth~\cite{Kostov:2000ed,Zabrodin:2004cc}, as well as to the matrix model formulation of 2D gravity with various matter fields~\cite{Kazakov:1985ds,David:1984tx,Kazakov:1985ea,Kazakov:1986hu,Kazakov:1987qg,KOSTOV1989295}. It would be interesting to understand how the universal singularities in the complex matrix model influence the resulting LLM metric.  Moreover, the double scaling limit near such singularities~\cite{Brezin:1990rb,Douglas:1989ve,Gross:1989vs}, corresponding to the full 2D string field theory, could provide a window to study quantum properties of type IIB supergravity . The expectation is that after taking the double scaling limit around various singularities, one should obtain a simpler topological string theory as in \cite{Dijkgraaf:2002fc, Aganagic:2003qj}. From the chiral algebra point of view this is clear and it is related to the topological theory of gravity introduced in \cite{Bershadsky:1993cx}. The fact that we can tune the operators to obtain topological string theories means that we should think of these models as being embedded in the full-fledged type IIB theory within BPS sectors as advocated by the twisted holography program \cite{Costello:2018zrm}. The double scaling limit of an LLM geometry around a smooth droplet is generically a pp-wave geometry, so other types of singularities could provide interesting string backgrounds.

An obvious set of problems, certainly beyond of the scope of this paper, is to revisit the physics of light non-BPS operators on LLM backgrounds, including the singular ones described here \cite{Vazquez:2006id, deMelloKoch:2018ert, Chen:2007gh}. Most probably, such systems are generically non-integrable, both in weak and strong coupling regimes. However, there may exist particular backgrounds with certain integrability properties.


We also presented the computation of  Huge-Huge-Giant (HHG) structure constant,
where by ``Giant" operator we mean that $\Delta\sim N$. The computation is rather straightforward for the huge operators given by characters with Young diagrams built out of a few large rectangles. We worked out the explicit answer for the rectangular YT. It would be very interesting to reproduce this result with a supergravity calculation, where the determinant operator sources a giant graviton D-brane. The fact that the asymptotics of the field theory result is determined by a saddle point calculation gives support to the idea that the dual gravity calculation is also determined semi-classically. However because the probe operator has a dimension comparable with $N$ the usual holographic renormalization methods are inapplicable, and we should instead look for supersymmetric D3-instanton configurations in LLM backgrounds. The scaling of the HHG correlator is consistent with that of a D-instanton effect \cite{Shenker:1990uf} and reproducing the field theory prediction using string theory methods would give a very non-trivial check of AdS/CFT in the non-perturbative regime. 

We also presented in this work new results for particular $\frac12$-BPS HHH three-point functions. In particular, we give the exact saddle point solution for three exponential operators with cubic potentials $V(Z)$. A similar solution is known in relation to the $Q$-state Potts model on random planar graphs~\cite{Kazakov:1987qg,Daul:1994qy,Eynard:1992cn,Eynard:1995nv,Eynard:1995zv,Eynard:1999gp} ($Q=3$ in our case). 
We have calculated numerically the behavior of the three-point function of such operators as a function of the coupling constants. 
Another example of exact and explicit computation in this work concerns the HHH structure constant of three character-type operators of rectangular YTs.  Interestingly, in this case the two cut solution appears to give the dominating saddle point. Understanding the geometric dual to the case of general HHH structure constants would be a huge step forward to understanding the AdS/CFT duality for the BPS sector. 
One can expect that geometry duals for $\frac{1}{2}$-BPS HHH three-point functions would be given by configurations with $S^2 \times S^2$ isometry, in analogy with the $\frac{1}{4}$-BPS and $\frac{1}{8}$-BPS  backgrounds constructed in \cite{Chen:2007du}.
This task seems to be difficult but realistic, regarding the simplicity and explicitness of the complex matrix representation of these quantities, and the large amount of supersymmetry of the problem. Knowing the density of eigenvalues for HHH background gives us information about protected higher point functions (four point and higher) in the chiral algebra sector, which might give a hint about the geometry in bulk after inserting three huge BPS operators.

Much less is known about the protected two-point and three-point correlation functions of $\frac{1}{4}$-BPS and $\frac{1}{8}$-BPS operators. Already the construction of such operators from complex scalars $X,Y,Z$ in SYM theory is a challenge. Generalizations of character operators to the $\frac{1}{4}$-BPS case 
were proposed in~\cite{Bhattacharyya:2008rb,Lewis-Brown:2020nmg}\footnote{see also \cite{deMelloKoch:2024sdf} for a pedagogical introduction to the subject}. This construction is difficult for practical use in computing the aforementioned correlators. Much simpler looking operators are the so-called coherent state operators~\eqref{CSoperators} proposed in~\cite{Berenstein:2022srd}.  These are believed (but not strictly proven for $\frac{1}{4}$-BPS and $\frac{1}{8}$-BPS cases) to be protected. The parameter space of such operators is very rich and it is not easy to find their gravity duals (though some proposals of the corresponding $\frac{1}{4}$-BPS and $\frac{1}{8}$-BPS bubbling  geometries exist \cite{Berenstein:2005aa, Donos:2006iy,Chen:2007du}). In the case of  $\frac{1}{4}$-BPS and $\frac{1}{8}$-BPS operators, even the two-point correlators
boil down to generically unsolvable matrix integrals\footnote{For instance, they cannot be reduced to eigenvalues in any obvious way.}. An interesting, but probably a remote perspective on these problems could open if we clearly identify the gravity duals of such operators. This problem arises even for $\frac12$-BPS case. 
We found that in that case the problem is related to special classes of domains called quadrature domains. 
Many density theorems exist for such domains which reflect the fact that the coherent states operators are able to create any configuration of droplets. We provide some examples that demonstrate how much control we have over the density of eigenvalues. It would be good to try extend these results to the cases of $\frac{1}{4}$- and $\frac{1}{8}$-BPS coherent states. 


Finally, we make a curious observation: the correlation function of two BPS coherent state operators  boils down (for particular, uniform distribution of parameters) 
  to the partition function of the quenched Eguchi-Kawai reduction of the principal chiral model (PCM). 
  This could open a perspective on employing numerous results of study of PCM, using its quantum integrability~\cite{Polyakov:1983tt,Polyakov:1984et,Wiegmann:1984pcm,Wiegmann:1984ec,Fateev:1994ai,Fateev:1994dp,Kazakov:2019laa,Kazakov:2023imu}, to the study of such $\frac14$-BPS correlators. It is not clear whether the quenched version of  Eguchi-Kawai (EK) reduction of PCM can   describe the physical  behavior of 2D PCM (characterized by asymptotic freedom in weak coupling and by the know mass spectrum), or it stays for all couplings in a ``wrong" phase\footnote{A pathological behavior of the quenched EK reduction of the lattice gauge theory was noticed in~\cite{Bringoltz:2008ek}. But their argument is not directly applicable to the quenched EK reduced  PCM}. 
If it does, the ultimate, and very exciting goal here would be to use the gravity duals of such correlators to study the large $N$ PCM in 2, or even 3 dimensions (in case of a similar $\frac14$-BPS CS operator). It seems of course a long way to go before we understand the basic properties of AdS/CFT correspondence for these BPS objects and connect them to the bubbling geometries like those described in~\cite{Donos:2006iy,Chen:2007du}. Maybe, at first we have to try to study some of them using numerical methods: Monte-Carlo simulations or matrix bootstrap approach~\cite{Lin:2020mme,Kazakov:2021lel}.

\section*{ Acknowledgments}

We thank Francesco Aprile, Constantin Bachas, Iosif Bena, Kasia Budzik, Davide Gaiotto, Shota Komatsu, Ivan Kostov, Juan Maldacena, Joeseph Minahan, Shiraz Minwalla, Pedro Vieira, Paul Wiegmann and Anton Zabrodin for useful discussions. The work of A.H. is supported by ERC-2022-CoG - FAIM 101088193. V.K. thanks the Perimeter Institute for Theoretical Physics and the
Simons Center for Geometry and Physics, where a part of this work has been done. 
   
\newpage

\centerline{\Huge {\bf Appendices}}

\appendix
\appendixpage
\addappheadtotoc
\addtocontents{toc}{\protect\setcounter{tocdepth}{-1}}
\section{Exponential operators and orthogonal polynomials}\label{coord_exp}
\paragraph{Hermitian matrices:} It is given through Hermitian 2-matrix integral
\begin{align*}
K_{V,W}\left(x_{1}-x_{2}\right)&\equiv\langle O^{V}\left(\Phi_{1}(x_{1})\right) O^{W}\left(\Phi_{2}(x_{2})\right)\rangle=\\
&=\frac{1}{\mathcal{Z}} \int \mathcal{D}\Phi_1\mathcal{D}\Phi_2  \,
 \exp\left[\tr\left(-\frac{1}{\kappa}\Phi_{1}\Phi_{2}+V(\Phi_{1})+W(\Phi_{2})\right)\right],
\end{align*} 
where \(\kappa=\frac{2(x_{1}-x_{2})^{2}}{(y_{1}-y_{2})^{2}}\). In terms of e.v.   it is
\begin{align*}
K_{V,W}=\frac{1}{\mathcal{Z}} \int  \prod_{k=1}^{N} da_{k}db_{k}\, e^{-\frac{1}{\kappa}a_{k}b_{k}+V(a_{k})+W(b_{k})}\Delta(a)\Delta(b)=\prod_{j=1}^{N}\frac{h_{j}}{\kappa^j (j-1)!},
\end{align*}
where \(h_{k}  \) is the normalization factor of  bi-orthogonal polynomials 
\begin{align*}
\int  \frac{dadb}{\pi}\,e^{-\frac{1}{\kappa}ab+V(a)+W(b)}P_{m}(a)Q_{n}(b)=h_{n}\delta_{mn}.
\end{align*}
We can now define the most general system of bi-orthogonal conformal operators:
\begin{align}
O^{V}_{\{m\}}=e^{\tr V(\Phi)}\,\,\Upsilon_{\{m\}}(\Phi),\qquad\text{where}\,\,\, \Upsilon_{\{m\}}(\Phi)= \frac{\underset{1<k,j<N}{\det} P_{m_{k}}(a_{j})}{\Delta(a)} \end{align}
and similarly for \(O^{W}_{\{n\}}(\Phi)\). Obviously, the bi-orthogonality relation is 
\begin{align*}
\langle O^{V}_{\{m\}}\left(\Phi_{1}(x_{1})\right) O^{W}_{\{n\}}\left(\Phi_{2}(x_{2})\right)\rangle=\delta_{\{m\},\{n\}}\prod_{j=1}^{N} \frac{h_{m_{j}}}{\kappa^j (j-1)!}\,,
\end{align*}
where \(\delta_{\{m\},\{n\}}\) is the Kronecker symbol between two sets of increasing non-negative integers \(\{m\}=\{m_{1},\dots,m_{N}\}\) and \(\{n\}=\{n_{1},\dots,n_{N}\} \). For \(V=W\) these operators obey the usual orthogonality relation.

Unlike  the Schur operators~\cite{Berenstein:2004kk} these operators do not have the radial  symmetry in \((a,b)\) space. And since they are not homogeneous functions of the field they have a non-trivial space and \(R\)-symmetry coordinates dependence.

\paragraph{Complex matrix:}In terms of complex matrices   the correlator reads\begin{align*}
K=\langle \bar{O}_{}(\bar{Z}) O_{}(Z)\rangle=\frac{1}{\mathcal{Z}}\int \mathcal{D}^{2}Z \, \exp \tr\left[-\,\bar{ Z}Z+V(Z)+\bar{V}(\bar{Z})\right].
\end{align*}

Schur decomposition  \(Z=\Omega^{\dagger}(z+u)\Omega\), where \(z=\text{diag}(z_{1},\dots,z_{N})\) and \(u\) is a complex upper triangular matrix, gives
\begin{align*}
K=\frac{1}{\mathcal{Z}}\int  \prod_{k=1}^{N} d^{2}z_{k}\, e^{-\,\bar z_{k}z_{k}+V(z_{k})+
\bar{V}(\bar{z}_{k})}\,\left\vert\Delta(z)\right\vert^{2}=\prod_{j=1}^{N} \frac{h_{j}}{\kappa^j (j-1)!} \end{align*}
where 
\begin{align*}
h_{j}=\int \frac{d^{2}z}{\pi}\, e^{-\,\bar zz+V(z)+
\bar{V}(\bar{z})}\,(\bar{z}z)^{j-1}.
\end{align*}

The orthogonal polynomials are simply the monomials 
\begin{align*}
P_{m}(z)=z^{m},
\end{align*}
so that
the operators
\begin{align*}
O^{V}_{\{m\}}=e^{\tr \,V(Z)}\,\,\Upsilon_{\{m\}}(Z),\qquad\text{where}\,\, \Upsilon_{\{m\}}(Z)= \frac{\underset{1<k,j<N}{\det} z_{j}^{m_{k}}}{\Delta(z)}, \end{align*}
form an orthogonal basis
\begin{align*}
\langle O^{V}_{\{m\}}\left(Z(x_{1})\right) O^{\bar V}_{\{n\}}\left(\bar{Z}(x_{2})\right)\rangle=\delta_{\{m\},\{n\}}\prod_{j=1}^{N} \frac{h_{m_j}}{\kappa^j (j-1)!}\,.
\end{align*}

\section{Normal Ordered Operators}\label{QtransformApp}
\subsection{Single Trace Operators}
When computing three point functions involving light single trace operators using the complex matrix model we often encounter moments of normal ordered functions such as 
\begin{equation}
  \mathcal{T}_{\Delta}( q)=  \;:\tr[(q Z + \bar{q} \bar{Z})^\Delta]:\;= \frac{1}{\mathcal{Z}_\phi}\int d \phi \;e^{-\tr\phi^2}\tr[(q Z + \bar{q} \bar{Z}+ 2 i \phi)^\Delta].
\end{equation}
To compare with the existing gravity calculations we need expressions for $\Delta=1,\dots, 4$; additionally we need to reduce these expressions so that they only depend on the eigenvalues of $Z$. For completeness we list the reduced operators up to dimension $4$ at finite here. The rules for the Wick contractions are 
\begin{equation}
\begin{aligned}
 \langle \phi_{ij} \phi_{kl}\rangle&= \frac{1}{2}\delta_{il}\delta_{jk}\\
    \langle T_{ij}T_{kl}^\dagger \rangle&= \delta_{il}\delta_{jk} \;\;\;\texttt{for}\;\;\; i<j ,\; l<k
\end{aligned}
\end{equation}
and some useful identities are 
\begin{equation}
\begin{aligned}
    \big\langle \tr[A \phi^2]\big\rangle&= \frac{N}{2}\tr A\\
     \big\langle \tr[A \phi B \phi]\big\rangle&= \frac{1}{2}\tr A\, \tr B\\
 \big\langle \tr[A \phi^4]\big\rangle&= \frac{2N^2+1}{4}\,\tr A
\end{aligned}
\end{equation}
Explicit calculation gives:
\begin{equation}
\begin{aligned}
\tau_1(q)&=\tr m\\
    \tau_2(q)&= \tr\left[m^2 -\left(1+\frac{1}{N}\right)\mathbb{I}\right]\\
     \tau_3(q)&= \tr\left[m^3- 3(1+\frac{1}{N}) m\right]\\
      \tau_4(q)&= \tr\left[m^4 -4\left(1+\frac{2}{3N}\right)m^2 + 2\left(1+\frac{1}{N}\right)\left(1+\frac{3}{N}\right)\mathbb{I}\right]-\frac{2}{N} (\tr m)^2
\end{aligned}
\end{equation}
where we used the short-hand notation $m=qz + \bar{q} \bar{z}$.

Finally to obtain operator insertions with definite charge we integrate over $q=e^{i\psi}$ with the appropriate weight. The unnormalized operators up to dimension 4 at finite $N$ are given by
\begin{equation}
\begin{aligned}
\tau_{\Delta, \Delta}&= \tr[z^\Delta]\\
   \tau_{2,0}&= 2\tr\left[z \bar{z} -\frac{1}{2}\left(1+\frac{1}{N}\right)\mathbb{I}\right]\\
     \tau_{3,1}&= 3\tr\left[z^2 \bar{z}- \left(1+\frac{1}{N}\right) z\right]\\
      \tau_{4,2}&= 4\tr\left[z^3\bar{z} -\left(1+\frac{2}{3N}\right)z^2 \right]-\frac{2}{N} (\tr z)^2\\
       \tau_{4,0}&= 2\tr\left[3(z\bar{z})^2 -4\left(1+\frac{2}{3N}\right)z\bar{z} + \left(1+\frac{1}{N}\right)\left(1+\frac{3}{N}\right)\mathbb{I}\right]-\frac{4}{N} (\tr z)(\tr\bar{z})
\end{aligned}
\end{equation}
and the remaining half-BPS operators of dimension up four can be obtained by complex conjugation of these. In the next section we give a more efficient algorith for determinig there in the large $N$ limit.
\subsubsection{Single Trace Operators from HCIZ Hopf Flows}
Here we present a nice way to compute the Q-transforms of single trace operators without having to do tedious wick contractions. See \cite{Kazakov:2024ald} for a review and a recent application of this method. Consider the following integral
\beq
    \mathcal{Z}[X] = \frac{\int dM\, e^{-\frac N2 \tr M^2+N\tr MX}}{\int dM\, e^{-\frac N2\tr M^2}}\nn
\eeq
Of course this integral can be performed exactly and we get $\mathcal{Z}[X] = e^{\frac N2\tr X^2}$. We are interested in computing moments of $M$, which one can obtain by differentiation -- $\tr(\frac{\partial}{\partial X})^n\mathcal{Z}[X]$. We will instead present a computationally simpler method to compute these moments. First, we diagonalize $M$ to get
\beq
    \int \prod_i dm_i\ e^{-\frac N2 \sum_i m_i^2} \int dU e^{N\tr mUxU^\dagger}
\eeq
where $m$ and $x$ are eigenvalues of $M$ and $X$. The integral over unitary $U$ is the famous Harish-Chandra-Itzykson-Zuber (HCIZ) integral. There is a very nice large $N$ limit of this integral in terms of a one-dimensional fluid flow \cite{Matytsin_1994}. Let the density and velocity of the fluid be denoted by $\rho(x,t)$ and $v(x,t)$. The density at the initial and final times is given by
\beq
    \rho(x,t=0) = \texttt{density of m's}\equiv \sigma_1(x)\ ,\qquad \rho(x,t=1) = \texttt{density of x's}\equiv \sigma_2(x) \nn
\eeq
By varying $\mathcal{Z}[X]$ with respect to the density of $X$ we obtain
\beq
    x - v(x,1) - \fint \frac{\sigma_2(z)}{x-z} = x
\eeq
Which tells us what the final velocity of the flow must be. Now, the Hopf flow is integrable and the conserved charges imply the following
\beq
    \langle\tr M^n\rangle = \frac{1}{n+1}\oint \frac{dz}{2\pi i}\, G_-(x)^{n+1}
\eeq
where $G_-(x)$ is defined as
\beqa
    G_-(x) &= x - v(x,1) - i\pi \sigma_2(x)\nn\\
    &= x - R_2(x)^*\nn
\eeqa
where in the second line, we plugged in the velocity $v(x,1)$ and combined the principal value integral with the density to obtain the full resolvent $R_2(x) = \int dz\frac{\sigma_2(z)}{x-z}$. Putting it all together, we get
\beq
    \langle \tr M^n\rangle = \frac{-1}{n+1} \oint \frac{dx}{2\pi i} \left(x+\sum_{k=0}^\infty \frac{\tr X^k}{x^{k+1}}\right)^{n+1}
\eeq
Which one can easily evaluate by deforming the contour to infinity and picking up the coefficient of $\frac1x$. For instance, we have
\beq
\begin{aligned}
    \langle\tr M\rangle&=\tr X\\
    \langle\tr M^2\rangle&=1+\tr X^2\\
    \langle\tr M^3\rangle&=\tr X^3+3\,\tr X\\
    \langle\tr M^4\rangle&=\tr X^4+4\, \tr X^2+2\, \tr X^2+2\\
    \langle\tr M^5\rangle &= \tr X^5 + 5\,\tr X^3+5\,\tr X^2\,\tr X+10\,\tr X.
\end{aligned}
\eeq
These moments are related to the complex matrix model reduction of the single trace operators by the relation 
\begin{equation}
    \tau_\Delta(q)= (-i)^\Delta \;\big\langle \tr M^\Delta \big\rangle_{X\rightarrow i(qz+ \bar{q}\bar{z})}.
\end{equation}

\subsection{Giant Graviton Operators}
We will also need normal ordered versions of fully symmetric and fully anti-symmetric Schur polynomials. Starting with the symmetric case, it is convinient to use a coherent generating function for the single row Schur polynomial:
\begin{equation}
\begin{aligned}
\mathcal{S}_\alpha(M)&= \int_{\mathbb{CP}^{N-1}} d\varphi^\dagger d\varphi\; e^{-\alpha\varphi^\dagger M\varphi}= \sum_{k} \frac{(-\alpha)^k}{(N+k-1)!}\chi_{\texttt{sym}_k}(M)
\end{aligned}
\end{equation}
where $M= q Z + \bar{q} \bar{Z}$ as before. The integration is performed over $\mathbb{CP}^{N-1}$ of unit radius (i.e. $\varphi^\dagger \varphi=1)$.  The normal ordered version of this generating function is given by
\begin{equation}
\begin{aligned}
\mathbb{Q}[\mathcal{S}_\alpha(M)]&= \frac{1}{\mathcal{Z}_\phi}\int d\phi \int_{\mathbb{CP}^{N-1}} d\varphi^\dagger d\varphi\; e^{-\tr \phi^2 -2i\alpha \varphi^\dagger \phi \varphi} e^{-\alpha\varphi^\dagger M\varphi}= e^{-\alpha^2}\int_{\mathbb{CP}^{N-1}} d\varphi^\dagger d\varphi\; e^{-\alpha\varphi^\dagger M\varphi}.
\end{aligned}
\end{equation}
Then the integration over the triangular parts can be done in a straight forward way since the integral is also Gaussian:
\begin{equation}
\begin{aligned}
\langle\mathbb{Q}[\mathcal{S}_\alpha(M)]\rangle_{T, T^\dagger}
&= \frac{e^{\frac{1}{2}\tr m^2}}{(Z_{\sigma})^N}\int \prod_k d\sigma_k e^{-\frac{1}{2}\sigma_k^2-i m_k \sigma_k}\,e^{-\frac{1}{2}\alpha^2}\int_{\mathbb{CP}^{N-1}}d\varphi^\dagger d\varphi\; e^{-i\alpha\varphi^\dagger  \sigma\varphi}
\end{aligned}
\end{equation}
The last line is normal ordered in the diagonal variables $m$. We wrote it in the same form as the generating function of single traces in the previous section. 
The last step is to perform the  integration over projective space. For a moment let us assume the values of $\sigma_k$ are generic, in which case the integrant is invariant under a $U(1)^N$ action generated by $\varphi^\dagger \sigma \varphi$. This is enough to apply the Berline–Vergne localization formula. 
\begin{equation}
\begin{aligned}
    \int_{\mathbb{CP}^{N-1}}d\varphi^\dagger d\varphi\; e^{-i\alpha\varphi^\dagger  \sigma\varphi}&= \sum_{j=1}^N \frac{e^{-i \alpha \sigma_j}}{(-i\alpha)^{N-1}\prod_{l\neq j}(\sigma_j-\sigma_l )}= \frac{1}{\alpha^{N-1}}\oint ds \,\frac{e^{\alpha s}}{\prod_j(s+i \sigma_j)}
    \end{aligned}
\end{equation}
where the contour of integration is taken to enclose all of the poles of the denominator, or equivalently the pole at infinity. For non generic $\sigma_k$ the Berline–Vergne formula fails since the critical points are degenerate but the residue formula holds.
To project to a particular operator we integrate over $\alpha$ giving an integral of the form
\begin{equation}
(N+\Delta-1)! \oint \frac{d\alpha}{\alpha^{\Delta+1}}\,\frac{e^{-\frac{1}{2}\alpha^2 + s \alpha}}{\alpha^{N-1}}= \text{He}_{\Delta+N-1}(s).
\end{equation}
Putting everything together we get 
\begin{equation}
  \mathbb{Q}[\chi_{\texttt{sym}_k}(M)]=(-1)^{N-1}\oint d s\;\text{He}_{N+k-1}(s) \times  \frac{e^{\frac{1}{2}\tr m^2}}{(Z_{\sigma})^N}\int \prod_k d\sigma_k \frac{e^{-\frac{1}{2}\sigma_k^2-i m_k \sigma_k}}{ i\sigma_k -s}.
\end{equation}
The overall sign in front is depends on whether the integration is closed around infinity or around the poles (i.e. the orientation of the contour).

For determinant operators we instead use the following generating function:
\begin{equation}
\begin{aligned}
\det(M-\alpha)&=\int d\psi^\dagger d\psi \;e^{\psi^\dagger(M-\alpha)\psi}=\sum_{k=0}^N (-\alpha)^{N-k} \chi_{\texttt{asym}_k}(M)\\
    : \det(M-\alpha): 
    &=\int d\psi^\dagger d\psi \;e^{\psi^\dagger(M-\alpha)\psi}\; e^{\frac{1}{N}( \psi^\dagger \psi)^2}
\end{aligned}
\end{equation}
Integrating over the triangular parts of the matrices gives
\begin{equation}
\begin{aligned}
   \big\langle  : \det(M-\alpha):\big \rangle_{T, T^\dagger}
   &= \sum_{n=0}^\infty \frac{(\sqrt{N}\alpha)^n}{n!}\frac{1}{\mathcal{Z}_\sigma} \int d\sigma \;e^{-\frac{N}{2}\sigma^2}\, \text{He}_n(\sqrt{N}\sigma)\;\det(m-\sigma\,\mathbb{I})
\end{aligned}
\end{equation}
where we used the fact that $\psi_i^2=0$ to eliminate the last term in the second line. To project into a particular operator we integrate over $\alpha$ as before giving
\begin{align}
 \langle:\chi_{\texttt{asym}_\Delta}(M):\rangle_{T, T^\dagger}
 &=\frac{(\sqrt{N})^{N-\Delta}}{(N-\Delta)!}\frac{1}{\mathcal{Z}_\sigma}\int d\sigma \;e^{-\frac{N}{2}\sigma^2}\, \text{He}_{N-\Delta}(\sqrt{N}\sigma)\;\det(m+\sigma\,\mathbb{I})
\end{align}

\subsection{Rectangular Young Tableaux}
We compute here the Q-transform for a rectangular Young Tableau. One way to obtain this is using the formula a general YT, derived in \cite{Kazakov:2024ald}. Here, we present an alternative derivation using the Hubbard Stratonovich (HS) trick. We want to evaluate the following integral
\beq
    \mathbb{Q}_{\det^K}(X) \equiv Q_K(X) = \int dM\, e^{-\frac N2 \tr M^2}\det (X+iM)^K\,,
\eeq
where $M$ and $X$ are $N\times N$ matrices and $K$ is $O(N)$. We now introduce fermions $\psi^i_a,\bar\psi^i_a$ where $i=1\ldots N$ and $a=1\ldots K$, in terms of which the integral is
\beq
    \int dM\, d\bar\psi\,d\psi\ \exp\left[-\frac N2\tr M^2+\sum_a\bar \psi_a\cdot(X+iM)\cdot\psi_a\right]\,.
\eeq
The integral over $M$ is now gaussian and integrating it out we get
\begin{equation}
    \int d\bar\psi\,d\psi \exp\left[\frac1N \sum_{\substack{a,b\\i,j}}(\bar\psi_a^j\psi_b^j)(\bar\psi^i_b\psi^i_a) + \sum_a \bar\psi_a \cdot X\cdot \psi_a \right]
\end{equation}
Now we introduce an auxiliary Hermitian $K\times K$ matrix $\rho$ to simplify the four fermion term
\begin{multline}
    \int d\rho\,d\bar\psi\,d\psi\ \exp\left[-\frac1N\left(\tr\rho^2 + 2\rho_{ab} \bar\psi_b\cdot\psi_a\right)+\sum_a\bar\psi_a\cdot X\cdot \psi_a\right]\\
    \propto\int d\rho\,d\bar\psi\,d\psi\ \exp\left[-\frac N2 \tr\rho^2 + \bar\psi_a^j\left(X_{jk}\delta_{ab}-\delta_{jk}\rho_{ab}\right) \psi^k_b\right]
\end{multline}
Integrating over the fermions now gives us the determinant $\det (X\otimes \mathbb{I}_K-\mathbb{I}_N\otimes\rho) = \prod_{i,a}(x_i-y_a)$, where $\{x_i\}$ and $\{y_a\}$ are eigenvalues of $X$ and $\rho$ respectively. Therefore, upto normalization factors, we find
\begin{equation}
    Q_K(X) = \int \prod_{a=1}^Kdy_i\, e^{-\frac N2\sum_a y_a^2} \prod_{\substack{1\le i\le N\\1\le a\le K}} (x_i-y_a)
\end{equation}

\section{Densities and resolvents for LLM Fermi sea  for various operators } \label{App-densities-LLM}

Suppose we have a  general 2D Fermi sea describing a particular $\frac12$-BPS operator. It can be any of the types of operators considered in this paper (character, exponential, CS operators).
It is described by a uniform density inside a domain \(\mathcal{D}\):
\begin{align*}
\rho_{\mathcal{D}}(z,\bar{z})=\begin{cases} 1/\pi, & z\in \mathcal{D}\ \\
0, & z\notin \mathcal{D}\,, \\
\end{cases}
\end{align*}
and the form of the domain $\cal{D}$ is fixed by the choice of the operator. Note, that the density is normalized as $\int_{\mathcal{D}}d^2z \,\rho_{\cal{D}}(z,\bar{z})=1$.

In the Type IIB supergravity the most general $\frac12$-BPS configuration with $R \times SO(4) \times SO(4)$ isometry is given by Lin-Lunin-Maldacena background \cite{Lin:2004nb}:
\begin{equation}
    ds^2 = -h^{-2} (dt + V_i dx^i)^2 + h^2 (dy^2 + dx^i dx^i ) + ye^G d\Omega_3^2 + ye^{-G} d\Tilde{\Omega}_3^2,
\end{equation}
where $i=1,2$, and
\begin{equation}
    h^{-2} = 2y \cosh G, \qquad \zeta = \frac{1}{2} \tanh G, \qquad y \partial_y V_i = \epsilon_{ij} \partial_j \zeta,
\end{equation}
or
\begin{equation}
    h^{-2} = \frac{2y}{\sqrt{1-4\zeta^2}}.
\end{equation}
The 5-form field strength is given by
\begin{equation}
    \begin{aligned}
        F_{(5)} = F_{\mu \nu} dx^{\mu} \wedge dx^{\nu} \wedge d\Omega_3 + \Tilde{F}_{\mu \nu} dx^{\mu} \wedge dx^{\nu} \wedge d \Tilde{\Omega}_3, \\
        F = d B_t \wedge (dt + V) + B_t dV + d \hat{B},\\
        \Tilde{F} = d \Tilde{B}_t \wedge (dt+V) + \Tilde{B}_t dV + d \hat{\Tilde{B}}, \\
        B_t = - \frac{1}{4} y^2 e^{2G}, \qquad \Tilde{B}_t = - \frac{1}{4} y^2 e^{-2G}, \\
        d \hat{B} = - \frac{1}{4} y^3 \star_3 d \bigg( \frac{\zeta+1/2}{y^2} \bigg), \qquad d \hat{\Tilde{B}} = - \frac{1}{4} y^3 \star_3 d \bigg(\frac{\zeta-1/2}{y^2}  \bigg).
    \end{aligned}
\end{equation}
The  function $\zeta(x_1,x_2,y)$ obeys $6d$ Laplace equation \cite{Lin:2004nb}, from which one derives that
\begin{align}
    &\zeta(x_1,x_2,y) = \frac{y^2}{\pi} \int_{\mathcal{D}} \frac{\zeta(x'_1,x'_2,0) dx'_1 dx'_2}{[(\Vec{x}-\Vec{x}')^2+y^2]^2},\\
    &V_i = \frac{\epsilon_{ij}}{\pi} \int_{ \mathcal{D}} \frac{\zeta(x_1',x_2',0) (x_j-x_j')  dx_1' dx_2'}{[(\Vec{x}-\Vec{x}')^2+y^2]^2} .
\end{align}
As one can see all the components of the metric and the 5-form field strength are uniquely determined by the specifying boundary conditions $\zeta(x_1,x_2,y=0)$. Non-singular geometries correspond to boundary conditions 
\begin{align*}
\zeta(x_1,x_2,0)=\begin{cases} -\frac{1}{2}, & (x_1,x_2)\in \mathcal{D}\ \\
\frac{1}{2}, & (x_1,x_2)\notin \mathcal{D}\,. \\
\end{cases}
\end{align*}
For example, one can reconstruct the usual $AdS_5 \times S^5$ background by taking the droplet $\mathcal{D}$ to be a unit disc. 

We identify the density $\rho_{\mathcal{D}}(z,\bar{z})$ that we find using SPEs in the Complex Matrix Model of two- and three-point functions of various operators in $\mathcal{N}=4$ SYM with the function $\zeta(x_1,x_2,0)$ that defines $\frac12$-BPS configuration in $10d$ SUGRA in the following way
\begin{eqnarray}
    \rho_{\mathcal{D}}(z,\bar{z}) = - \frac{1}{\pi} \bigg[ \zeta\bigg(\frac{z+\bar{z}}{2},\frac{z-\bar{z}}{2i},0 \bigg)- \frac{1}{2} \bigg].
\end{eqnarray}

For the purpose of HHG calculations, it is convenient to introduce the phase space coordinates \(x=\frac{1}{2}(z+\bar{z}),\quad p=\frac{1}{2i}(z-\bar{z})\) and a one-dimensional density given by
\begin{align*}
\rho(x)=\int_{-\infty}^{\infty} dp\,\rho_{\mathcal{D}}(x+ip,x-ip)\,, 
\end{align*}
which is normalized as $\int_{-\infty}^{\infty}dx \rho(x)=1$.
Let us consider a few important examples.

Let us consider calculation of one-dimensional density $\rho(x)$ for Schur polynomial operators. Consider a Young tableau consisting of a few rectangles. Fig.\ref{fig_YT-Discs} shows how to put the YT into one-to-one correspondence with the  the Fermi sea picture.  In principle, it is easy to build the resolvent and density for any such YT. The relation is built on observations~\eqref{YTrings},\eqref{hrrings}.  Some examples are below and at the end we give  the general formula for density and resolvent.

\paragraph{Disc (AdS)} Let us consider an empty YT, which in gravity dual corresponds to $AdS_5 \times S^5$ space. Plugging $\lambda_i=0$ into \eqref{hrrings}, we see that this corresponds to the LLM droplet given by a disc of unit radius. Therefore, the we can an explicit formula for the density using the Heaviside theta-function as
\begin{align*}
\rho_{\mathcal{D}}(\bar{z},z)= \frac{1}{\pi}\Theta \left(1-z\bar{z}\right).
\end{align*}
Integrating over \(p\) we get the normalized density in \(x\)-space:
\begin{align*}
\rho(x)=\frac{2}{ \pi }\sqrt{1-x^2}.
\end{align*}
The corresponding resolvent is simply that of the semi-circle law:
\begin{align*}
G(x)=2 \big( x- \sqrt{x^2-1 } \big).
\end{align*}

\paragraph{Annulus (simplest LLM)}

The YT is a rectangle of size \(N\times K\). 
From \eqref{hrrings} we see that the LLM droplet is an annulus of inner radius $r_1$ and outer raduis $r_2$, where 
\begin{equation}
    r_1^2 = \frac{K}{N}, \qquad r_2^2 = \frac{K+N}{N}.
\end{equation}
Then, the density of eigenvalues is
\begin{align*}
\rho_\mathcal{D}(z,\bar{z})=\frac{1}{\pi} \bigg[ \Theta \left(r_2^2 - z \bar{z} \right)- \Theta \left(r_1^2 - z \bar{z} \right) \bigg].
\end{align*}
The density in \(x\)-space is then
\begin{align}
\rho(x)=\frac{2}{\pi} \bigg( \sqrt{r_2^2 -x^2 }- \sqrt{r_1^2-x^2} \, \Theta(r_1^2-x^2) \bigg).\label{annulusResolvent}
\end{align}
The resolvent is:
\begin{align*}
G(x)=2 \bigg( \sqrt{x^2-r_1^2 }- \sqrt{x^2-r_2^2} \bigg)
\end{align*}


\paragraph{Annulus with a disc}

\begin{figure}[H]
\begin{subfigure}{.5\textwidth}
  \centering
  \includegraphics[width=.7\linewidth]{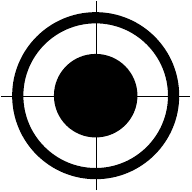}
  \caption{Two dimensional density $\rho_{\mathcal{D}}(z,\bar{z})$.}
\end{subfigure}%
\begin{subfigure}{.5\textwidth}
  \centering
  \includegraphics[width=1.0\linewidth]{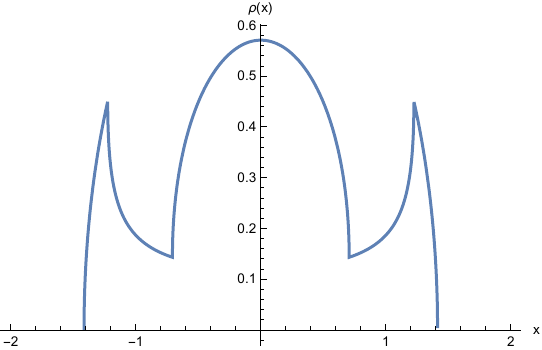}
  \caption{Projection onto the $x$-space.}
\end{subfigure}
\caption{Two-dimensional density of eigenvalues $\rho_{\mathcal{D}}(z,\bar{z})$, corresponding to the Schur polynomial operator with rectangular Young tableau with $N/2$ rows and $N$ columns, and its projection onto the $x$-axis. }
\label{fig:annulus-disc}
\end{figure}

Consider a YT with $L$ rows and $K$ columns. From \eqref{hrrings}, we see that the LLM droplet is an annulus with a disc in center. For the radius of the disc one has
\begin{equation}
    r_1^2 = \frac{N-L}{N},
\end{equation}
while the inner and outer radii of the annulus are 
\begin{equation}
    r_2^2 = \frac{N-L+K}{N}, \qquad r_3^2 = \frac{N+K}{N}.
\end{equation}
The density in the phase space is:
\begin{align*}
\rho_\mathcal{D}(\bar{z},z)=\frac{1}{\pi} \bigg[ \Theta \left(r_1^2-z\bar{z}\right)-\Theta \left(r_2^2-z\bar{z}\right)+\Theta \left(r_3^2-z\bar{z}\right) \bigg].
\end{align*}
The density in \(x\)-space (see the Fig.\ref{fig:annulus-disc} for its plot)
is
\begin{align}\label{DenDiskRing}
\rho(x)=\frac{2}{\pi} \bigg( \sqrt{r_3^2-x^2} - \sqrt{r_2^2-x^2} \, \Theta(r_2^2-x^2)+\sqrt{r_1^2-x^2} \, \Theta(r_1^2-x^2) \bigg).
\end{align}
The resolvent is:
\begin{align*}
G(x)=2 \bigg( x- \sqrt{x^2-r_3^2} + \sqrt{x^2-r_2^2} - \sqrt{x^2-r_1^2} \bigg).
\end{align*}

\paragraph{General YT}

Let YT consist of $m$ rectangles, see Fig. \ref{fig_YT-Discs}.
Then, as in previous examples we find the radii of the concentric rings from \eqref{hrrings}, and we find for the inner and outer radii of the annuli respectively
\begin{equation}
    R_i^2 = \frac{L_{i-1}+K_i}{N}, \qquad r_i^2 = \frac{L_i+K_i}{N},
\end{equation}
with $L_{-1}$ being zero.
Then, we write the density in the phase space as:
\begin{align*}
\rho_\mathcal{D}(\bar{z},z)=\frac{1}{\pi} \bigg[  \sum_{i=1}^{m} \Theta \left(r_i^2-z \bar{z} \right) - \sum_{i=1}^{m} \Theta \left(R_i^2-z \bar{z} \right) \bigg].
\end{align*}
The \(x\)-space density is 
\begin{align*}
\rho(x)=\frac{2}{\pi} \bigg[  \sum_{i=1}^{m}\,\sqrt{r_i^2 -x^2 } \, \Theta(r_i^2-x^2) - \sum_{i=1}^{m}\,\sqrt{R_i^2 -x^2 } \, \Theta(R_i^2-x^2) \bigg].
\end{align*}
The \(x\)-space resolvent is different
for even and odd \(m\):
\begin{align*}
G(x)=\begin{cases}-2 \sum_{i=1}^{m}\,\sqrt{x^2-r_i^2} + 2 \sum_{i=1}^{m}\,\sqrt{x^2-R_i^2} & (\text{even}\,\,m)  \\
2x-2\sum_{i=1}^{m}\,\sqrt{x^2-r_i^2 }+2\sum_{i=1}^{m}\,\sqrt{x^2-R_i^2} & (\text{odd}\,\,m). \\
\end{cases}
\end{align*}

\section{Cubic Potential and Criticality}
\label{cubicApp}
\begin{figure}[H]
    \centering
    \includegraphics[width=\linewidth]{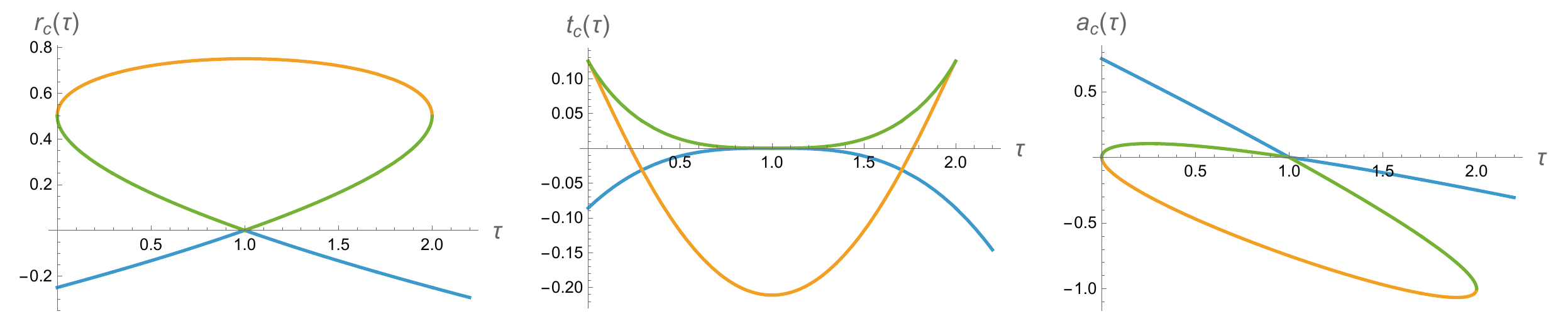}
    \caption{Plots of critical parameters for the (first order) cusp at $z'(1)=0$. We have a one parameter family of criticality.}
    \label{criticalParamsCubic}
\end{figure}

\begin{figure}[H]
    \centering
    \includegraphics[width=0.8\linewidth]{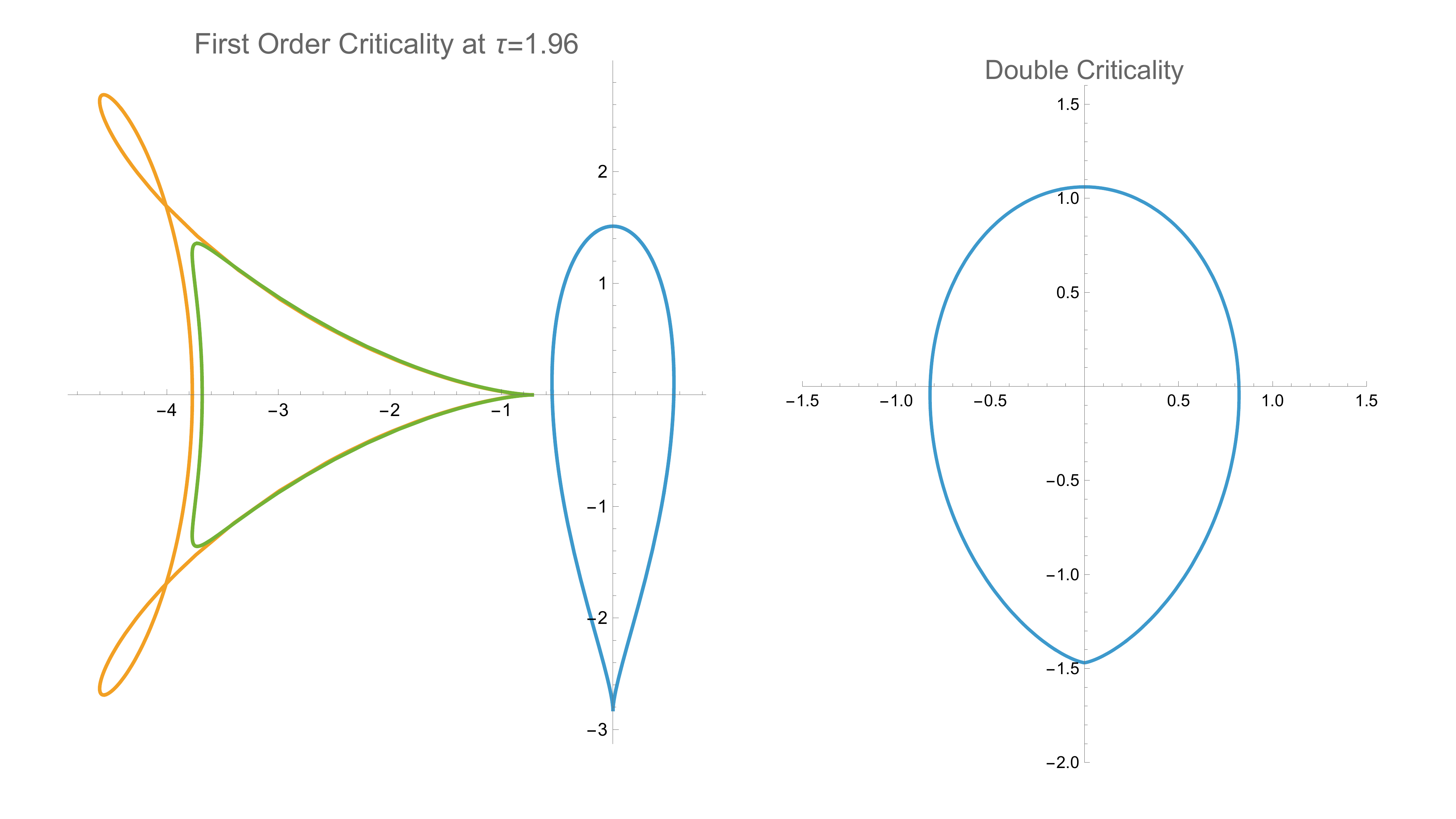}
    \caption{Shapes of  the domain at first order and second order criticality for the cubic potential. Here, we rescaled the domains such that they have unit area. For the first order criticality, the three curves correspond to the three branches of figure \ref{criticalParamsCubic}. Only the green and blue curves make sense physically and are free of self-intersections.}
    \label{fig:criticalShapes}
\end{figure}

Here we study the complex matrix model with cubic potential $V(z)=\frac\tau2z^2+\frac{1}3z^3$. This is the most general cubic potential because we can always translate and rescale to kill the linear term in $z$ and set the coefficient of $z^3$ to $\frac13$. Recall that the coefficients of the potential $t_k$ are given by
\begin{equation}
    t_k = \frac1{2\pi ik}\oint_{\p \mathcal{D}} dz\, \bar z(z) z^{-k}\label{tksCubic}
\end{equation}
A convenient parametrization of the conformal map is as follows
\begin{equation}
    z(w) = r w + a + \frac rw(6a+2\tau)+3\frac{r^2}{w^2}
\end{equation}
In this form, \eqref{tksCubic} is already satisfied for $t_3=\frac13$ and $t_2=\frac\tau 2$. The remaining parameters of the conformal map $a$ and $r$ are fixed by $t_1=0$ and the constraint that $\text{area}=t$,
\begin{equation}
    \begin{aligned}
        0&=a^2+a \left(4 r^2+\tau \right)+2 r^2 \tau -a\\
        t&= r^2 \left(1-(2 a+\tau )^2-2 r^2\right)
    \end{aligned}\label{constraintsCubicPot}
\end{equation}
This implies the following relation between the area $t$ and radius $r$,
\begin{equation*}
    128 r^{10}-124 r^8+4 r^6 ((\tau -2) \tau +9)+r^4 (\tau -3) (\tau -1)^2 (\tau +1)+t \left(64 r^6-28 r^4+2 r^2 (\tau -1)^2\right)+t^2=0
\end{equation*}
Like in the quartic example we considered in the main text, we have criticality when $t'(r)=0$ or equivalently when the curve develops cusps or beaks:
\begin{equation}
    z'(1) = -r (2 a+2 r+\tau -1)=0
\end{equation}
Solving this criticality condition together with \eqref{constraintsCubicPot} gives us three solutions each for the critical parameters $a_c(\tau),r_c(\tau)$ and $t_c(\tau)$. For the sake of brevity, we will not reproduce these expressions here. See figure \ref{criticalParamsCubic} for a plot of these parameters. As seen from figure \ref{fig:criticalShapes}, the orange branch has self-intersections and is super-critical while the green and blue ones are well behaved. Note also that beyond $\tau=2$ there is only one branch with real area.
Along the critical line of $\tau$'s, we can tune further to get to the doubly critical point
\begin{equation}
    z''(1)=2 r (2 a+\tau )+6 r^2=0
\end{equation}
which leads to 
\begin{eqnarray*}
    a_*=1\pm\sqrt 7\,,\quad&&\quad \tau_*=3-2(1\pm\sqrt  7)\\
    r_*=-1\,,\quad&&\quad t_* =-10
\end{eqnarray*}
The algebraic curve, shown in figure \ref{fig:criticalShapes} takes the following simple form at the double critical point
\begin{equation*}
    z(w)=-w+\left(1-\sqrt 7\right)-\frac3w+\frac{1}{w^2}
\end{equation*}
We can compute the free energy from its second derivative \eqref{FreeE} as we did for the quartic potential and they take the form
\begin{equation*}
    \mathcal F_m=\begin{cases}
        \text{reg}(t) + \sqrt{\frac2{t''(r_c)}}\frac8{15r_c}(t_c-t)^{\frac52}+\ldots\qquad \text{for }m=2\\
        \text{reg}(t) + \sqrt[3]{\frac6{t^{(3)}(r_*)}}\frac9{14r_*}(t_*-t)^{\frac73}+\ldots\qquad \text{for }m=3
    \end{cases}
\end{equation*}

\section{Saddle Point Equations for three point function of Rectangles}
\label{ThreeRectanglesAppendix}
In this appendix, we simplify the saddle point equations for the three point function of three equal rectangular YTs with $N$ rows and $K$ columns. This is a review of the manipulations already performed in \cite{Kazakov:2024ald}. We start with the pair of SPEs \eqref{OGspeRect}
\begin{equation*}
    \begin{aligned}
        -y + \frac2\gamma \text{Re}\,g(y) + G(y) &= 0\\
        -x + 2\, \text{Re}\,G(x)+\frac3\gamma g(x)&=0
    \end{aligned}
\end{equation*}
where $g(y)$ and $G(x)$ are the resolvents for the $y$ and $x$ eigenvalues respectively. In the above equation, $\gamma=\frac NK$\footnote{Note that this is the inverse of the $\gamma$ defined in \cite{Kazakov:2024ald}}. The SPE for $y$ can be solved using standard manipulations, assuming there is only one cut in $y$ say in the interval $[\alpha,\beta]$
\begin{equation}
    g(y) = -\frac{\gamma}{2\pi i}\int_\alpha^\beta dz\, \frac{z-G(z)}{y-z}\sqrt{\frac{(y-\alpha)(y-\beta)}{(z-\alpha)(z-\beta)}} 
\end{equation}
The square-roots in this expression can be uniformized by using a Joukowsky map $y(z)$
\begin{equation}
    \begin{aligned}
        y(z) &= \frac{\alpha+\beta}2 + \frac{\alpha-\beta}4\left(z+\frac1z\right)\\
        \frac1\gamma g(y(z)) &= \frac{\alpha+\beta}{4} + \frac{\alpha-\beta}{4 z} -\frac{F(-1)}{\alpha-\beta} \frac{z-1}{z+1} - \frac{F(1)}{\alpha-\beta} \frac{z+1}{z-1} + \frac{4F(\tfrac1z)}{(z^2-1)(\alpha-\beta)}
\end{aligned}
\end{equation}
where we introduced a new resolvent, 
\beq
    F(z)=\int d\xi\, \frac{\rho(y(\xi)) y'(\xi)}{z-\xi}
\eeq
Note that $F(z)$ lives on the double cover, but only has cuts in the region $|z|> 1$, which maps to the first sheet in $y$. Furthermore, since the $x$ and $y$ cuts touch each other as seen in figure \ref{fig:discreteSPESols}, the cuts in $F(z)$ reach $\pm1$. For the one-cut case, we have the cut on $(1,z_0)$ and in the two cut case, we have cuts on $(1,z_1)\cup(-z_2,-1)$.

Now, the asymptotics of $g(y)\sim\frac1y$ implies,
\begin{equation}
    \begin{aligned}
        F(1) &= \frac{1}{16} (\alpha -\beta ) (3 \alpha +\beta )-\frac{1}{\gamma }\\
        F(-1) &= \frac{1}{16} (\alpha -\beta ) (\alpha +3 \beta )+\frac{1}{\gamma }  \label{Fpm1}      
    \end{aligned}
\end{equation}
Plugging in $g(y)$ into the SPE for $x$, we get,
\begin{equation*}
    z \text{ Re}F(z) +\frac1{2z}F\left(\frac1z\right) = \frac{1}{8} (\alpha^2 -\beta^2 ) \left(\frac1z+\frac z2\right)+\frac{1}{16} (\alpha -\beta )^2\left(\frac1{z^2}+\frac{z^2}2\right)-\frac{3}{2\gamma}
\end{equation*}
we can simplify SPE by introducing a new function,
\beq
    \tilde f(z) = z\, F(z)+ \frac{1}{\gamma }-\frac{(\alpha -\beta )^2}{16 z^2}+\frac{\beta ^2-\alpha ^2}{8 z}
\eeq
In terms of this function, the SPE takes the simple form,
\beq
    \tilde f(z+i\epsilon)+\tilde f(z-i\epsilon) = -\tilde f(\tfrac1z)
\eeq
It is convenient to make one final change of variables, 
\beq
    w = \frac{z-1}{z+1}
\eeq
which brings the SPE to a standard form,
\beq
    f(w+i\epsilon)+f(w-i\epsilon) = -f(-w)
\eeq
\section{Three-point function of exponential operators} \label{app-3pt-exp}
\subsection{Solution of the SPE in general case}

If we define the free energy as
\begin{equation}
    e^{N F} \equiv \int dA \,  e^{N \, \tr (XA - \frac{1}{2\alpha^2} A^2 -\frac{{g}}{3\alpha^3} A^3)},
\end{equation}
using loop equations we can write it (up to an arbitrary $g$-dependent function) in terms of the density of eigenvalues of $X$ $\rho(x)$ (which has a support in the interval $[a;b]$) as follows \cite{Gross:1991aj}:
\begin{multline}
    \frac{F}{N} = -\frac{1}{2} \int dx dy \rho(x) \rho(y) \log \big( \sqrt{\alpha x+c}+\sqrt{\alpha y+c} \, \big) - \frac{\alpha}{2{g}} \int dx \rho(x) x + \frac{2}{3 {g}^{1/2}} \int dx \rho(x) (\alpha x+c)^{\frac{3}{2}} + \\+\frac{1}{{g}^{1/2}} \bigg( \frac{1}{4{g}}-c  \bigg) \int dx \rho(x) \sqrt{\alpha x+c}  - \frac{1}{4}\log {g}  - \frac{1}{12{g}^2},
\end{multline}
where we have picked the integration constant to match with the $g \to 0$ limit. The constant $c$ is fixed from the condition\footnote{Note that this condition implies that the integral $\int dy\, {\rho(y)}/{\sqrt{\alpha y+c}}$ is a constant as a functional of $\rho(x)$.}
\begin{equation}
    c = \frac{1}{4{g}} - {g}^{\frac{1}{2}} \int dy \frac{\rho(y)}{\sqrt{\alpha y+c}}.
\end{equation}

The saddle-point equation is
\begin{equation}
    x = 2 \fint dy \frac{\rho(y)}{x-y} - \frac{3\alpha}{2g} + \frac{3\alpha }{ \sqrt{g}} \sqrt{\alpha x+c} + \frac{3\alpha}{2} \int dy \frac{\rho(y)}{\sqrt{\alpha y+c}} \frac{1}{\sqrt{\alpha x+c}+\sqrt{\alpha y+c}} .
\end{equation}
Changing the variables as
\begin{equation}
    p = \sqrt{\alpha y+c}, \qquad \pi(p) = \rho(y(p)),
\end{equation}
and defining
\begin{equation}
    G_0(z) \equiv \int^B_A dp \frac{\pi(p)}{z-p}, \qquad z = \sqrt{\alpha x+c},
\end{equation}
the SPE is rewritten as
\begin{equation}
    G_0(z+i0)+G_0(z-i0) = G_0(-z) + V'(z),
\end{equation}
where we define
\begin{equation}
    V'(z) \equiv \frac{z^2-c}{\alpha} + \frac{3\alpha}{2g}- \frac{3\alpha z}{\sqrt{g}}.
\end{equation}
Note that the normalization condition for $\rho(x)$ is rewritten as
\begin{equation} \label{exp_norm}
    \int dp \, \pi(p) \, p = \frac{\alpha}{2}. 
\end{equation}
The SPE is the same as for the $O(n)$ model, solution for which was given in \cite{Eynard:1995zv,Eynard:1992cn}.
Next, we split the resolvent in regular and non-analytic part
\begin{equation}
    G_0(z) = G_r(z) + G(z),
\end{equation}
where the regular part is
\begin{equation}
    G_r(z) = \frac{1}{3} (2V'(z)+V'(-z))=\frac{z^2}{\alpha}-\frac{\alpha z}{\sqrt{g}} + \frac{3\alpha}{2g} - \frac{c}{\alpha},
\end{equation}
and the non-analytic part satisfies the homogeneous equation
\begin{equation}
    G(z+i0)+G(z-i0) = G(-z),
\end{equation}
and it has asymptotic behavior
\begin{equation} \label{exp_asympt}
    G(z) \to -  \frac{z^2}{\alpha} + \frac{\alpha z}{ \sqrt{{g}}} +  \frac{c}{\alpha}-\frac{3\alpha}{2{g}}  +  \frac{\alpha}{2 \sqrt{{g}}}  \bigg( \frac{1}{4{g}}-c \bigg) \frac{1}{z} + \frac{\alpha}{2 z^2}  + O(z^{-3}), \quad z \to \infty.
\end{equation}

We define
\begin{equation}
    \begin{aligned}
        G_+(z) = \frac{i}{\sqrt{3}} \big[  e^{\frac{i\pi}{6}} G(z) - e^{\frac{-i\pi}{6}} G(-z) \big],\\
        G_-(z) = -\frac{i}{\sqrt{3}} \big[  e^{-\frac{i\pi}{6}} G(z) - e^{\frac{i\pi}{6}} G(-z) \big],
    \end{aligned}
\end{equation}
with the inverse
\begin{equation}
    G(z) = - ( e^{\frac{i \pi}{6}} G_+(z) + e^{-\frac{i \pi}{6}} G_-(z)).
\end{equation}
The SPE is satisfied if
\begin{equation}
    G_{\pm} (z-i0) = e^{\pm \frac{i\pi}{3}} G_{\mp} (z+i0), \qquad z\in [A,B].
\end{equation}
We can write the solution in the following form \eqref{non_crit_sol}. We parametrize the polynomials as
\begin{equation}
    S(z) = s_0 + s_2 z^2 + z^4, \qquad R(z) = r_1 z + r_3 z^3 + r_5 z^5.
\end{equation}
The requirement $G_+(z) G_-(z)$ to be polynomial:
\begin{equation}
    S(z)^2 (z^2-A^2)(z^2-B^2)+R(z)^2 = (z^2-\lambda^2)^3 (z^2-\mu^2)^3,
\end{equation}
together with the requirement for the solution \eqref{non_crit_sol} to have the correct asymptotic behavior  give 4 additional constraints for 10 unknown parameters $\lambda,\mu,A,B,s_0,s_2,r_1,r_3,r_5,c$ in terms of $\alpha, {g}$.

\subsection{Critical solution}
Let us repeat the solution in the critical case 
\begin{equation}
    \begin{aligned}
    G_0(z)= G(z)+\frac{z^2}{\alpha}-\frac{\alpha z}{\sqrt{g}} + \frac{3\alpha}{2g} - \frac{c}{\alpha}, \\
    G(z) = - ( e^{\frac{i \pi}{6}} G_+(z) + e^{-\frac{i \pi}{6}} G_-(z)), \\
    G_{\pm} (z)= \bigg(\sqrt{1-\frac{B^2}{z^2}} \pm \frac{iB}{z}  \bigg)^{-\frac{2}{3}} \bigg[ \frac{z^2}{\sqrt{3}\alpha}   \sqrt{1-\frac{B^2}{z^2}} \pm  i B \, b_2 z   \bigg].
    \end{aligned}
\end{equation}
Then, matching with the asymptotics \eqref{exp_asympt}, we obtain the following set of equations:
\begin{align}
    B \left(b_2-\frac{2}{3 \sqrt{3} \alpha }\right)=\frac{\alpha }{\sqrt{g}}, \nonumber\\
    {B^2 \left(13-12 \sqrt{3} \alpha  b_2\right)}={18c}-\frac{27 \alpha^2 }{g}, \nonumber\\
    { B^3 \left(11 \sqrt{3}-27 \alpha  b_2 \right)}=\frac{243\alpha^2 }{4 \sqrt{g}}\left(\frac{1}{4 g}-c\right), \nonumber\\
    {5 B^4 \left(31-24 \sqrt{3} \alpha  b_2 \right)}={972\alpha^2 }.
\end{align}
These 4 equations relate 4 unknowns $b_2,b,c,g$ to $\alpha$. Note that since we are looking into critical solution, these equations define a critical line $g=g_c(\alpha)$.

Solving this system of equations, we obtain the following solution:
\begin{equation}
    b_2 = \frac{2 \left(4 \sqrt{3} \alpha ^5+6 \sqrt{3} \alpha ^3-\sqrt{3} \alpha \right)}{9 \left(6 \alpha ^4-\alpha ^2\right)}-\frac{\sqrt[3]{2} Q}{3\ 3^{2/3} \left(6 \alpha ^4-\alpha ^2\right)}-\frac{2^{2/3} \left(32 \alpha ^{10}+18 \alpha ^8-3 \alpha ^6\right)}{3 \sqrt[3]{3} \left(6 \alpha ^4-\alpha ^2\right) Q},
\end{equation}
where
\begin{multline}
    Q \equiv \big( -256 \sqrt{3} \alpha^{15}-216 \sqrt{3} \alpha^{13}+756 \sqrt{3} \alpha^{11}-240 \sqrt{3} \alpha ^9+20 \sqrt{3} \alpha^7+\\ \sqrt{2} \sqrt{-576288 \alpha^{26}-291960 \alpha^{24}+1003620 \alpha^{22}-558738 \alpha^{20}+131841 \alpha^{18}-14400 \alpha^{16}+600 \alpha^{14} } \big)^{1/3},
\end{multline}
and the critical line is
\begin{equation}
    g_c(\alpha) = \frac{\sqrt{15} \sqrt{24 \sqrt{3} \alpha ^7 b_2-31 \alpha ^6}}{2 \sqrt{-729 \alpha ^4 b_2^4+648 \sqrt{3} \alpha ^3 b_2^3-648 \alpha ^2 b_2^2+96 \sqrt{3} \alpha  b_2-16}}.
\end{equation}

The density $\pi(z)$ for the critical solution is
\begin{multline}
    \pi(z) = \frac{\sqrt{3}}{2\pi} \bigg[ \bigg( \frac{B}{z} + \frac{\sqrt{(B-z)(z+B)}}{z} \bigg)^{-2/3} \bigg( \frac{z}{\sqrt{3}\alpha} \sqrt{(B-z)(z+B)} +  B b_2 z    \bigg)  -\\- \bigg( \frac{B}{z} - \frac{\sqrt{(B-z)(z+B)}}{z} \bigg)^{-2/3} \bigg( -\frac{z}{\sqrt{3}\alpha} \sqrt{(B-z)(z+B)} +  B b_2 z    \bigg) \bigg].
\end{multline}

\section{\(\gamma\)-deformed \(\mathcal{N}=4\) SYM action}
\label{sec:gammaSYM}


In \(\mathcal{N}=1\) superspace the theory can be written in terms of one real vector superfield \(V\) and three chiral superfields \(\Phi_i\) (with \(i=1,2,3\)). The gamma deformation modifies only the superpotential by introducing phase factors into the products of the chiral superfields. One defines a “\(q\)–commutator” by
\begin{align*}
[\Phi_i,\Phi_j]_q \equiv q\,\Phi_i\,\Phi_j -
 q^{-1}\,\Phi_j\,\Phi_i\,.
\end{align*}
In the \(\gamma\)-deformed action we replace the usual (matrix) product of  the superfields by a star product defined by
\begin{align*}
\Phi_i\star\Phi_j\equiv e^{i\pi\epsilon^{ijk}\gamma_{k}}\,\Phi_i\,\Phi_j\,.
\end{align*}
Thus, the deformed superpotential becomes
\begin{align*}
W_\gamma(\Phi) = \sqrt{2}\,g\,\operatorname{Tr}\Bigl( \,\Phi_1\,\star \Phi_2\,\star \Phi_3 -\,\Phi_1\,\star \Phi_3\,\star \Phi_2\Bigr)\,.
\end{align*}
The full gamma–deformed action is then
\begin{align}\label{N=4twisted}
S_\gamma = \frac{1}{g^2}\Bigg\{ 
&\int d^4x\, d^2\theta\, \operatorname{Tr}\Bigl( \mathcal{W}^\alpha \mathcal{W}_\alpha\Bigr)
+ \int d^4x\, d^4\theta\, \sum_{i=1}^{3}
\operatorname{Tr}\Bigl( e^{-V}\,\bar{\Phi}_i\,e^{V}\,\Phi_i \Bigr)\notag\\
&+\int d^4x\, d^2\theta\, \left(W_\gamma(\Phi) + \text{h.c.}\right)
\Bigg\}\,
\end{align}
Here the chiral field strength \(W_\alpha\) is defined as
\begin{align*}
\mathcal{W}_\alpha = -\frac{1}{4}\,\bar{D}^2\Bigl(e^{-V}\,D_\alpha\,e^{V}\Bigr)\,.
\end{align*}





\section{Integrability for $\frac14$-BPS correlator}
\label{sec:Integrability1/4BPS}

Let us study the classical integrability properties of the action in matrix integral~\eqref{2pointf} which we write in the form (up to the trivial constant \(\tr\mathcal{P} \bar{\mathcal{P}}\)) as follows
\begin{align*}
S=\frac{1}{2}\tr\Big( [\mathcal{P},g]\,[\bar{\mathcal{P}}g^{\dagger }]+[\mathcal{P},g^{\dagger }]\,[\bar{\mathcal{P}}g]\,\Big),\qquad \mathcal{P}=P_{1}+iP_{2}\,,\quad \bar{\mathcal{P}}=P_{1}-iP_{2}\,.
\end{align*}
We remind that   \([\mathcal{P}, \bar{\mathcal{P}}]=0\).

Introduce the current
\begin{align*}
 \,\, j
  =g^{\dagger}[\mathcal{P},g],\qquad \bar{j}
  =g^{\dagger}[\bar{\mathcal{P}},g]. \,
\end{align*}
The e.o.m. is
\begin{align*}
\,\,\,
 [\bar{\mathcal{P}},j ]+[\mathcal{P},\bar{j}
  ]=0
\end{align*}
The unitarity   \(g^{\dagger}g=\mathbb{I}\) imposes:
\begin{align*}
\,\, \epsilon_{\mu\nu}
[p_{\mu}+g^{\dagger}[p_{\mu},g],p_{\nu}+g^{\dagger}[p_{\nu},g]]=0
\end{align*}

Both relations can be packed into the ``classical" integrability statement -- the existence of Lax pair (in complex notations):
\begin{align*} J(u)=\frac{j}{1+u},\qquad  \bar J(u)=\frac{\bar j}{1-u} \end{align*}
Conservation laws can be encoded into the ``zero curvature" condition:
\begin{align*}i[\bar{\mathcal{P}},J(u) ]-i[\mathcal{P},\bar{J}(u)]-[J(u),
\bar J(u)]=0
\end{align*}
It is not clear whether this classical integrability  can be promoted to the full quantum integrability, which would help to compute interesting physical quantities for $\frac{1}{4}$-BPS physics.
But it might be useful for computing these quantities in the classical limit \(\mathcal{P}\to\infty\).

\newpage

\bibliographystyle{JHEP}
\bibliography{BPS_MM}

\end{document}

%% file: tikz/YDblocks.tex
\usetikzlibrary{matrix, positioning, arrows.meta, calc}

\begin{tikzpicture}[scale=0.6, every node/.style={scale=0.6}, xshift=5cm,  every left delimiter/.style={xshift=-2.em, scale=1.6}, every right delimiter/.style={xshift=+2.em, scale=1.6},add paren/.style={
      left delimiter={(},
      right delimiter={)}, 
    },trim left=-1.5cm]
 \matrix[matrix of math nodes, 
       add paren, 
        row sep=0.1 cm, column sep=0.1cm,
        inner sep=0pt,
        nodes={minimum size=0.8cm}] (m) {
  x & L_3+1  &   &   &   &   &   & &\\
   1 &  x & L_3+2  &   &   &   &   & &\\
    &   1&  x & L_3+3  &   &   &   & &\\
    &   &   1&  x &   &   &   & &\\
    &   &   &   &  x &L_2+1   &   & &\\
    &   &   &   &   1&  x &  L_2+2 & &\\
    &   &   &   &   &   1&  x & &\\
    &   &   &   &   &   &   &x&L_1+1 \\
    &   &   &   &   &   &   & 1& x \\
};

  \draw[red, thick, rounded corners] 
    ($(m-1-1.north west)+(-0.2,0.2)$) rectangle ($(m-4-4.south east)+(0.2,-0.2)$);
  \draw[blue, thick, rounded corners] 
    ($(m-5-5.north west)+(-0.2,0.2)$) rectangle ($(m-7-7.south east)+(0.2,-0.2)$);
  \draw[green!60!black, thick, rounded corners] 
    ($(m-8-8.north west)+(-0.2,0.2)$) rectangle ($(m-9-9.south east)+(0.2,-0.2)$);

  \coordinate (B1) at ($(m-2-2.east)+(0.4,-.4)$);
  \coordinate (B2) at ($(m-6-6.east)+(0.4,-.4)$);
  \coordinate (B3) at ($(m-8-8.east)+(0.4,-.4)$);

  \begin{scope}[xshift=10cm, yshift=5cm]

    \foreach \i in {0,1,2,3} {
      \foreach \j in {0,...,8}
        \node[draw=red, thick, minimum size=1cm, anchor=north west] at (\j, -\i) {};
    }

    \foreach \i in {4,5,6} {
      \foreach \j in {0,...,5}
        \node[draw=blue, thick, minimum size=1cm, anchor=north west] at (\j, -\i) {};
    }

  \foreach \i in {7,8} {
      \foreach \j in {0,...,2}
        \node[draw=green!60!black, thick, minimum size=1cm, anchor=north west] at (\j, -\i) {};
    }
    \coordinate (Y1) at (2.5, -1.5);  
    \coordinate (Y2) at (1.5, -5.5);  
    \coordinate (Y3) at (0.5, -7.5);  

  \end{scope}

  \draw[<->, thick, red] (B1) -- (Y1);
  \draw[<->, thick, blue] (B2) -- (Y2);
  \draw[<->, thick, green!60!black] (B3) -- (Y3);

\end{tikzpicture}